\documentclass[12pt,a4paper]{article}
\pdfoutput=1
\usepackage{jheppub}
\usepackage{setspace,verbatim} 
\usepackage{soul}
\usepackage[normalem]{ulem}
\usepackage{amsmath,amsfonts,amssymb,amsthm,graphicx,color}
\usepackage{verbatim}
\usepackage{sidecap}
\usepackage{enumerate}
\numberwithin{equation}{section}
\usepackage{bbm,bm}
\usepackage{braket,mathrsfs}
\usepackage{xcolor}
\usepackage{framed}
\usepackage{tikz}
\usetikzlibrary{decorations.pathmorphing,arrows.meta,bending,snakes}
\usepackage{caption}
\usepackage{subcaption}
\usepackage{dsfont}

\usepackage[normalem]{ulem}
\newcommand{\stkout}[1]{\ifmmode\text{\sout{\ensuremath{#1}}}\else\sout{#1}\fi}
\definecolor{darkgreen}{rgb}{0,0.4,0}
\definecolor{darkred}{rgb}{0.4,0,0}
\definecolor{darkblue}{rgb}{0,0,0.4}

\def\undertilde#1{\mathord{\vtop{\ialign{##\crcr
$\hfil\displaystyle{#1}\hfil$\crcr\noalign{\kern1.5pt\nointerlineskip}
$\hfil\tilde{}\hfil$\crcr\noalign{\kern1.5pt}}}}}

\usepackage[framemethod=TikZ]{mdframed}
\definecolor{darkcerulean}{rgb}{0.03, 0.27, 0.49}
	
\newcounter{ex}[section]

\makeatletter
\def\@xfootnote[#1]{%
  \protected@xdef\@thefnmark{#1}%
  \@footnotemark\@footnotetext}
\makeatother

\newcommand{\eb}{\bar{\epsilon}}
\newcommand{\hb}{\bar{h}}
\newcommand{\db}{\bar{\delta}}
\newcommand{\bb}{\bar{\beta}}
\newcommand{\tb}{\bar{t}}
\newcommand{\sinc}{\text{sinc}}

\usepackage[most]{tcolorbox}

\tcbset{colback=yellow!10!white, colframe=red!50!black, 
        highlight math style= {enhanced, 
            colframe=red,colback=red!10!white,boxsep=0pt}
        }

\begin{document}
\title{Tauberian-Cardy formula with spin}
\author[1,2]{Sridip Pal}
\author[1]{\& Zhengdi Sun}

\affiliation[1]{Department of Physics, 
University of California, San Diego\\
La Jolla, CA 92093, USA}

\affiliation[2]{School of Natural Sciences, Institute for Advanced Study\\ Princeton, NJ 08540, USA}

\emailAdd{sridip@ias.edu}
\emailAdd{z5sun@ucsd.edu}

\abstract
{We prove a $2$ dimensional Tauberian theorem in context of $2$ dimensional conformal field theory. The asymptotic density of states with conformal weight $(h,\hb)\to (\infty,\infty)$ for any arbitrary spin is derived using the theorem. We further rigorously show that the error term is controlled by the twist parameter and insensitive to spin. The sensitivity of the leading piece towards spin is discussed. We identify a universal piece in microcanonical entropy when the averaging window is large. An asymptotic spectral gap on $(h,\hb)$ plane, hence the asymptotic twist gap is derived. We prove an universal inequality stating that in a compact unitary $2$D CFT without any conserved current $Ag\leq \frac{\pi(c-1)r^2}{24}$ is satisfied, where $g$ is the twist gap over vacuum and $A$ is the minimal ``areal gap", generalizing the minimal gap in dimension to $(h',\hb')$ plane and $r=\frac{4\sqrt{3}}{\pi}\simeq 2.21$. We investigate density of states in the regime where spin is parametrically larger than twist with both going to infinity. Moreover, the large central charge regime is studied. We also probe finite twist, large spin behavior of density of states.}

\maketitle
%


\section{Summary \& Discussion}

The Cardy formula \cite{cardy1986operator} for the asymptotic density of states has recently been rigorously derived with an estimate for the error term in \cite{Baur,Ganguly:2019ksp}. A natural question is to ask whether one can generalize the formalism so as to make it sensitive to the spin or equivalently to the conformal weights $h,\hb$ separately. This necessitates working out a $2$ dimensional Tauberian theorem, which we achieve here. The motivations for investigating Cardy formula on $(h',\hb')$ plane are several. First of all, the notion of infinity on a $2$d plane is richer than $\Delta\to\infty$ limit. We will see that the finer details of the Cardy formula actually depend on how infinity is approached unless one makes extra assumption about the spectrum. Furthermore, there have been interesting developments in the direction of lightcone bootstrap in recent times \cite{Kusuki:2018wpa,Kusuki:2019gjs,Collier:2018exn,Maxfield:2019hdt,Benjamin:2019stq}, our analysis puts some of these results on rigorous footing. Another amazing feature is the ability to investigate the ``areal" notion of spectral gap. If we probe the $(h',\hb')$ plane with circular areas of radius $R$, centered at $(h,\hb)$, then we find the optimal value of $R$ which guarantees that the area contains at least one state. Again unless we put in extra assumption, the value of $R$ depends on how infinity is approached and thus showing a richer asymptotic behavior. If we one assumes existence of twist gap, it turns out that the twist gap is complementary to asymptotic spectral gap in some sense, which we will make precise in due course. \\

The naive Cardy like analysis provides us with an expression for the asymptotic density of states where $h$ and $\hb$ are of the same order. One can re-express this as a function of dimension $\Delta$ and spin $J$ with $\Delta \simeq J$. Now a natural question is to ask whether the result is valid when $\Delta$ and $J$ is not of the same order. For example, in the large charge expansion literature \cite{hellerman2015cft,hellerman2017note,monin2017semiclassics,Cuomo:2017vzg,Cuomo:2019ejv,Banerjee:2019jpw,Orlando:2019hte,Favrod:2018xov,Kravec:2018qnu,Kravec:2019djc}, the regime where $J \simeq \Delta^{1/n}$ with $n>1$ is being probed. It turns out that only a part of the answer coming from the naive Cardy like analysis is meaningful while the rest of it is comparable to the error term. We emphasize that the analysis is only possible because now we have a rigorous estimate of the error term due to the Tauberian theorem that we prove in this paper.\\

With our rigorous treatment, it is possible to address issues regarding whether we can trust the naive Cardy formula when $h$ and $\hb$ are not of the same order. It turns out that the answer to this question is intimately connected with the existence of twist gap. We show that we can trust the naive Cardy formula for all the operators when $\text{max}(h,\hb)=\text{min}(h,\hb)^{\Upsilon}$ with $1\leq \Upsilon <2$. It is also shown that with the assumption of twist gap, the validity of Cardy formula for primaries for $c>1$ CFTs does not require any restriction on $\Upsilon$. \\

The another motivation for taking up a rigorous study of Cardy formula is to be able to probe the large central charge ($c$) sector, to be specific, to derive the density of states when $h/c,\hb/c$ are finite but  $c$ is very large. This part is in the spirit of result derived in \cite{HKS}. A nice feature that reveals itself through the rigorous treatment is a curious connection between validity of Cardy regime and the twist gap above the vacuum. These features are important in the context of holography.\\

The plan of the paper is to quote the main results here in the beginning and discuss its consequences in terms of CFT data, such that the current section can be thought of as mostly self contained. The next section \S\ref{setup} gives some intuitive understanding of the technical stuff that follows. From \S\ref{lemma0} onwards, we plunge into technical proofs with a healthy relaxing intermission in \S\ref{verify}, where we numerically verify our results. For readers going for a really quick ride, we have highlighted the main equations and results in what follows.

\subsection{Integrated density of states}
We prove a $2$ dimensional Tauberian theorem in context of $2$ dimensional conformal field theory. The asymptotic density of states with conformal weight $(h,\hb)\to (\infty,\infty)$ is derived using the theorem. We find that the error term is controlled by the twist parameter. We note that as $(h,\hb)\to (\infty,\infty)$, the twist also goes to $\infty$. We remark that the regime of validity depends on whether we put in the assumption of having a twist gap.\\

\textit{\label{def}Definition: by finite twist gap, we mean there exists a number $\tau_*>0$ such that there is no operator with twist $\tau \in (0,\tau_*)$ and there are finite number of zero twist operators\footnote{Usually, by finite twist gap, it is assumed that there is no zero twist primaries except the Identity. Here we are using it in a slightly different manner, so one needs to be careful about using bounds on twist gap, such as the one appearing in \cite{Collier:2016cls}.} with dimension less than $c/12$.}\\

We make two remarks: a) the fact that there are finite number of zero twist operators with dimension less than $c/12$ is always true since there are finite number of operators with dimension less than $c/12$ for finite central charge, b)  Not having any operator with twist $\tau \in (0,\tau_*)$ disallows having $0$ as twist accumulation point.

\subsubsection{Main theorems on integrated density of states}
\paragraph{No assumption on twist gap:}
We show that for finite central charge $c$, the number of states with conformal weights less than or equal to some specified large conformal weight $h,\hb$ is given by:
\begin{equation}\label{masterequation}
\tcboxmath{\begin{aligned}
F(h,\hb)&\equiv \int_{0}^{h}\text{d}h' \int_{0}^{\hb}\text{d}\hb' \rho(h',\hb')\\
&\underset{\overset{h/\hb=O(1)}{h,\hb \to \infty}}{=}\frac{1}{4\pi^2}\left(\frac{36}{c^2h\hb}\right)^{1/4}\exp\left[2\pi\left(\sqrt{\frac{ch}{6}}+\sqrt{\frac{c\hb}{6}}\right)\right]\left[1+O\left(\tau^{-1/4}\right)\right]\,.
\end{aligned}}
\end{equation}
where $\tau$ is the twist of the state with $h,\hb$ and given by $\tau=2\text{min}\{h,\hb\}$. Here we have assumed that $h/\hb=O(1)$ number\footnote{If we say $f =O(1)$, we mean $|f|<M$ for a fixed positive number $M$.}. As a result one could have written the error term as $O\left(h^{-1/4}\right)$ or $O\left(\hb^{-1/4}\right)$.\\

\paragraph{Assuming a twist gap:}
It turns out that if we assume a finite twist gap, we can trust eq.~\eqref{masterequation} even when $h$ and $\hb$ are not of the same order but $h=\hb^{\upsilon}$ with $1/2<\upsilon<2$. In such a scenario, the error term becomes $O(\tau^{\frac{\Upsilon}{4}-1/2})$, where $\Upsilon=\text{max}\left(\upsilon,1/\upsilon\right)$. The $\Upsilon$ characterizes how $h$ and $\hb$ are of different order asymptotically in a symmetrized fashion, for example, if we approach the infinity along the curve $h=\hb^{1.1}$ or $\hb=h^{1.1}$, we have $\Upsilon=1.1$. Thus our error estimation is symmetric if we reflect the line of approach to infinity about $h=\hb$ line.\\

We have for $1\leq \Upsilon <2$,
\begin{equation}\label{masterequationupsilon}
\tcboxmath{\begin{aligned}
&F(h,\hb)\equiv \int_{0}^{h}\text{d}h' \int_{0}^{\hb}\text{d}\hb' \rho(h',\hb')\\
&\!\!\!\!\!\!\underset{\underset{\tfrac{1}{2}<\tfrac{\log h}{\log \hb}<2}{h,\hb \to \infty}}{=}\frac{1}{4\pi^2}\left(\frac{36}{c^2h\hb}\right)^{1/4}\exp\left[2\pi\left(\sqrt{\frac{ch}{6}}+\sqrt{\frac{c\hb}{6}}\right)\right]\left[1+O\left(\tau^{\frac{\Upsilon}{4}-1/2}\right)\right]\,,\quad \Upsilon<2\,.
\end{aligned}}
\end{equation}

The eq. \eqref{masterequation} and eq. \eqref{masterequationupsilon} are two of the central results obtained in this paper.  If $h$ and $\hb$ are not of the same order, we basically probe the large spin sector of density of states\footnote{A cautionary remark is that here in this paper unless otherwise mentioned, the twist is \textbf{NOT} kept finite while taking this limit. This can be contrasted to the scenario in the usual large spin expansion \cite{Alday:2007mf}, where one keeps the twist finite.}, to be precise, the regime where spin is parametrically larger than the twist but both goes to infinity.\\

The basic structure of both the eq. \eqref{masterequation} and eq. \eqref{masterequationupsilon} is that they have leading exponential piece multiplied with a subleading polynomial suppression. The error term is then further suppressed by a polynomial piece. Now if $\Upsilon\geq 2$, one can see the error term in \eqref{masterequationupsilon} is not really suppressed, hence is not in fact an error term. Thus we can not trust the polynomially suppressed terms. In this regime, we are able to show that
\begin{equation}\label{largespin}
\tcboxmath{\begin{aligned}
F(h,\hb)\underset{h,\hb\to\infty}{=}& \exp\left[2\pi\sqrt{\frac{ch}{6}}+2\pi\sqrt{\frac{c\hb}{6}}\right]O\left(\tau^{-3/4}\right)\,,\quad \Upsilon\geq 2\,.
\end{aligned}}
\end{equation}

We further remark that for CFTs where the partition function nicely factorizes into holomorphic and antiholomorphic pieces, the leading result directly follows from the analogous result for large $\Delta=h+\hb$, proven in \cite{Baur}, nonetheless the error term in analogues of eq.~\eqref{masterequation} and eq.~\eqref{masterequationupsilon} goes like $O(h^{-1/2})$, hence, in such a case, we have more control over the approximation.

\subsubsection{Corollaries of the theorems [Eq.~\eqref{masterequation} and Eq.~\eqref{masterequationupsilon}] on integrated density of states}
Below we will digress a bit and touch upon some of the interesting results that can be extracted from the above before coming back to summazing our main results in the next subsection \S\ref{primarydurga}.
\paragraph{Rich structure of asymptotic approach:}
The integrated density of states show distinct leading behavior depending on how the asymptotic infinity is approached. In \cite{Baur}, it has been shown that as $\Delta\to\infty$, we have
\begin{equation}
\begin{aligned}
F^{\text{MZ}}(\Delta)&\equiv \int_{0}^{\Delta}\text{d}\Delta'  \rho(\Delta^\prime)\\
&\underset{\Delta\to\infty}{=}\frac{1}{2\pi}\left(\frac{3}{c\Delta}\right)^{1/4}\exp\left[2\pi\sqrt{\frac{c\Delta}{3}}\right]\left[1+O\left(\Delta^{-1/2}\right)\right]\,.
\end{aligned}
\end{equation}

\begin{figure}[!ht]
\centering
\includegraphics[scale=0.5]{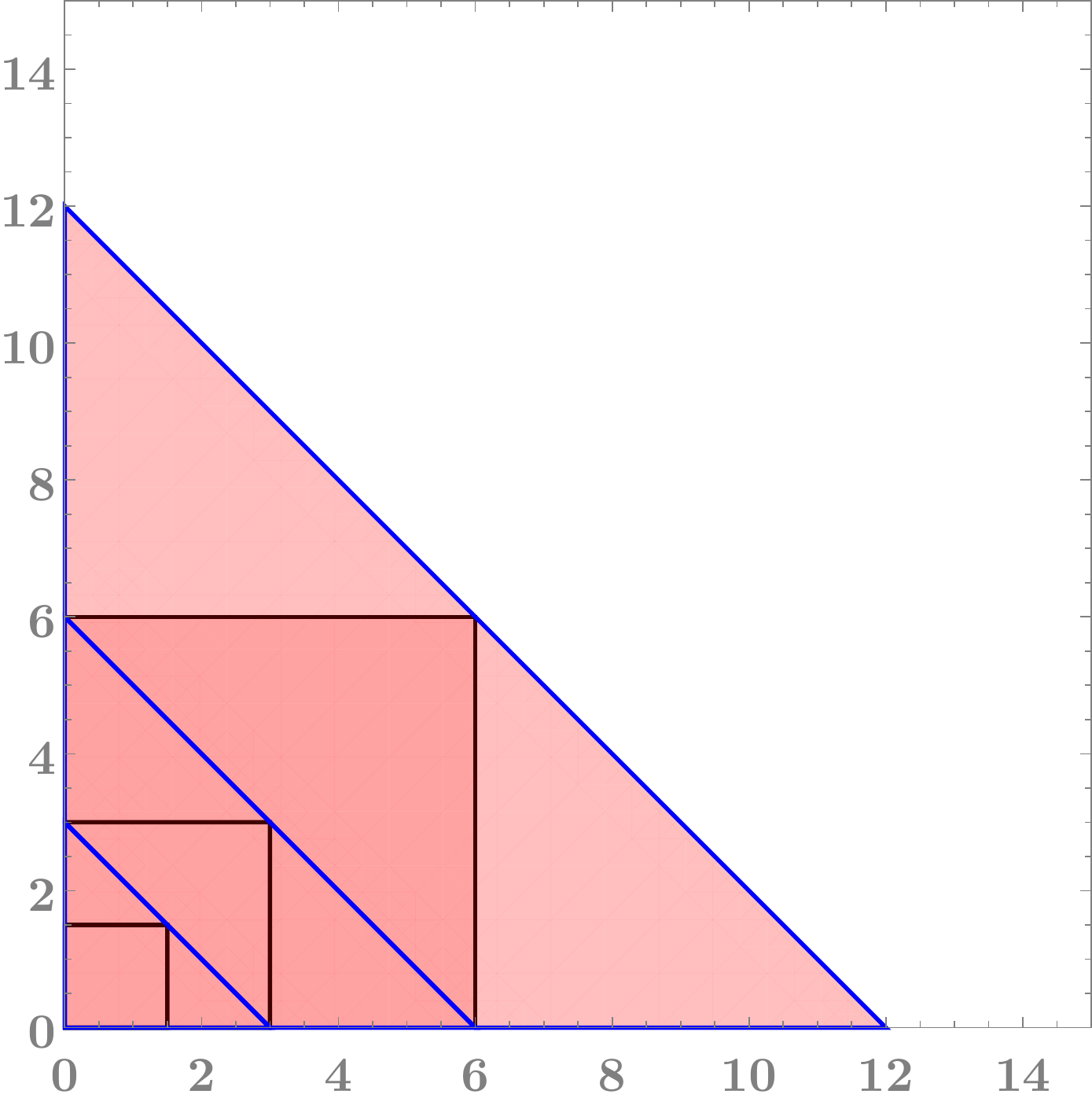}
\caption{Approaching to infinity on $(h',\hb')$ plane: The blue lines denote how the number of operators with $(h',\hb')$ is counted such that $h'+\hb'\leq \Delta$ and then we let $\Delta\to\infty$, this is given by the function $F^{\text{MZ}}(\Delta)$, originally calculated in \cite{Baur}. On the other hand, the black lines denote how the number of operator is counted upto some value of $h=\hb=\Delta/2$ i.e. $h'\leq\Delta/2,\hb'\leq\Delta/2$ and then we let $\Delta\to\infty$. This approach to infinity is captured by the function $F(\Delta/2,\Delta/2)$, calculated in this paper. We see that the square of size $\Delta/2$ with one vertex at origin and another one at $(\Delta/2,\Delta/2)$ are always contained within the rightangled triangular region, created by $h'$ axis, $\hb'$ axis and $h'+\hb'=\Delta$ line. This is consistent with the observation that leading behavior of $F(\Delta/2,\Delta/2)$ is suppressed compared to $F^{\text{MZ}}(\Delta)$.}
\label{approachtoinfinity}
\end{figure}

\begin{figure}[!ht]
\centering
\includegraphics[scale=0.5]{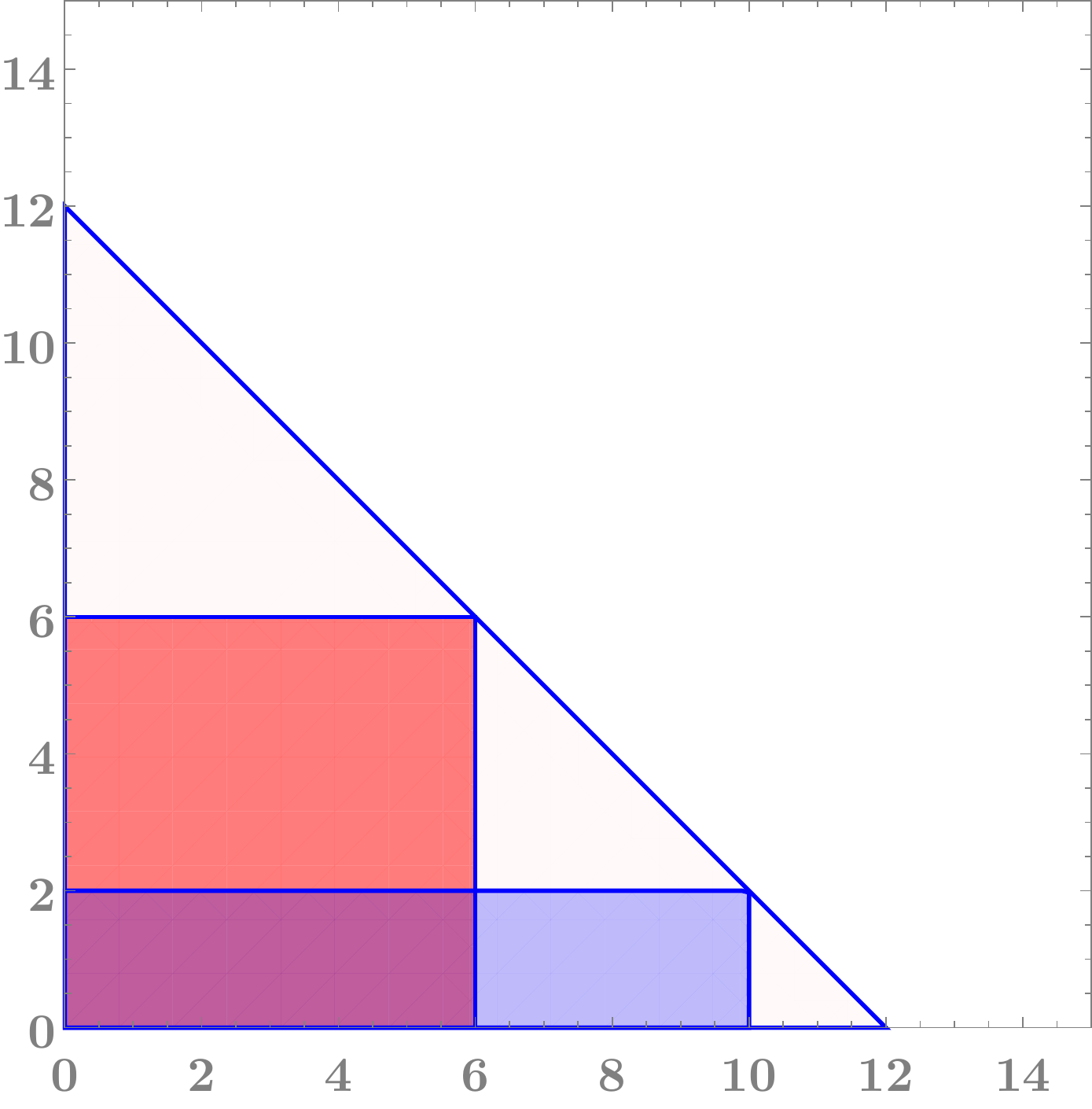}
\caption{$(h',\hb')$ plane : asymptotically, the rectangular region (blue shaded) contains exponentially less number of operators compared to square region (red shaded). They are contained within the rightangled triangle, created by $h'$ axis, $\hb'$ axis and $h'+\hb'=\Delta$ line. Here $\Delta=12$.}
\label{squarevsrectangle}
\end{figure}

We remark that in the asymptotic limit, both $F^{\text{MZ}}(\Delta\to\infty)$ and $F(h\to\infty,\hb\to\infty)$ count the total number of operators. But these functions approach infinity in a different manner (see the figure~\ref{approachtoinfinity}). 
To be concrete, let us choose $h=\hb=\Delta/2$, thus we have
\begin{equation}
\begin{aligned}
F(\Delta/2,\Delta/2)&\underset{\Delta\to\infty}{=}\frac{1}{2\pi^2}\left(\frac{3}{c\Delta}\right)^{1/2}\exp\left[2\pi\sqrt{\frac{c\Delta}{3}}\right]\left[1+O\left(\Delta^{-1/4}\right)\right]\,.
\end{aligned}
\end{equation}
So we can see that $\underset{\Delta\to\infty}{\text{lim}}F(\Delta/2,\Delta/2)$ is power law suppressed compared to $\underset{\Delta\to\infty}{\text{lim}}F^{\text{MZ}}(\Delta)$ i.e. $$\underset{\Delta\to\infty}{\text{lim}}\left(\frac{F(\Delta/2,\Delta/2)}{F^{\text{MZ}}(\Delta)}\right)=O(\Delta^{-1/4})\,.$$ We see that the square $\mathbb{Sq}$ of size $\Delta/2$ with one vertex at origin and another one at $(\Delta/2,\Delta/2)$ is always contained within the rightangled triangular region $\mathbb{T}$, created by $h'$ axis, $\hb'$ axis and $h'+\hb'=\Delta$ line. This is consistent with the observation that leading behavior of $F(\Delta/2,\Delta/2)$ is suppressed compared to $F^{\text{MZ}}(\Delta)$. In fact, one can similarly study the distribution of the operators in rectangular (or square) areas such that the rectangle is contained within $\mathbb{T}$, and one vertex is on the line $h'+\hb'=\Delta$ (see the figure~\ref{squarevsrectangle}).
This study reveals that the among such areas, the square $\mathbb{Sq}$ contains the most number of operators while any other rectangular region contains fewer number of operators, in fact the number is exponentially suppressed compared to that of the square $\mathbb{Sq}$.

\paragraph{Spin sensitivity of the asymptotics:} One can make a detailed analysis of spin sensitivity of the above result, which we expound on \S\ref{spinsense}.

\paragraph{Windowed entropy with respect to $h$ and $\hb$:} An immediate consequence of the eq.~\eqref{masterequation} is the expression for ``windowed" entropy $S_{\delta,\db}$. The windowed entropy is defined as logarithm of number of states within a rectangular window of side length $2\delta$ and $2\db$, centered at $(h,\hb)$. This is analogous to entropy defined as in microcanonical ensemble by proper ``binning", where the bin size is dictated by $\delta,\db$. As we take $h\to\infty ,\hb\to\infty$, we can keep the bin size $\delta,\db$ order one or let them scale like $h^{\alpha}$ and $\hb^\alpha$ respectively. We find that 
\begin{align}
\nonumber S_{\delta,\db}&\equiv \log\left(\int_{h-\delta}^{h+\delta}\text{d}h'\int_{\hb-\db}^{\hb+\db}\text{d}\hb' \rho(h',\hb')\right)\\
&\underset{h,\hb\to\infty}{=}2\pi\left(\sqrt{\frac{ch}{6}}+\sqrt{\frac{c\hb}{6}}\right)+\frac{1}{4}\log\left[\frac{c^2\delta^4\db^4}{36h^3\hb^3}\right]+s(\delta,\db,h,\hb)\,,
\end{align}
where for $3/8<\alpha\leq 1/2$, we have, :
\begin{align}
\begin{cases}
\delta\simeq h^{\alpha}\\
\db \simeq \hb^{\alpha}
\end{cases}&s(\delta,\db,h,\hb)=\log\left(\frac{\sinh\left(\pi\sqrt{\frac{c}{6}}\frac{\delta}{\sqrt{h}}\right)}{\pi\sqrt{\frac{c}{6}}\frac{\delta}{\sqrt{h}}}\right)+\log\left(\frac{\sinh\left(\pi\sqrt{\frac{c}{6}}\frac{\db}{\sqrt{\hb}}\right)}{\pi\sqrt{\frac{c}{6}}\frac{\db}{\sqrt{\hb}}}\right)+O(\tau^{3/4-2\alpha})\,,\\
\delta,\db\simeq O(1)&\quad \quad s_-(\delta,\db) \leq s(\delta,\db,h,\hb) \leq s_+(\delta,\db) 
\end{align}
where the functions $s_{\pm}(\delta,\db)$ are determined in the section \S\ref{lemma0}, in particular, we have $s_{\pm}\equiv \exp(c_{\pm})$, and $c_\pm$ is given by \eqref{defsplusminus}. We remark that when the bin size is large, there is a universal correction to Cardy formula given by the sinhyperbolic functions. This is analogous to what is found in \cite{Baur} from the analysis sensitive to dimension only.

\paragraph{Windowed entropy with respect to $\Delta+J$:}
One can define a microcanonical entropy with respect to $\Delta+J=2\max\{h,\hb\}$ (name this parameter $\kappa$) as
\begin{align}
S^{\kappa}_{\delta}\equiv \log\left[F(\kappa/2+\delta,\kappa/2+\delta)-F(\kappa/2-\delta,\kappa/2-\delta)\right]\,,
\end{align}

The asymptotic behavior of $S^{\kappa}_{\delta}$ is given by
\begin{align}
S^{\kappa}_{\delta}= 4\pi \sqrt{\frac{c\kappa}{12}}+\log\left(\frac{2\delta}{\pi\kappa}\right)+s(\delta,\kappa)\,,
\end{align}
where for large enough bin size ($\delta\simeq \kappa^{\alpha}$) we have 
\begin{align}
\delta\simeq \kappa^{\alpha}: &\ s(\delta,\tau)=\log\left(\frac{\sinh\left(2\pi\sqrt{\frac{c}{3}}\frac{\delta}{\sqrt{\kappa}}\right)}{2\pi\sqrt{\frac{c}{3}}\frac{\delta}{\sqrt{\kappa}}}\right)+O\left(\kappa^{1/4-\alpha}\right)\,,\ 1/4<\alpha\leq 1/2\,.
\end{align}


\subsection{$c>1$ CFTs-results specific for primaries}\label{primarydurga}
 One can make the results in the previous subsection specific to Virasoro primaries only, in fact do better. This boils down essentially repeating the argument presented in \S\ref{lemma0},\S\ref{lemma1} and \S\ref{mainproof} with minor modification. The idea of extending the argument from \S\ref{lemma0},\S\ref{lemma1} and \S\ref{mainproof} to this case is similar in spirit and practice to how \cite{Baur} obtained the specific results for primary using methods suitable to study all the operators. The details can be found in \S\ref{lemma0}, specifically eq.~\eqref{primary} onwards. Without much ado, here is the result: 
for finite central charge $c$, we find the integrated density of states specific for primaries behave like (from now on, we will be using the superscript ``Vir" to denote the result specific for primaries):
\begin{equation}\label{virmainresult}
\tcboxmath{\begin{aligned}
&F^{\text{Vir}}(h,\hb)\equiv\int_{0}^{h}\text{d}h' \int_{0}^{\hb}\text{d}\hb'\ \rho^{\text{Vir}}(h',\hb')\\
&\underset{h,\hb \to \infty}{=}\frac{1}{\pi^2}\left(\frac{3}{c-1}\right)\exp\left[2\pi\left(\sqrt{\frac{(c-1)h}{6}}+\sqrt{\frac{(c-1)\hb}{6}}\right)\right]\left[1+O\left(\tau^{-1/4}\right)\right]\,.
\end{aligned}}
\end{equation}

The ``windowed" entropy (we have considered bin of size $2\delta$ by $2\db$ just like what we did for the analysis of all the operators) for Virasoro primaries is given by
\begin{align}
\nonumber S^{\text{Vir}}_{\delta,\db}&\equiv\log\left(\int_{h-\delta}^{h+\delta}\text{d}h'\int_{\hb-\db}^{\hb+\db}\text{d}\hb' \rho^{\text{Vir}}(h',\hb')\right)\\
&\underset{h,\hb\to\infty}{=}2\pi\left(\sqrt{\frac{(c-1)h}{6}}+\sqrt{\frac{(c-1)\hb}{6}}\right)-\frac{1}{2}\log\left[\frac{h\hb}{4\delta^2\db^2}\right]+s^{\text{Vir}}(\delta,\db,h,\hb)\,,
\end{align}
where for $1/8<\alpha\leq 1/2$, we have :
\begin{equation}
\begin{aligned}
\begin{cases}
\delta\simeq h^{\alpha}\\
\db \simeq \hb^{\alpha}
\end{cases}&:s^{\text{Vir}}(\delta,\db,h,\hb)\\
&=\log\left(\frac{\sinh\left(\pi\sqrt{\frac{c-1}{6}}\frac{\delta}{\sqrt{h}}\right)}{\pi\sqrt{\frac{c-1}{6}}\frac{\delta}{\sqrt{h}}}\right)+\log\left(\frac{\sinh\left(\pi\sqrt{\frac{c-1}{6}}\frac{\db}{\sqrt{\hb}}\right)}{\pi\sqrt{\frac{c-1}{6}}\frac{\db}{\sqrt{\hb}}}\right)+O(\tau^{1/4-2\alpha})\,,\\
\delta,\db\simeq O(1)&: s_-(\delta,\db) \leq s^{\text{Vir}}(\delta,\db,h,\hb) \leq s_+(\delta,\db) 
\end{aligned}
\end{equation}
where the functions $s_{\pm}(\delta,\db)$ are the same functions that appear in the analysis for all the operators.

\paragraph{Large spin, large twist sector for primaries:} If we assume a finite twist gap (as defined in \{\ref{def}\}), the result given in eq.~\eqref{virmainresult} is true irrespective of whether $h$ and $\hb$ are of the order one or not. Thus unlike the case for all the operators, here we can trust the polynomially suppressed correction for all values of $\upsilon$, where $h=\hb^{\upsilon}$. 

\subsection{Large spin, finite twist sector}
The large spin, finite twist sector is not entirely asymptotic regime since the quantity knows about low lying spectrum in one of the weights. It turns out we can only put an upper bound in this case. There is an $O(1)$ error in the estimation. While for the upper bound this does not cause any trouble, for the lower bound, it makes thing tricky. In particular, the lower bound on the density of states, appropriately integrated, contains an exponential piece as expected from extended Cardy formula \cite{Kusuki:2018wpa,Kusuki:2019gjs,Maxfield:2019hdt,Benjamin:2019stq} but it comes with a multiplicative order one number, which can become negative unless proven otherwise. 

\paragraph{Analysis for all the operators:} In what follows, we will keep $h$ finite and let $\hb\to\infty$, the windowed entropy $S^{\text{ft}}_{\delta,\db}$ is found to be bounded above by 
\begin{equation}
\tcboxmath{\begin{aligned}
S^{\text{ft}}_{\delta,\db} \leq \mathbb{S}^{\text{ft}}_{h,\delta,\db}&\leq 2\pi\sqrt{\frac{c\hb}{6}}-\frac{1}{4}\log\left(\frac{\hb^3}{16\db^4}\right)+M\,,
\end{aligned}}
\end{equation}
where $M$ is an order one number. Here $S^{\text{ft}}_{\delta,\db}$ and $\mathbb{S}^{\text{ft}}_{h,\delta,\db}$ are defined as
\begin{align*}
\exp\left[S^{\text{ft}}_{\delta,\db}\right]&\equiv\int_{h-\delta}^{h+\delta} \text{d}h' \int_{\hb-\db}^{\hb+\db}\ \text{d}\hb'\ \rho(h',\hb')\,, \quad \exp\left[\mathbb{S}^{\text{ft}}_{h,\delta,\db}\right]\equiv\int_{0}^{h+\delta} \text{d}h' \int_{\hb-\db}^{\hb+\db}\ \text{d}\hb'\ \rho(h',\hb')\,.
\end{align*}
The number $M$ is given (or estimated) by 
\begin{align}
M&=2\pi \left(h+\delta-\frac{c}{24}\right)+ \log\left[c_+\sum_{\tilde{h}}\chi_{\tilde{h}}(e^{-2\pi})\right]\,.
\end{align}
where $\chi_{\tilde{h}}$ is the character for the conserved current with weight $(\tilde{h},0)$, $\tilde{h}\geq 0$ (including the Identity) and $c_{\pm}$ is an order one $h,\hb$ independent number, defined in \S\ref{lsft}. $M$ is a finite number as the absolute value of the sum over $\tilde{h}$ is bounded above by the partition function evaluated at $\beta=\bb=2\pi$, which is a finite number.\\

\paragraph{Analysis for primaries with/without conserved currents:}The above result can also be made specific to primaries:
\begin{equation}
\tcboxmath{\begin{aligned}
S^{\text{Vir,ft}}_{\delta,\db} \leq \mathbb{S}^{\text{Vir,ft}}_{h,\delta,\db}\leq 2\pi\sqrt{\frac{(c-1)\hb}{6}}-\frac{1}{2}\log\left(\frac{\hb}{4\db^2}\right)+M^{\text{Vir}}\,.
\end{aligned} }
\end{equation}
Here $S^{\text{Vir,ft}}_{\delta,\db}$ and $\mathbb{S}^{\text{Vir,ft}}_{h,\delta,\db}$ are defined as
\begin{align*}
\exp\left[\mathbb{S}^{\text{Vir,ft}}_{h,\delta,\db}\right] &\equiv \int_{0}^{h+\delta} \text{d}h' \int_{\hb-\db}^{\hb+\db}\ \text{d}\hb'\ \rho^{\text{Vir}}(h',\hb')\,,\\
 \exp\left[S^{\text{Vir,ft}}_{\delta,\db}\right]& \equiv \int_{h-\delta}^{h+\delta} \text{d}h' \int_{\hb-\db}^{\hb+\db}\ \text{d}\hb'\ \rho^{\text{Vir}}(h',\hb')\,.
\end{align*}
and $M^{\text{Vir}}$ is an order one number, given by
\begin{align}
M^{\text{Vir}} =2\pi \left(h+\delta-\frac{c-1}{24}\right) +\log\left[c_+\sum_{\tilde{h}}e^{-2\pi\left(\tilde{h}-\frac{c-1}{24}\right)}\right]\,,
\end{align}
where the zero twist primaries have weight $(\tilde{h},0)$, $\tilde{h}\geq 0$ (including the Identity) and $c_+$ is an order one $h,\hb$ independent number, defined in \S\ref{lsft}. $M^{\text{Vir}}$ is a finite number since the sum inside the log is convergent. This happens because the absolute value of the sum is bounded by the partition function evaluated at $\beta=\bb=2\pi$, which is a finite number.

\paragraph{Analysis for primaries for CFT with no nontrivial conserved current:} 
If we assume that there is no nontrivial conserved current i.e the only zero twist primary is the Identity and there is a finite twist gap (the finite twist gap as defined in \{\ref{def}\}, combined with the absence of nontrivial conserved current implies the usual twist gap condition used in the literature, for example in \cite{Collier:2016cls}). we show that 
\begin{equation}
\tcboxmath{\begin{aligned}
S^{\text{Vir,ft}}_{\delta,\db} \leq \mathbb{S}^{\text{Vir,ft}}_{h,\delta,\db}&\leq 2\pi\sqrt{\frac{(c-1)\hb}{6}}-\frac{1}{2}\log\left(\frac{\hb}{4\db^2}\right)+ \log\left(\sqrt{h+\delta-\frac{c-1}{24}}\right)\\
&\quad +\frac{\pi^2}{6}(c-1)\left(h+\delta+\frac{c-1}{24}\right)+\log\left(1-e^{-4\pi^2\left(h+\delta-\frac{c-1}{24}\right)}\right)+ M^\prime\,,
\end{aligned}}
\end{equation}
where $M^\prime$ is an order one $h$ independent number. If we assume that $(h+\delta-\frac{c-1}{24})$ is a very small number compared to $\frac{1}{c-1}$, this matches with the leading result appeared in the lightcone bootstrap program \cite{Kusuki:2018wpa,Kusuki:2019gjs,Collier:2018exn,Maxfield:2019hdt,Benjamin:2019stq} i.e.
\begin{equation}
\tcboxmath{\begin{aligned}
S^{\text{Vir,ft}}_{\delta,\db} \leq \mathbb{S}^{\text{Vir,ft}}_{h,\delta,\db}&\leq 2\pi\sqrt{\frac{(c-1)\hb}{6}}-\frac{1}{2}\log\left(\frac{\hb}{4\db^2}\right)+ \frac{3}{2}\log\left(h+\delta-\frac{c-1}{24}\right)+ \tilde{M}^\prime\,,
\end{aligned}}
\end{equation}
where $\tilde{M}^\prime=M+\log(4\pi^2)$. We remark that the limit is very subtle here. There are several scales. The scale set by $\hb$ is the largest one and we are seeking an asymptotic behavior in $\hb$. Then there are two fixed parameters $h$ and $c$. We are probing the regime where $(h+\delta-\frac{c-1}{24})$ is a very small number compared to $\frac{1}{c-1}$. The details of the calculation can be found at the end of \S\ref{lsft}.
\subsection{Asymptotic spectral gap}
The idea about deriving an upper bound on spectral gap comes from binning the states. If we make the bin size very small, we can not prove a positive lower bound on the number of states in that bin, because the bin might not have any state at all. As we increase the bin size, the chances are more that we find such positive lower bound. If we find a positive lower bound for a specific bin size centered at some large $h,\hb$; that would immediately imply existence of an upper bound on the asymptotic spectral gap.
\subsubsection{Probing spectral gap via Circle of order one area}
\paragraph{With/without twist gap:} Here we do not put any assumption on twist gap. 
\begin{tcolorbox}
Let us consider a square $\mathbb{S}$ of side $\frac{4\sqrt{3}\gamma }{\pi}+\epsilon_g$ centered at $(h,\hb)$ on $(h',\hb')$ plane. Here $\epsilon_g$ can be any arbitrarily small positive number. In the limit $h\to\infty,\hb\to\infty$ we have 
\begin{align}
\int_{\mathbb{S}}\text{d}h'\ \text{d}\hb'\ \rho^{\text{Vir}}(h',\hb') >0\,,
\end{align}
where the asymptotic region is reached along a curve for which $\text{max}(h,\hb)\simeq \frac{\gamma^4\tau}{2}$.
Thus the spectral gap along this curve is bounded above by a circle of radius $\frac{\gamma r}{\sqrt{2}}$  and 
the best possible value of $r$ that we find is $r=\frac{4\sqrt{3}}{\pi}$, this being the circle circumscribing the square.
\end{tcolorbox}

An immediate corollary is that the asymptotic twist gap is upper bounded by $\frac{8\sqrt{3}\gamma}{\pi}\simeq 4.42\gamma$. For $c>1$, the argument can be made specific for primaries, hence the asymptotic gap. This in some sense complements the bound on primary twist gap over the vacuum\footnote{In \cite{Collier:2016cls}, it is mentioned that the argument is due to Tom Hartman.} \cite{Collier:2016cls, Benjamin:2019stq}. \\

\begin{figure}[!ht]
\centering
\includegraphics[scale=0.6]{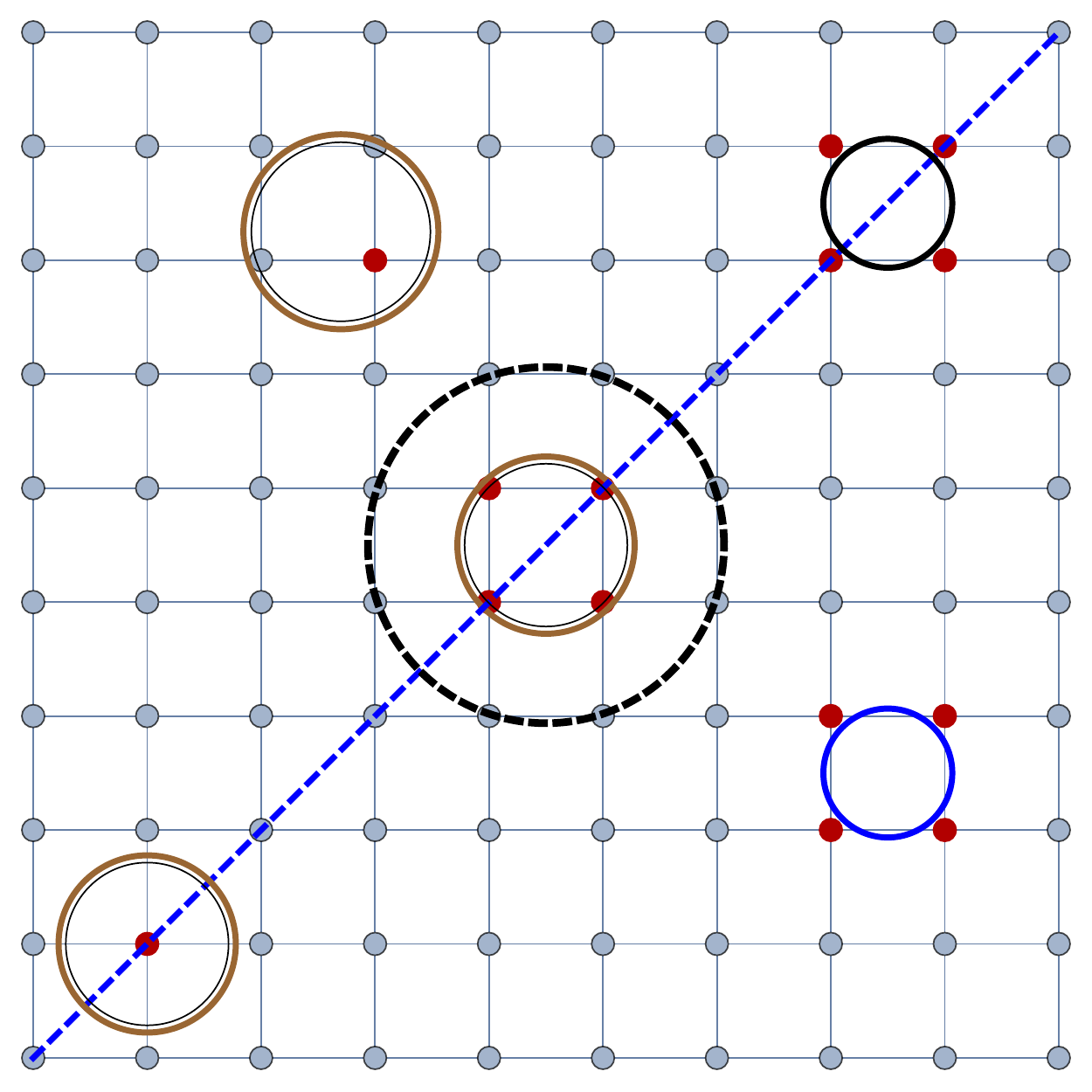}
\caption{Operator spectrum of chiral Monster CFT tensored with its antichiral avatar on $(h',\hb')$ plane: each vertex in the lattice represents the presence of operators. One can always circumscribe a square of unit length by a circle of radius $\frac{1}{\sqrt{2}}$. Thus any circle centered at $(h,\hb)$ and of radius $\frac{1}{\sqrt{2}}+\epsilon_g$ with $\epsilon_g>0$  would contain at least an operator. This is consistent with our result that asymptotically any circle of radius $\frac{\gamma r}{\sqrt{2}}+\epsilon_g$ with $\gamma\geq1$ will contain at least one operator where $\gamma$ is defined as the fourth root of asymptotic ratio of $\text{max}(h,\hb)$ and $\text{min}(h,\hb)$. We show $r=\frac{4\sqrt{3}}{\pi}>1$ and suspect that $r$ can be made to $1$. Here we have shown three such examples of containment. For each choice, two circles with different radius have been drawn to show that a circle of $\frac{\gamma}{\sqrt{2}}+\epsilon_g$ will contain at least one operator, as long as $\epsilon_g>0$. On the diagonal we have $\gamma=1$. The red dots denote the contained operators. The black circle shows that one can not ensure containment with circle of radius less than $\frac{1}{\sqrt{2}}$, showing optimality along the $h=\hb$ curve if one can show $r=1$. We remark that as we have only shown that $r=\frac{4\sqrt{3}}{\pi}>1$ (the dashed black circle), we have not yet reached the optimal bound. The blue circle suggests that the optimal bound should be $\frac{1}{\sqrt{2}}$ even for the case where $h\neq\hb$. Assuming a twist gap makes this bound insensitive to the line of approach towards infinity, but sensitive to the gap. See fig.~\ref{fig:monster2}.}
\label{fig:monster}
\end{figure}

We suspect that either by suitable choice of function or by the better estimate of heavy sector of the partition function, $r$ and/or length of a side of the bounding square can be made to $1$. If this can be done, then the bound becomes optimal for $\gamma=1$, (assuming we always consider circular/square region) since tensoring chiral Monster CFT with antichiral Monster CFT saturates the bound. One can see the saturation by circumscribing a square of unit length by a circle of radius $\frac{1}{\sqrt{2}}$ on $(h',\hb')$ plane $[$See the fig.~\ref{fig:monster}$]$. Nonetheless, the optimality along a curve for $\gamma\neq1$ is not guaranteed. An immediate corollary of finding such a circle is that the asymptotic twist gap is upper bounded by $2\sqrt{2}$ along $h=\hb$ curve. The same bound holds for asymptotic primary twist gap. We remark that in terms of twist, the above gap might not be optimal, since if we tensor chiral Monster CFT with anti-chiral monster CFT, the asymptotic twist gap is $2$. If one can find a bounding square of side length given by $1$, that would reproduce the optimal twist gap $2$.\\

The above result and the conjectures can not be applied to a scenario, where infinity is approached along a curve where $h$ is of widely different order compared to $\hb$, in particular, say, if we want to approach the infinity along the curve $h^\upsilon=\hb$ with $\upsilon \neq 1$. To circumnavigate this issue, we assume existence of twist gap $g$. We remind the readers that by finite twist gap, we mean there exists a number $\tau_*>0$ such that there is no operator with twist $\tau \in (0,\tau_*)$ and there are finite number of zero twist operators with dimension less than $c/12$, and $g\geq \tau_*$. Moreover, assuming existence of $g$ helps us to get rid of dependence on $\gamma$.

\begin{figure}[!ht]
\centering
\includegraphics[scale=0.6]{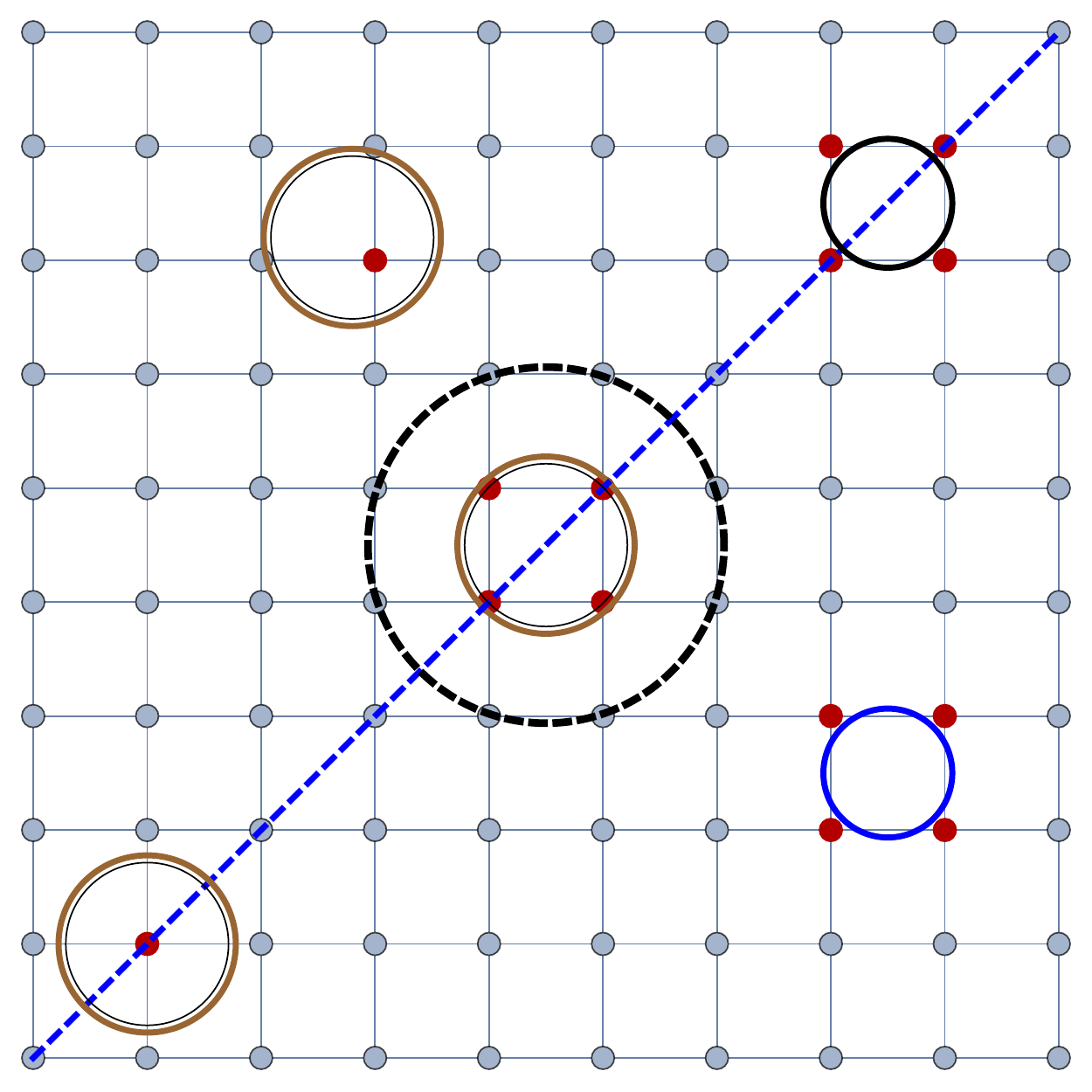}
\caption{Assuming twist gap $g$: operator spectrum of chiral Monster CFT tensored with its antichiral avatar on $(h',\hb')$ plane: each vertex in the lattice represents the presence of operators. One can always circumscribe a square of unit length by a circle of radius $\frac{1}{\sqrt{2}}$. Thus any circle centered at $(h,\hb)$ and of radius $\frac{\kappa}{\sqrt{2}}+\epsilon_g$ with $\epsilon_g>0,\kappa\geq 1$  would contain at least an operator. This is consistent with our result that asymptotically any square of side length $\frac{4\sqrt{3}}{\pi}$, hence any circle of radius $\frac{\sigma r}{\sqrt{2}}+\epsilon_g$ will contain at least one operator, where $\sigma=\text{max}\left(\sqrt{\frac{c-1}{12g}},1\right)$ and $r=\frac{4\sqrt{3}}{\pi}>1$. Again we suspect that $r$ can be made to $1$. In the Monster example $g=4$, hence $\sigma=1$.}
\label{fig:monster2}
\end{figure}

\paragraph{CFT with twist gap $g$:} Now we assume that the CFT has a twist gap as defined in \{\ref{def}\}.
\begin{tcolorbox}
For a CFT with twist gap $g$ (as defined in \{\ref{def}\}) and central charge $c>1$ , one can have a bounding circle $\mathbb{C}$ specific to primaries having a radius $\frac{\sigma r}{\sqrt{2}}+\epsilon_g$ with $\epsilon_g>0$, where 
\begin{align}\label{def:sigma}
\sigma=\text{max}\left(1,\sqrt{\frac{c-1}{12g}}\right)\,,\quad r=\frac{4\sqrt{3}}{\pi}\,.
\end{align}
irrespective of how infinity is approached, such that the bounding circle contains at least one operator. \end{tcolorbox}
Thus for such a scenario there exists $h_*$ and $\hb_*$, two order one numbers such that 
\begin{align}
\int_{\mathbb{C}}\text{d}h'\ \text{d}\hb'\ \rho^{\text{Vir}}(h',\hb') >0\,,\ \forall\ h>h_*,\hb>\hb_*\,.
\end{align}
Again this is obtained by circumscribing the appropriate bounding square $[$See the fig.~\ref{fig:monster2}$]$. The superscript ``Vir'' on $\rho^{\text{Vir}}$ denotes that it is density of primary operators as opposed to all the operators. In a compact unitary $2$D CFT without any conserved current, one can use  the upper bound of twist gap due to Hartman, appearing in \cite{Collier:2016cls}, to deduce 
\begin{align}\label{baser}
\sigma=\text{max}\left(1,\sqrt{\frac{c-1}{12g}}\right)=\sqrt{\frac{c-1}{12g}}\,.
\end{align}

Now we will explain the sense in which the minimal gap is complementary to twist gap:\\ 

\begin{tcolorbox}
Suppose we consider a $2$D compact unitary CFT with twist gap $g$ such that it does not have any zero twist primaries (conserved currents) except the Identity: if asymptotically there exists a circle of minimal area\footnote{One can imagine that operators having conformal weight $(h,\hb)$ are denoted by the point $(h,\hb)$ on $(h',\hb')$ plane. We name this set to be $\mathbb{S}$. Now consider the set $\mathbb{S}_d\equiv \{ d(a,b): a,b\in\mathbb{S}\}$, where $d(a,b)$ is the Euclidean distance on the plane between the points $a$ and $b$. Existence of a circle with minimal area means $\text{Inf}\ \mathbb{S}_d>0$. Asymptotically minimal area means that we look at the plane for $h'>h_*\ \&\ \hb'>\hb_*$ and construct the set $\mathbb{S}_d$ and consider its infimum.} $A$ on $(h',\hb')$ plane such that it does not contain any operator, we immediately deduce the following inequality
\begin{align}
A g\leq \frac{\pi(c-1)r^2}{24}\,.
\end{align}
\end{tcolorbox}

This can be thought of as an upper bound on twist gap if the minimal areal gap $A$ is known (note that minimal areal gap obtained from the full spectra has to be less than or equal to the asymptotic minimal gap). If one can make $r=1$ and show that $\frac{\pi}{2}<A=\frac{k\pi}{2}$ then it is possible to lower the upper bound on twist gap from $\frac{c-1}{12}$ to $\frac{c-1}{12k}$ with $k>1$. This might be of importance for proving the proposed upper bound $\frac{c-1}{16}$ in \cite{Benjamin:2019stq}, if it is true. To rephrase, if one can show that the minimal area $A=\frac{2\pi}{3}$ (thus the diameter of the circle would be $2\sqrt{\frac{2}{3}}$), it would imply the proposed bound. Similarly, any lower bound on twist gap translates to upper bound on minimal areal gap $A$.\\

The similar analysis can be done for all the operators assuming a twist gap. The only difference is that $\sigma$ would be given by $\sigma=\text{max}\left(1,\sqrt{\frac{c}{12g}}\right)$.
We elucidate on these bounds in the \S\ref{lemma0}.

\subsubsection{Probing spectral gap via Strips}\label{stripping}
\begin{figure}[!ht]
\centering
\includegraphics[scale=0.7]{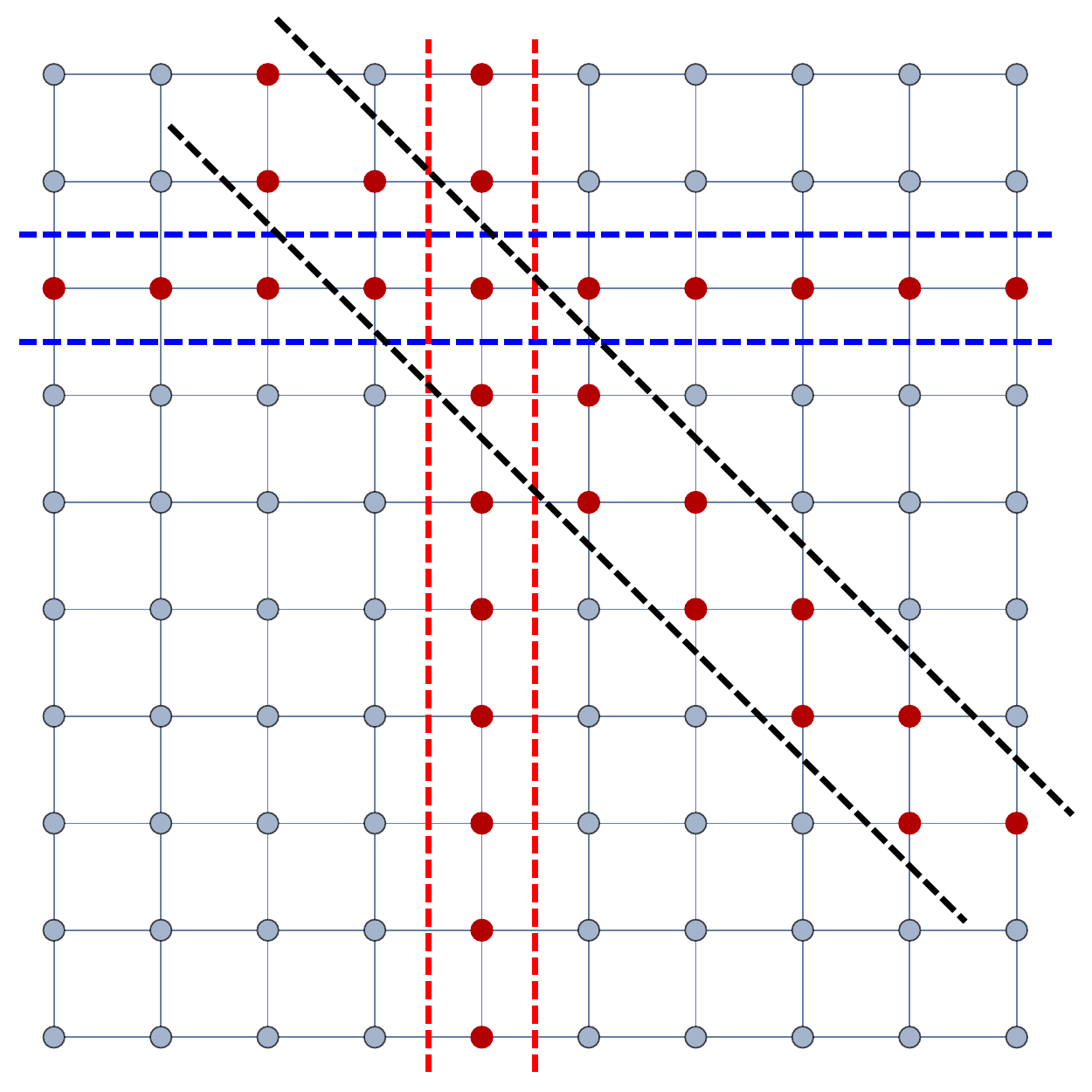}
\caption{We consider three kind of strips: the red vertical one is $H_2(\hb)$, the blue horizontal one is $H_1(h)$ and the black one is $H_3(\Delta)$. In each cases, we see that there is a minimum width of the strip such that the strips contain at least one operator. If we tensor chiral Monster CFT with its anti-chiral avatar, we can not make the width of the horizontal and vertical strip less than $1$. Thus the bound we find (which is $\sqrt{2}$) might not be optimal. A technical remark is that if it is not optimal, that would hint that one could have estimated the heavy part of the partition function in a better way.} 
\label{strips}
\end{figure}

Instead of squares , we can think of covering the $(h',\hb')$ plane via strips of finite width and ask what is the minimum width of the strip that guarantees existence of at least one state (one can make the analysis sensitive to primaries such that the state is Virasoro primary) in the strip. We can consider three kind of strips (see figure~\ref{strips}) :
\begin{align}\label{h1h2def}
H_1(h)&\equiv \{(h',\hb'): h' \in [h-\delta_1,h+\delta_1], \hb' \geq 0\}\,,\\
H_2(\hb)&\equiv \{(h',\hb'): \hb' \in [\hb-\delta_2,\hb+\delta_2]\,, h'\geq 0\},\\
H_3(\Delta)&\equiv \{(h',\hb'): h'+\hb' \in [\Delta-\delta_3,\Delta+\delta_3], h',\hb'\geq 0\}\,.
\end{align}

It is shown in \cite{Ganguly:2019ksp} that if $\delta_3>\frac{1}{2}$, we have 
\begin{align}
\int_{H_{3}(\Delta\to\infty)}\text{d}h'\ \text{d}\hb'\ \rho^{\text{Vir}}(h',\hb') >0
\end{align}
Thus the asymptotic spectral gap is bounded above by $1$.\\

In this work we show that 
\begin{equation}
\tcboxmath{\begin{aligned}
\int_{H_1(h \to\infty)}\text{d}h'\ \text{d}\hb'\ \rho^{\text{Vir}}(h',\hb') >0\ &\text{for}\ \delta_1>\frac{1}{\sqrt{2}}
\end{aligned}}
\end{equation}
The same result holds true for the $H_2$ strip with $h$ replaced by $\hb$, where $H_1$, $H_2$ are defined in \eqref{h1h2def}. This comes from putting a positive lower bound on the right hand side of the following inequality:
\begin{align}
\int_{H_1(h \to\infty)}\text{d}h'\ \text{d}\hb'\ \rho^{\text{Vir}}(h',\hb')\geq \int_{H_1(h \to\infty)}\text{d}h'\ \text{d}\hb'\ \rho^{\text{Vir}}(h',\hb') e^{-\bb\hb}
\end{align}
We achieve this as a corollary of the lemma proven in \S\ref{lemma1} (a similar lemma can be proven for primaries only and then we use the above inequality) . This shows that \\

\begin{tcolorbox}
Asymptotically on the $(h',\hb')$ plane, if the width of the horizontal or vertical strip is bigger than $\sqrt{2}$, it contains at least one Virasoro primary. This might not be optimal since the gap we find by tensoring chiral Monster CFT with its anti-chiral avatar is $1$.
\end{tcolorbox}

\subsection{Analysis at large central charge}
 We consider the $c\to\infty $ limit and parametrize the conformal weights in following way:
\begin{equation}
h=c\left(\frac{1}{24}+\epsilon\right)\,, \hb=c\left(\frac{1}{24}+\bar\epsilon\right)\,;\quad \epsilon,\bar{\epsilon}\ \text{fixed}\,.
\end{equation} 
Let us define $\epsilon_*=\text{min}\left(\epsilon,\bar\epsilon\right)$ and $\epsilon^*=\text{max}\left(\epsilon,\bar\epsilon\right)$. 

\paragraph{With/without twist gap:}
We show that\footnote{It might be possible to extend the region of validity beyond this, in particular, following \cite{HKS}, one might expect it to be valid for $\epsilon\bar\epsilon>\frac{1}{24^2}$!} 
\begin{tcolorbox} For $\text{min}\left\{(h/c-1/24), (\hb/c-1/24)\right\}=\epsilon_*>\frac{1}{6}$, the microcanonical entropy for order one window $\delta,\db\simeq O(1)$ is given by
\begin{equation}\label{Cardyc}
\tcboxmath{S_{\delta,\db}\underset{c\to\infty}{\simeq} 2\pi\left(\sqrt{\frac{c}{6}\left(h-\frac{c}{24}\right)}+\sqrt{\frac{c}{6}\left(\hb-\frac{c}{24}\right)}\right)-\log(c)+O(1)\,,}
\end{equation}
\end{tcolorbox}
while for $\delta,\db \simeq c^{\alpha}$ with $0<\alpha\leq 1$, we have 
\begin{equation}
\tcboxmath{S_{\delta}\underset{c\to\infty}{\simeq} 2\pi\left(\sqrt{\frac{c}{6}\left(h+\delta-\frac{c}{24}\right)}+\sqrt{\frac{c}{6}\left(\hb+\db-\frac{c}{24}\right)}\right)-\log(c)+O(1)\,.}
\end{equation}

\paragraph{Assuming finite twist gap:} If we assume a finite twist gap $g$ (as defined in \{\ref{def}\}), then 

\begin{tcolorbox}
The regime of validity of the Cardy result as in \eqref{Cardyc} can be extended to 
\begin{equation}
\tcboxmath{
\begin{aligned}\label{hksq}
\epsilon_*>\frac{1}{6}\left[\text{max}\left\{\frac{1}{2}\,,\left(1-\frac{6g}{c}\right)^2\right\}\right]\,,\quad \epsilon_*\equiv \text{min}\left\{(h/c-1/24), (\hb/c-1/24)\right\}\,. 
\end{aligned}}
\end{equation}
\end{tcolorbox}

\subsection*{Relevant recent work:} There has been recent surge in analyzing the asymptotics of CFT data on a rigorous footing. The results have been obtained \cite{pappadopulo2012operator,Qiao:2017xif} using techniques borrowed from a part of mathematics literature, which goes by the name of Tauberian theorems. The appendix C of \cite{dattadaspal} emphasizes the importance of Ingham theorem \cite{ingham1941tauberian} in analyzing Cardy's result \cite{cardy1986operator} for the asymptotic density of states in 2D CFT. Subsequently, the complex Tauberian theorems, as appeared originally in \cite{subhankulov1976tauberian} is utilized in the work of  \cite{mukhametzhanov2018analytic}. A complete rigorous treatment of Cardy formula appeared in the work \cite{Baur}, where they figured out the density of states in $\Delta\to\infty$ limit with a rigorous optimal estimate of the error term. The improvement of the result along with a proof of the conjecture made in \cite{Baur} has been put forward in \cite{Ganguly:2019ksp}. An rigorous analysis of the asymptotics of three point coefficients \cite{KM} appeared recently \cite{Pal:2019yhz} where the main challenge was to circumnavigate the negativity issue for the analysis of three point coefficients.

\section{Set up}\label{setup}
We consider a $2$D CFT with spectrum of operators having conformal weights $(h',\hb')$. We assign different real temperatures $\beta,\bar{\beta}$ to the left-moving and the right-moving sectors respectively. The partition function $Z(\beta,\bar{\beta})$ is given by:

\begin{equation}
    Z(\beta,\bar{\beta}) = \sum_{h',\hb'} e^{-\beta(h'-c/24)-\bar{\beta}(\hb'-c/24)}\,.
\end{equation}
The modular invariance of the partition function yields: 
\begin{equation}
    Z(\beta,\bar{\beta}) = Z(\beta',\bar{\beta}'), \,\,\,\, \beta' = \frac{4\pi^2}{\beta}, \,\,\,\, \bb' = \frac{4\pi^2}{\bb}\,.
\end{equation}
We further define the following measure
\begin{align}
\text{d}F(h^\prime,\bar h^\prime)=\sum d(h_i,\bar{h}_i)\delta(h^\prime-h_i)\delta(\bar h^\prime-\bar h_i)\,,
\end{align}
where $d(h_i,\bar{h}_i)$ is the degeneracy of the state with conformal weight $(h_i,\bar h_i)$. Our goal is to estimate the integral of the measure $\text{d}F$ over different regions.

{
\subsection{A semi technical glimpse of the subtleties} 
One of the key step in deriving the Cardy formula is the intuitive understanding that at high temperature, the partition function is dominated by the heavy states, thus doing an inverse Laplace transform of the high temperature behavior of the partition function should produce the asymptotic density of states. Schematically, 
\begin{align}
Z(\beta \to 0) &= \underbrace{\text{Leading Term}}_{\text{Produces asymptotic density of states}} +\ \text{Error}\,,\\
\label{jus}\text{Inverse Laplace}\left[\text{Error}\right]&\overset{???}{=} \text{Error in asymptotic density of states}\,.
\end{align}
The underlying assumption while doing the above is that the inverse Laplace transform of the error term is bounded as well, thus producing an error compared to the leading behavior of the asymptotic density of states. The Tauberian formalism justifies this step by carefully estimating the error terms. The way it works is following: one bounds the number of states within an order an window centered at some heavy $\Delta$, from above and below by some convolution ($\circledast$) of partition function at high temperature ($\beta$) and bandlimited function $\phi_\pm$, schematically this looks like 
\begin{equation}
\begin{aligned}
\left[Z(\beta+it)\circledast\phi_-(t)\right] \leq \#\text{states in window} \leq \left[Z(\beta+it)\circledast\phi_+(t)\right]\,.
\end{aligned}
\end{equation} 
Intuitively, at this stage, we know that heavy states contribute to this partition function at high temperature. Now we implement modular transformation, the partition function at high temperature becomes partition function at low temperature. Schematically we have
\begin{equation}
\begin{aligned}
\left[Z\left(\frac{4\pi^2}{\beta+it}\right)\circledast\phi_-(t)\right] \leq \#\text{states in window} \leq\left[Z\left(\frac{4\pi^2}{\beta+it}\right)\circledast\phi_+(t)\right]\,.
\end{aligned}
\end{equation} 
 At low temperature, low lying states contribute the most i.e $Z\left(\frac{4\pi^2}{\beta}\to\infty\right)$ is dominated by low lying states. So, following \cite{HKS}, we separate the low lying states from heavy states; the low lying states constitute the ``light" part, while the ``heavy" part is complement of that. The ``light" part contains finite number of operators at finite central charge. So we do the inverse Laplace transform of this ``Light" part to get the leading answer $\rho_*$. Thus $\rho_*$ reproduces the $Z\left(\frac{4\pi^2}{\beta}\to\infty\right)$ behavior upon doing Laplace transformation. We are still left with the ``heavy" part contribution of $Z\left(\frac{4\pi^2}{\beta+it}\right)$. This part can be shown to produce a subleading correction to the leading piece of asymptotic density of states, thus justifying \eqref{jus} using the bound proven in \cite{HKS}. This requires relating the ``heavy" part at temperature $\beta$ to the ``light" part at temperature $\beta^\prime$. The upshot of this discussion is that we have a full control of the error term and its inverse Laplace transformation. In practice, we only consider the Identity operator among all the operators in the ``light" region. So one might worry about the error coming from that but since there are finite number of operators in the ``Light" region, one can do inverse Laplace transformation term by term and show that each of them is exponentially suppressed and hence the finite sum of them. We emphasize the ``finiteness" of finite sum is really very important for this and this is precisely why we need to treat heavy part separately\footnote{One could have imagined doing inverse Laplace transformation term by term including the operators from ``heavy" sector, even if each of them is suppressed, there's no guarantee that the infinite sum is suppressed.}. In fact, we remind the readers that in the large central charge limit, we have infinite number of operators even in the ``light" sector, hence we need an extra assumption of \textit{sparseness}, as done in \cite{HKS}.\\

The immediate generalization of this technique used in the \cite{Baur} has obstacles because of the cross terms present in the analysis, for example, say contribution from the states where $h'$ is large but $\hb'$ is not that large. The most obvious way to generalize the argument is to use the generalized HKS \cite{HKS} cut: dividing the $(h',\hb')$ plane into two regions, where the ``light" region (call it $\mathbb{L}$) contains all the operators with one of $h'$ or $\hb'$ being less than $c/24$ while the heavy region is the complement of them. It is possible to make a similar statement about this ``heavy" region, relating it to the ``light" part using HKS like argument. Nonetheless, one then stumbles upon the issue of defining $\rho_*$, which is supposed to reproduce the leading contribution to the partition function at high temperature ($\beta\to 0,\beta^\prime\to\infty$), to be precise, the light part of the partition function at temperature $\beta^\prime$. Now the issue is that there are infinite number of operators in this region $\mathbb{L}$. Unlike the case in \cite{Baur}, we just can not take the Identity operator to prove that this produces the leading behavior and say the rest are suppressed. In particular, the previous argument of term by term exponential suppression fails because there are infinite of them. The take home message is that it is not a priori clear whether just considering the vacuum to calculate $\rho_*$ is good enough, because infinite number of other operators might conspire to spoil the ``leading contribution", even if each one of them is exponentially suppressed. We reemphasize that \cite{Baur} did not face this problem, since in their case, the light region was $\Delta'<c/12$ and the region has finite number of operators and everything is under control, so in principle their $\rho_*$ was defined having contribution from all those states with $\Delta'<c/12$ and in practice derivable from the vacuum. To circumnavigate this problem, in \S\ref{lemma0}, we use the original HKS cut i.e. we define the ``light" region to be the one where $h'+\hb'<c/12$ and the ``heavy" region is the complement of the light region. See fig.~\ref{fig:hks}.
\begin{figure}[!htb]
\centering
\includegraphics[scale=0.4]{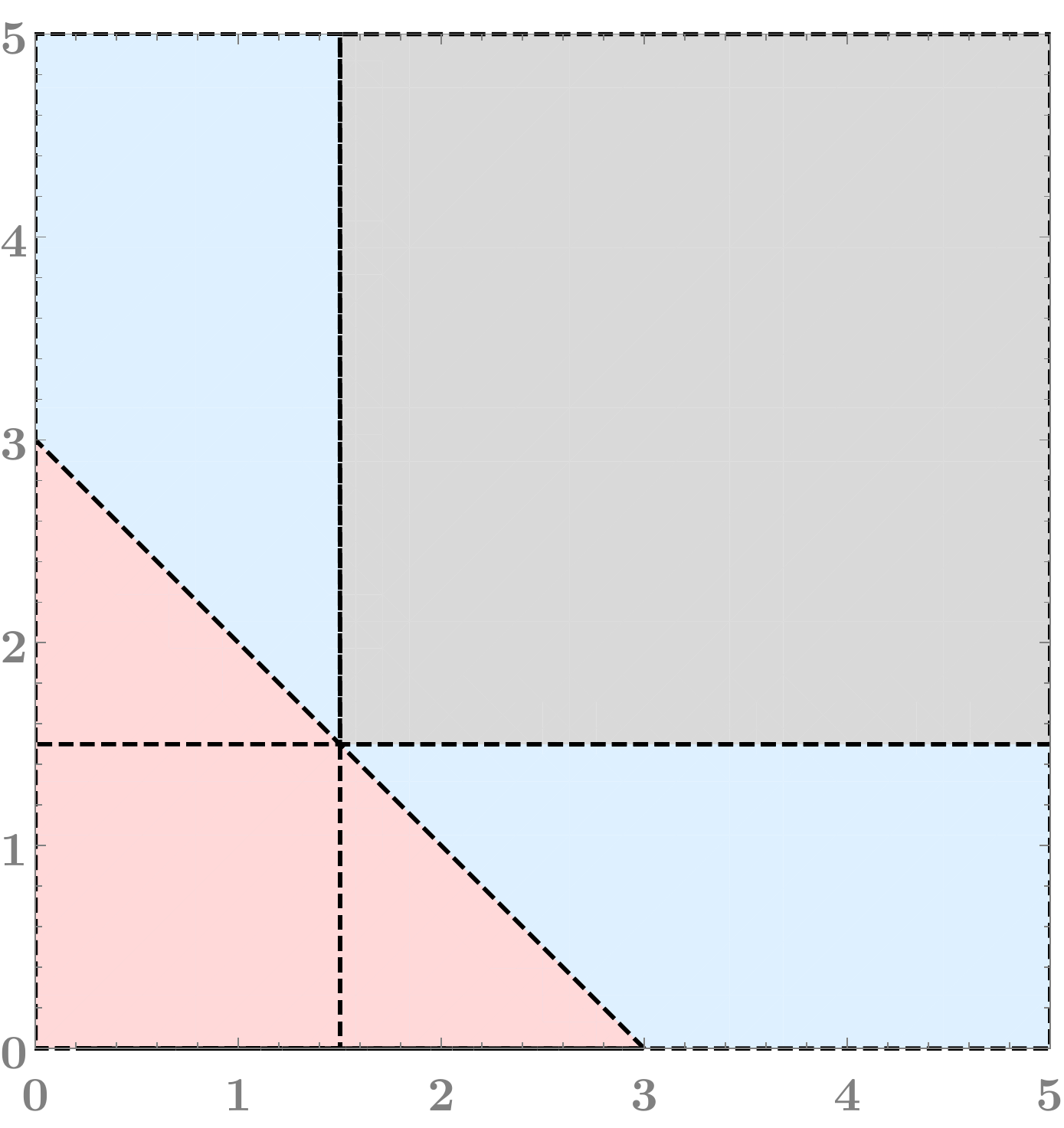}
\includegraphics[scale=0.4]{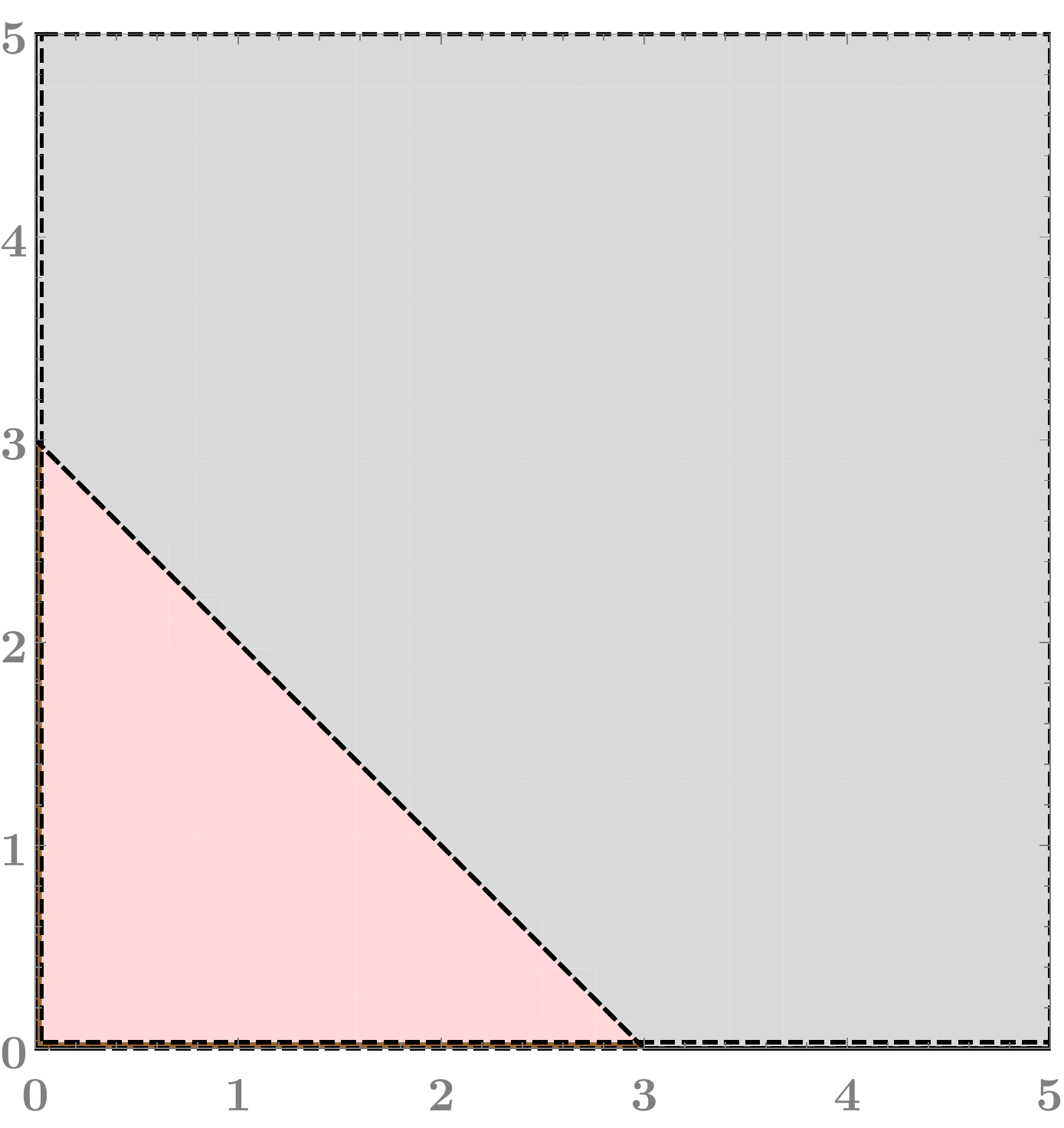}
\caption{The $(h',\hb')$ plane: the light red shaded region contributes to $Z_{L}$ and given by the states with $h'+\hb'<c/12$. The complement of this contributes to $Z_{H}$. The naive HKS cut  makes the union of red and blue shaded region to be the ``Light" region. But this brings in the problem of having infinite number of operators in the light region. This cut is applicable to the analysis sensitive to primaries as well with a shifted central charge $c\mapsto c-1$. The picture on the right emphasizes that one needs to treat the zero twist operators separately for some part of the analysis.}
\label{fig:hks}
\end{figure}
The immediate cost for doing this is that we can make comment only when $h$ and $\hb$ is of the same order asymptotically. The details depend on how infinity is approached. But this is expected because intuitively the infinity can be reached several ways on a plane. This issue persists for the analysis sensitive to primary as well. It turns out that one can bypass this restriction of $h$ and $\hb$ being of the same order, if one assumes a twist gap (as defined in \{\ref{def}\} we require finite number of conserved currents with dimension less than $c/12$ to be precise). In that scenario, the results in \S\ref{lemma0} holds true even if $h$ and $\hb$ are not of the same order asymptotically. The estimation of the ``heavy" region becomes really involved in this case and carried out in details in \S\ref{lemma0}. One has to further separate out the ``heavy" zero twist operators and estimate their contribution separately (see the right fig.~\ref{fig:hks}).\\

We further point out that  estimating the integrated density of states require us to prove another lemma, which is special to $2$ D Tauberian theorem. We achieve this in \S\ref{lemma1}. The lemma is then fed into the main proof in \S\ref{mainproof}. The result for the integrated density of states also requires an estimate of number of states within an order one area on $(h',\hb')$ plane, which is achieved in \S\ref{lemma0}. In \S\ref{verify}, we verify our results using $2$D Ising model. The large spin, finite twist is discussed in \S\ref{lsft}. The large central charge regime is discussed in \S\ref{largec}. We conclude with a list of open of problems along with a brief discussion.

}

\section{$O(1)$ rectangular window}\label{lemma0}
In this section, we study the number of states lying on an order one rectangular/square area of $(h',\hb')$ plane, centered at some large $(h,\hb)$ and having sides of length $2\delta$ and $2\db$. We are interested in $h\to\infty, \hb\to\infty$ limit with $\delta,\db$ being fixed order one numbers. We do it in two ways, in the first subsection, we do it generically without any assumption on twist gap while in the next subsection, we assume existence of a twist gap. The assumption of the twist gap (as defined in~\{\ref{def}\}) facilitates studying the regime where $h$ is not of the same order as $\hb$. 
\subsection{Generic analysis: with/without Twist gap}
Following \cite{Baur}, let us choose functions $\Phi_\pm (h,\bar{h})$ such that
\begin{equation}\label{eq:masterin}
    \Phi_-(h',\hb') \leq \Theta_{h,\hb,\delta,\db}(h',\hb') \leq \Phi_+(h',\hb'),
\end{equation}
where $\Theta_{h,\hb,\delta,\delta'}(h',\hb')$ is the indicator function of the rectangle and defined as
\begin{align}
\Theta_{h,\hb,\delta,\delta'}(h',\hb')= \theta_{[h-\delta,h+\delta]}(h') \theta_{[\hb-\db,\hb+\db]}(\hb'),
\end{align}
In principle, one can choose the energy window for the microcanonical ensemble to be of a different shape, for example, a circle. But for now, we consider it to be a rectangle. Now, from eq.~\eqref{eq:masterin} we obtain
\begin{equation}
    e^{\beta(h-\delta)+\bb(\hb-\db)} e^{-\beta h' -\bb \hb'} \Phi_-(h',\hb') \leq \Theta_{h,\hb,\delta,\db}(h',\hb') \leq e^{\beta(h+\delta)+\bb(\hb+\db)} e^{-\beta h' -\bb \hb'} \Phi_+(h',\hb').
\end{equation}

Multiplying the above by the density of states $\rho(h',\hb')$ and integrating yields the following inequality: 
\begin{equation}\label{eq:master2}
    \begin{aligned}
    & e^{\beta(h-\delta)+\bb(\hb-\db)} \int \text{d}F(h',\hb') e^{-\beta h' - \bb \hb'} \Phi_-(h',\hb') \\
    & \leq \int_{h-\delta}^{h+\delta}\int_{\hb-\db}^{\hb+\db}  \text{d}F(h,\hb') \leq \\
    & e^{\beta(h+\delta)+\bb(\hb+\db)} \int  \text{d}F(h',\hb') e^{-\beta h' - \bb \hb'}\Phi_+(h',\hb').
    \end{aligned}
\end{equation}

At this point, we use the Fourier transform $\hat{\Phi}_{\pm}(t,\tb)$, defined as
\begin{align}
\Phi_{\pm}(h',\hb')\equiv \int_{-\infty}^{\infty}\  \text{d}t\ \int_{-\infty}^{\infty}\   \text{d}\tb\ \hat{\Phi}_{\pm}(t,\tb)e^{-i h't-i \hb' \tb}\,.
\end{align}
This facilitates us to rewrite the  inequality in \eqref{eq:master2} as
\begin{equation}
\begin{aligned}
    &\int \text{d}h'\ \text{d}\hb' \mathcal{L}_\rho(\beta+it,\bb+i\tb) \hat{\Phi}_-(t,\tb)\\
   &  \leq \int_{h-\delta}^{h+\delta}\int_{\hb-\db}^{\hb+\db} \text{d}F(h,\hb') \leq \\
    &\int \text{d}h'\ \text{d}\hb' \mathcal{L}_\rho(\beta+it,\bb+i\tb) \hat{\Phi}_+(t,\tb)\,,
    \end{aligned}
\end{equation}
where we have 
\begin{align}
\mathcal{L}_\rho(\beta,\bb) \equiv \int_0^\infty \text{d}h\ \int_0^\infty \text{d}\hb\ \rho(h,\hb) e^{-\beta h - \bb \hb}\,.
\end{align}
Next we need to split $Z = Z_L + Z_H$.  In \cite{HKS}, for the mixed temperature analysis, the light sector is chosen to be $\{(h,\hb)|h<\frac{c}{24}\,\, \text{or}\,\, \hb<\frac{c}{24}\}$ and the heavy sector is the complement of that. Here, we choose a different light sector $\{(h,\hb)|h+\hb<\frac{c}{12}\}$ which has finite size (for large central charge, this is not true and one needs to have an extra sparseness condition on the low lying spectra \cite{HKS,Baur}). So at least, in principle, we can choose $\rho_{*}(h,\hb)$ such that $\rho_{*}$ reproduce the contributions from all operators in the light sector. In practice, we take $\rho_{*}$ such that it reproduces only the vacuum state contribution, that is,

\begin{equation}
    \rho_{*}(h,\hb) = \bigg[\pi\sqrt{\frac{c}{6}}\frac{I_1\left(2\pi\sqrt{\frac{c}{6}(h-\frac{c}{24})}\right)}{\sqrt{h-\frac{c}{24}}}\theta\left(h-\frac{c}{24}\right) + \delta\left(h-\frac{c}{24}\right)\bigg]\times (h\rightarrow \hb),
\end{equation}
which has the asymptotic behavior
\begin{equation}
    \rho_{*}(h,\hb) \sim \sqrt{\frac{c}{96}}\bigg(\frac{1}{h^3\hb^3}\bigg)^{1/4} e^{2\pi\sqrt{\frac{ch}{6}}+2\pi\sqrt{\frac{c\hb}{6}}}\left[1+O\left(h^{-1/2}\right)+O\left(\hb^{-1/2}\right)\right].
\end{equation}
 Then we can write $e^{-(\beta+it+\bb+i\tb)c/24} Z_L\left(\frac{4\pi^2}{\beta+i t},\frac{4\pi^2}{\bb+i \tb}\right) = \mathcal{L}_{\rho_{*},L}(\beta+it,\bb+i\tb)$. So we get
 \begin{equation}\label{eq:n1}
    \begin{aligned}
    & e^{\beta(h-\delta)+\bb(\hb-\db)}\bigg(\int_{-\infty}^{\infty}\text{d}t\ \text{d}\tb\ \hat{\Phi}_-(t,\tb) \mathcal{L}_{\rho_{*},L}(\beta+it,\bb+i\tb)  \\  
    	& \,\,\,\,\,\,\,\,\,\,\,\,\,\,\,\,\,\,\,\,\,\,\,\,\,\,\,\,\,\,\,\,\,\,\,\,\,\,\,\,\,\,\,\,\,\,\,\,\,\,\,\,\,\,\,\,\,\,\,\,\,\,\,\,\, - \bigg|\int_{-\infty}^{\infty}\text{d}t \text{d}\tb\ e^{-(\beta+it+\bb+i\tb)c/24}\hat{\Phi}_-(t,\tb)Z_H\left(\frac{4\pi^2}{\beta+it},\frac{4\pi^2}{\bb+i\tb}\right)\bigg|\bigg) \\ & \leq  \int_{h-\delta}^{h+\delta}\int_{\hb-\db}^{\hb+\db} \text{d}F(h,\hb') \leq \\
    & e^{\beta(h+\delta)+\bb(\hb+\db)}\bigg(\int_{-\infty}^{\infty}\text{d}t\ \text{d}\tb\ \hat{\Phi}_+(t,\tb) \mathcal{L}_{\rho_{*},L}(\beta+it,\bb+i\tb) \\
    & \,\,\,\,\,\,\,\,\,\,\,\,\,\,\,\,\,\,\,\,\,\,\,\,\,\,\,\,\,\,\,\,\,\,\,\,\,\,\,\,\,\,\,\,\,\,\,\,\,\,\,\,\,\,\,\,\,\,\,\,\,\,\,\,\,\,\, + \bigg|\int_{-\infty}^{\infty}\text{d}t\ \text{d}\tb\ e^{-(\beta+it+\bb+i\tb)c/24}\hat{\Phi}_+(t,\tb) Z_H\left(\frac{4\pi^2}{\beta+it},\frac{4\pi^2}{\bb+i\tb}\right)\bigg|\bigg)\,,
    \end{aligned}
\end{equation}
Now we can estimate the contribution from the heavy sector using the HKS bound. We choose $\Phi_{\pm}$ such that $\hat{\Phi}_{\pm}(t,\tb)$ have finite support $[-\Lambda_\pm,\Lambda_\pm]$ for $t$ and $\tb$. A possible choice can be made via modifying the functions appearing in \cite{Baur, Ganguly:2019ksp} a little bit. To be concrete, let us make the following choices: 

\begin{align}
         \Phi_+(h',\hb') &= \frac{f_+(h-h')f_+(\hb-\hb')}{f_+(\delta)f_+(\db)}\,, \\
    \label{lower}     \Phi_-(h',\hb')& = f_-(h-h')f_-(\hb-\hb')\bigg(1-\bigg(\frac{h-h'}{\delta}\bigg)^2-\bigg(\frac{\hb-\hb'}{\db}\bigg)^2\bigg),
\end{align}
where we have
\begin{align}
f_{\pm}(x)=\bigg[\sinc\left(\frac{\Lambda_{\pm}x}{4}\right)\bigg]^4.
\end{align}
We remark that for $\Phi_-$, the locus $1-(a/\delta)^2-(b/\db)^2 = 0$ is inside the rectangle region, hence it is a valid choice conforming to the inequality \eqref{eq:masterin}. Here $a=h-h'$ and $b=\hb-\hb'$. At this point, our aim is to show that 
\begin{equation}
	I_\pm = e^{\beta(h\pm\delta)+\bb(\hb\pm\db)}\bigg|\int_{-\infty}^{\infty}\text{d}t\ \text{d}\tb\ e^{-(\beta+it+\bb+i\tb)c/24}\hat{\Phi}_\pm(t,\tb) Z_H\left(\frac{4\pi^2}{\beta+it},\frac{4\pi^2}{\bb+i\tb}\right)\bigg|
\end{equation}
is sub-leading. We will make use of the most basic HKS bound \cite{HKS} for $\Delta$ in a clever way. We remark that by requiring the saddle of the light sector is located at $(h,\hb)$, we find $\beta = \pi\sqrt{\frac{c}{6h}} << 1, \bb = \pi \sqrt{\frac{c}{6\hb}} << 1$, so some terms such as $\beta\delta$ can be dropped from the bounds as it goes to 0 for large $h,\hb$. Then using the fact that $\Phi_\pm$ is a bandlimited function, we have

\begin{equation}
I_\pm  \leq e^{\beta h+\bb \hb} \int \text{d}t\ \text{d}\tb\ Z_H\left(\frac{4\pi^2 \beta}{\beta^2+t^2},\frac{4\pi^2 \bb}{\bb^2+\tb^2}\right) \bigg| \hat{\Phi}_\pm(t,\tb) \bigg|\,.
\end{equation}
{Now $Z_H$ has contribution from states where either $h'$ or $\hb'$ is greater than $c/24$. Since the contributing states have $h'+\hb'>c/12$, both can not be less than or equal to $\frac{c}{24}$. We illustrate the case for $h'>c/24$.

\begin{equation}\label{trick1}
\begin{aligned}
Z_{H}\ni &\exp\left[-\frac{4\pi^2 \beta}{\beta^2+t^2}\left(h'-\frac{c}{24}\right)-\frac{4\pi^2 \bb}{\bb^2+\tb^2}\left(\hb'-\frac{c}{24}\right)\right] \\
&\leq e^{\frac{\pi^2 c}{6\bb}}\exp\left[-\frac{4\pi^2 \beta}{\beta^2+\Lambda_{\pm}^{2}}\left(h'-\frac{c}{24}\right)-\frac{4\pi^2 \bb}{\bb^2+\Lambda_{\pm}^2}\hb'\right]\\
&\leq e^{\frac{\pi^2 c}{6\bb}}e^{-\frac{\pi^2 \beta_* c}{6(\beta_*^2+\Lambda_{\pm}^2)}} \exp\left[-\frac{4\pi^2 \beta_*}{\beta_*^2+\Lambda_{\pm}^{2}}\left(h'+\hb'-\frac{c}{12}\right)\right]\,,
\end{aligned}
\end{equation}
where $\beta_*$ is defined as $\beta_*=\text{min}\left(\beta,\bb\right)$, hence we have (for small enough $\beta$ and $\bb$)
\begin{equation}
    \min\bigg(\frac{4\pi^2\beta}{\beta^2+\Lambda_\pm^2},\frac{4\pi^2\bb}{\bb^2+\Lambda_\pm^2}\bigg) = \frac{4\pi^2\beta_*}{\beta_*^{2}+\Lambda_\pm^2}\,,
\end{equation}
Thus we have
\begin{align}
I_{\pm} \leq e^{\beta h+\bb\hb}\left(e^{\frac{\pi^2 c}{6\bb}}+e^{\frac{\pi^2 c}{6\beta}}\right)e^{-\frac{\pi^2 \beta_* c}{6(\beta_*^2+\Lambda_{\pm}^2)}} Z_{H,\Delta}\left(\frac{4\pi^2 \beta_*}{\beta_*^2+\Lambda_{\pm}^{2}}\right)\,,
\end{align}
where the last $Z_{H,\Delta}=\sum_{\Delta>c/12}e^{-\beta(\Delta-c/12)}$ is the heavy contribution from the original HKS bound for $\Delta$. We will be showing that the above term is subleading. There are two pieces, one with $e^{\frac{\pi^2 c}{6\beta}}$ and another with $e^{\frac{\pi^2 c}{6\bb}}$. Let us illustrate the subleading nature of the term with $e^{\frac{\pi^2 c}{6\bb}}$. The other term can be treated similarly.

In $\beta,\bb\to 0$ limit, we have 

\begin{equation}\label{estimate}
    \begin{aligned}
         e^{\beta h + \bb \hb}e^{\frac{\pi^2 c}{6\bb}}e^{-\frac{\pi^2 \beta_* c}{6(\beta_*^2+\Lambda_{\pm}^2)}}&Z_{H,\Delta}\bigg(\frac{4\pi^2\beta_*}{\beta_*^{2}+\Lambda_\pm^2}\bigg)  
        \leq  e^{\beta h + \bb \hb}e^{\frac{\pi^2 c}{6\bb}} Z_{H,\Delta}\bigg(\frac{4\pi^2\beta_*}{\beta_*^{2}+\Lambda_\pm^2}\bigg) \\
        \sim \,\, & e^{\pi \sqrt{\frac{c h}{6}} + 2\pi \sqrt{\frac{c\hb}{6}} + 2\pi \sqrt{\frac{c h^*}{6}}\big(\frac{\Lambda_\pm}{2\pi}\big)^2} \\
        \leq  \,\, &  \begin{cases} h\geq \hb:\ e^{2\pi\sqrt{\frac{ch}{6}} + 2\pi\sqrt{\frac{c\hb}{6}} + 2\pi \sqrt{\frac{ch}{6}}\big(\big(\frac{\Lambda_\pm}{2\pi}\big)^2-\frac{1}{2}\big)}.\\
        h< \hb:\ e^{\pi\sqrt{\frac{ch}{6}} + 2\pi\sqrt{\frac{c\hb}{6}} + 2\pi \sqrt{\frac{c\hb}{6}}\big(\frac{\Lambda_\pm}{2\pi}\big)^2}.
       \end{cases}
    \end{aligned}
\end{equation}
where $h^*=\text{max}(h,\hb)$ and $\tau=2\ \text{min}(h,\hb)$. To make the contribution from the heavy sector sub-leading, we need 

\begin{align}
\Lambda_\pm < \text{min}\left(\frac{2\pi}{\sqrt{2}} ,\frac{2\pi}{\sqrt{2}\gamma} \right)\,,
\end{align}
where $\frac{\tau}{2}\gamma^4\simeq h^*$ (clearly, $\gamma>1$). This can be achieved by choosing
\begin{align}\label{condition}
\Lambda_\pm < \frac{\sqrt{2}\pi}{\gamma} \,.
\end{align}

}
Then at large $h,\hb$, we have the following bound, starting from \eqref{eq:n1} (the second term i.e. the term with the absolute value in \eqref{eq:n1} has already been shown to be subleading and we rewrite the first term as an integral over $h',\hb'$ below):
\begin{equation}\label{leading}
    \begin{aligned}
        & e^{\beta(h-\delta)+\bb(\hb-\db)}\int \text{d}h'\ \text{d}\hb' \rho_{*}(h',\hb') \Phi_-(h',\hb')e^{-\beta h' -\bb\hb'} \\
        & \leq \int_{h-\delta}^{h+\delta}\int_{\hb-\db}^{\hb+\db} \text{d}F(h,\hb') \leq \\
        & e^{\beta(h+\delta)+\bb(\hb+\db)} \int \text{d}h'\ \text{d}\hb' \rho_{*}(h',\hb') \Phi_+(h',\hb') e^{-\beta h' - \bb\hb'}\,.
    \end{aligned}
\end{equation}

We can evaluate this integral by saddle point approximation, 

\begin{equation}
    c_- \rho_{*}(h,\hb) \leq \frac{1}{4\delta\db} \int_{h-\delta}^{h+\delta}\int_{\hb-\db}^{\hb+\db} \text{d}F(h,\hb') \leq c_+ \rho_{*} (h,\hb)\,,
\end{equation}
where $c_{\pm}$ is defined as
\begin{align}\label{def:cpm}
c_\pm = \frac{1}{4} \int\ \text{d}x\ \text{d}y\ \Phi_\pm(h+\delta x,\hb+\db y)\,.
\end{align} 
With the previous choice of $\Phi_\pm$, we have 

\begin{equation}\label{defsplusminus}
    \begin{aligned}
    & c_+ = \frac{16\pi^2}{9} \frac{1}{\delta \db \Lambda_+^2 \sinc^4(\delta \Lambda_+/4)\sinc^4(\db \Lambda_+/4)}, \\
    & c_- = \frac{16\pi^2}{9}\frac{\delta^2 \db^2 \Lambda_-^2 - 12 \delta^2 - 12 \db^2}{\delta^3\db^3 \Lambda_-^4}.
    \end{aligned}
\end{equation}
where we must optimize over $0<\Lambda_\pm<\frac{\sqrt{2}\pi}{\gamma}$ to get the tightest bound while keeping $\delta,\db$ arbitrary.\\

The condition for the lower bound to be positive is 
\begin{equation}
    \frac{1}{\delta^2}+\frac{1}{\db^2} < \frac{\Lambda_-^2}{12}\,.
\end{equation}
The allowed region increases as we increase $\Lambda_-$. And the minimum area such that there has to be at least one operator is given by $4\delta\db = \frac{48\gamma^2}{\pi^2} = 4.86\gamma^2 $ at $\delta = \db = \frac{2\sqrt{3}\gamma}{\pi}$. This analysis can be made sensitive to primaries only, thus gives an asymptotic gap between primaries. We suspect that the above does not give the tightest bound for spectral gap (this intuition is coming from the similar analysis done for the spectral gap in $\Delta$, appearing in \cite{Baur,Ganguly:2019ksp})!

Next we can keep $\delta,\db$ arbitrary and optimize over $0<\Lambda_\pm<\frac{\sqrt{2}\pi}{\gamma}$ to get the tightest bound.\\

(a) For lower bound:
\begin{equation}
    \begin{aligned}
    &  \frac{1}{4\delta\db}\int_{h-\delta}^{h+\delta} \int_{\hb-\db}^{\hb+\db} \text{d}F(h,\hb') \geq \frac{\pi^2}{27} \frac{\delta \db}{\delta^2+\db^2} \rho_{*}(h,\hb), \,\,\,\, \frac{\sqrt{\delta^2+\db^2}}{\delta \db} < \frac{\pi}{2\sqrt{3}\gamma}, \\
    &  \frac{1}{4\delta\db}\int_{h-\delta}^{h+\delta} \int_{\hb-\db}^{\hb+\db} \text{d}F(h,\hb') \geq \frac{8\pi^2 \gamma^2 \delta^2 \db^2 - 48 \gamma^4( \delta^2 + \db^2)}{9\pi^2 \delta^3 \db^3}\rho_{*}(h,\hb), \,\,\,\, \frac{\sqrt{\delta^2+\db^2}}{\delta \db} \geq \frac{\pi}{2\sqrt{3}\gamma}.
    \end{aligned}
\end{equation}

(b) For upper bound:

\begin{equation}
	\begin{aligned}
	&	\frac{1}{4\delta\db}\int_{h-\delta}^{h+\delta} \int_{\hb-\db}^{\hb+\db} \text{d}F(h,\hb') \leq \frac{16\pi^2}{9\delta \db (\Lambda^{*}_+)^2} \frac{\rho_{*}(h,\hb)}{\sinc{\big(\frac{\delta \Lambda^*_+}{4}\big)}^4 \sinc{\big(\frac{\db \Lambda^*_+}{4}\big)}^4},  \,\,\, \Lambda^*_+ \leq \frac{\sqrt{2}\pi}{\gamma}, \\
	&   \frac{1}{4\delta\db}\int_{h-\delta}^{h+\delta} \int_{\hb-\db}^{\hb+\db} \text{d}F(h,\hb') \leq \frac{8 \gamma^2}{9\delta \db } \frac{\rho_{*}(h,\hb)}{\sinc{\big(\frac{\delta \sqrt{2}\pi}{4\gamma}\big)}^4 \sinc{\big(\frac{\db \sqrt{2}\pi}{4\gamma}\big)}^4},  \,\,\, \Lambda^*_+ > \frac{\sqrt{2}\pi}{\gamma}.
	\end{aligned}
\end{equation}

where $\Lambda_+^*(\delta,\db)$ is the non-zero least positive solution of 
\begin{equation}
	\delta \Lambda_+ \cot\left(\frac{\delta \Lambda_+}{4}\right) + \db \Lambda_+ \cot\left(\frac{\db\Lambda_+}{4}\right) = 6.
\end{equation}

\subsection{Analysis of $O(1)$ window assuming a twist gap} We have seen that the result proven in the previous subsection holds true when $h$ and $\hb$ are of the same order asymptotically. When we make the analysis sensitive to primary, this feature persists. Nonetheless, we can circumnavigate this issue by assuming an existence of twist gap (as defined in \{\ref{def}\}). One can also do this for the analysis sensitive to all the operators. The only catch is that one has to separately treat the zero twist operators with dimension greater than $c/12$. We revisit the analysis of suppression of heavy region in the light of the above discussion. At first, we take up the analysis for all the operators.

{
We go back to the eq.~\eqref{trick1} and redo the first part of the analysis. We separate out $Z_{H}$ into two pieces, one with zero twist heavy operators, we name it $Z^{(0)}_{H}$, while the other one contains all the heavy operators with non-zero twist, we name it $Z^{(\tau)}_{H}$. We start with analyzing $Z^{(\tau)}_{H}$, assuming a twist gap. We have two scenarios.\\

{\textbf{Scenario I:} If $g\leq \frac{c}{12}$, we have
\begin{equation}\label{trick}
\begin{aligned}
Z^{(\tau)}_{H}\ni &\exp\left[-\frac{4\pi^2 \beta}{\beta^2+t^2}\left(h'-\frac{c}{24}\right)-\frac{4\pi^2 \bb}{\bb^2+\tb^2}\left(\hb'-\frac{c}{24}\right)\right] \\
&\leq e^{\frac{\pi^2 c\left(1-\frac{12g}{c}\right)}{6\bb}+\frac{\pi^2 c\left(1-\frac{12g}{c}\right)}{6\beta}}\exp\left[-\frac{4\pi^2 \beta}{\beta^2+\Lambda_{\pm}^{2}}\left(h'-\frac{g}{2}\right)-\frac{4\pi^2 \bb}{\bb^2+\Lambda_{\pm}^2}\left(\hb'-\frac{g}{2}\right)\right]\\
&\leq e^{\frac{\pi^2 c\left(1-\frac{12g}{c}\right)}{6\bb}+\frac{\pi^2 c\left(1-\frac{12g}{c}\right)}{6\beta}}\exp\left[-\frac{4\pi^2 \beta_*}{\beta_*^2+\Lambda_{\pm}^{2}}\left(h'+\hb'-g\right)\right]\\
&\leq e^{\frac{\pi^2 c\left(1-\frac{12g}{c}\right)\left(\frac{1}{\beta}+\frac{1}{\bb}\right)}{6}} \exp\left[-\frac{4\pi^2 \beta_*}{\beta_*^2+\Lambda_{\pm}^{2}}\left(h'+\hb'-\frac{c}{12}\right)\right]\,.
\end{aligned}
\end{equation}
Here going from the first inequality to the second one, we used $2h'>g,2\hb'>g$; going from the second one to the third (last) one, we used $g\leq c/12$. The rest of the analysis goes in a similar manner, as done subsequently after \eqref{trick1} and we deduce:
\begin{align}\label{trickresult}
\Lambda_{\pm}< \frac{\sqrt{2}\pi}{\sigma}\,,\quad \sigma^2\equiv \left(\frac{c}{12g}\right)\,.
\end{align}

\textbf{Scenario II}: If $g\geq \frac{c}{12}$, we have $h'>c/24$ and $\hb'>c/24$, leading to
\begin{equation}
\begin{aligned}
Z^{(\tau)}_{H}\ni &\exp\left[-\frac{4\pi^2 \beta}{\beta^2+t^2}\left(h'-\frac{c}{24}\right)-\frac{4\pi^2 \bb}{\bb^2+\tb^2}\left(\hb'-\frac{c}{24}\right)\right] \\
&\leq \exp\left[-\frac{4\pi^2 \beta_*}{\beta_*^2+\Lambda_{\pm}^{2}}\left(h'+\hb'-\frac{c}{12}\right)\right]\,.
\end{aligned}
\end{equation}
The rest of the analysis goes in a similar manner and we deduce:
\begin{align}
\Lambda_{\pm}< \sqrt{2}\pi\,.
\end{align}
}
Now we come back to analyzing the zero twist heavy sector. For this sector, $h^*=\text{max}(h,\hb)=\Delta>c/12>c/24$, thus we have 
\begin{equation}
\begin{aligned}
&Z^{(0)}_{H}\\
&=\sum_{h'}\exp\left[-\frac{4\pi^2 \beta}{\beta^2+t^2}\left(h'-\frac{c}{24}\right)+\frac{\pi^2 \bb c}{6(\bb^2+\tb^2)}\right]+\sum_{\hb'}\exp\left[\frac{\pi^2 \beta c}{6(\beta^2+t^2)}-\frac{4\pi^2 \bb}{\bb^2+\tb^2}\left(\hb'-\frac{c}{24}\right)\right] \\
&\leq \sum_{h'}\exp\left[-\frac{4\pi^2 \beta}{\beta^2+\Lambda^2}\left(\Delta_{h',0}-\frac{c}{24}\right)+\frac{\pi^2c}{6\bb}\right]+\sum_{\hb'}\exp\left[\frac{\pi^2 c}{6\beta}-\frac{4\pi^2 \bb}{\bb^2+\Lambda^2}\left(\Delta_{0,\hb'}-\frac{c}{24}\right)\right]\\
&\leq e^{\frac{\pi^2c}{6\bb}}\sum_{h'}\exp\left[-\frac{4\pi^2 \beta}{\beta^2+\Lambda^2}\left(\Delta_{h',0}-\frac{c}{24}\right)\right]+e^{\frac{\pi^2c}{6\beta}}\sum_{\hb'}\exp\left[-\frac{4\pi^2 \bb}{\bb^2+\Lambda_{\pm}^2}\left(\Delta_{0,\hb'}-\frac{c}{24}\right)\right]\\
&\leq e^{\frac{\pi^2c}{6\bb}}\sum_{h'}\exp\left[-\frac{4\pi^2 \beta}{\beta^2+\Lambda_{\pm}^2}\left(\Delta_{h',0}-\frac{c}{12}\right)\right]+e^{\frac{\pi^2c}{6\beta}}\sum_{\hb'}\exp\left[-\frac{4\pi^2 \bb}{\bb^2+\Lambda_{\pm}^2}\left(\Delta_{0,\hb'}-\frac{c}{12}\right)\right]\\
&\leq e^{\frac{\pi^2c}{6\bb}}\sum_{h'}\exp\left[-\frac{4\pi^2 \beta}{\beta^2+\Lambda_{\pm}^2}\left(\Delta_{h',0}-\frac{c}{12}\right)\right]+e^{\frac{\pi^2c}{6\beta}}\sum_{\hb'}\exp\left[-\frac{4\pi^2 \bb}{\bb^2+\Lambda_{\pm}^2}\left(\Delta_{0,\hb'}-\frac{c}{12}\right)\right]\\
&\leq e^{\frac{\pi^2c}{6\bb}}\sum_{\Delta}\exp\left[-\frac{4\pi^2 \beta}{\beta^2+\Lambda_{\pm}^2}\left(\Delta-\frac{c}{12}\right)\right]+\sum_{\Delta}e^{\frac{\pi^2c}{6\beta}}\exp\left[-\frac{4\pi^2 \bb}{\bb^2+\Lambda_{\pm}^2}\left(\Delta-\frac{c}{12}\right)\right]\\
&\leq e^{\pi \sqrt{\frac{c\hb}{6}}} Z_{H,\Delta}\left(\frac{4\pi^2 \beta}{\beta^2+\Lambda_{\pm}^2}\right) + e^{\pi \sqrt{\frac{ch}{6}}} Z_{H,\Delta}\left(\frac{4\pi^2 \bb}{\bb^2+\Lambda_{\pm}^2}\right)\,.
\end{aligned}
\end{equation}
The subscript on $\Delta$ in the second line denotes the actual conformal weights of the operator and in the penultimate line, we have extended the sum to all the heavy operators. Now one can see the zero twist heavy sector is suppressed as long as we choose $\Lambda_{\pm}< \sqrt{2}\pi$. Thus, combining everything, we have 
\begin{equation}
\begin{aligned}
\Lambda_\pm &< \begin{cases}
\text{min}\left(\sqrt{2}\pi,\frac{\sqrt{2}\pi}{\zeta}\right)=\frac{\sqrt{2}\pi}{\zeta}\,, &g\leq c/12\\
\sqrt{2}\pi\,, &g\geq c/12
\end{cases} 
\end{aligned}
\end{equation}
where $\zeta^2=\frac{c}{12g}$. We can combine the above to write for all $g$, 
\begin{equation}
\Lambda_{\pm}< \text{min}\left(\sqrt{2}\pi,\frac{\sqrt{2}\pi}{\zeta}\right)\,.
\end{equation}
The above has immediate implication in terms of asymptotic gap for all the operators, in particular, the gap does not depend on the term $\gamma$ anymore. Nonetheless, as we already know existence of descendants asymptotically, the asymptotic gap for all the operators is not so illuminating, so we will not illustrate upon this. Rather we come back to this when we make our analysis sensitive to primaries and in that scenario, the result about asymptotic gap is indeed illuminating.

\paragraph{Analysis for primaries: twist gap complementary to asymptotic spectral gap:} 
The analysis for primaries proceeds in a similar manner.  We will be estimating the following object
\begin{align}\label{primary}
\exp\left[S^{\text{Vir}}_{\delta,\db}\right]=\int_{\hb-\db}^{\hb+\db} \int_{h-\delta}^{h+\delta} \text{d}h'\ \text{d}\hb' \rho^{\text{Vir}} (h',\hb')\,,
\end{align}
where $\rho^{\text{Vir}} (h',\hb')$ is the density of primaries. Instead of the partition function we consider the following object (for $c>1$, the expansion of this object is universal) 
\begin{equation}
\begin{aligned}
&Z_{\text{primary}}(\beta,\bb)\\
&\equiv \eta(\beta)\eta(\bb) Z(\beta) = e^{\beta\frac{c-1}{24}+\bb\frac{c-1}{24}}\left[\left(1-e^{-\beta}\right)\left(1-e^{-\bb}\right)+\sum_{h'\neq 0,\hb'\neq 0}d_{h',\hb'}e^{-\beta h'-\bb\hb'}\right]\,.
\end{aligned}
\end{equation}
Under modular transformation, we have
\begin{equation}
\begin{aligned}
Z_{\text{primary}}(\beta,\bb)=\sqrt{\frac{2\pi}{\beta}}\sqrt{\frac{2\pi}{\bb}}Z_{\text{primary}}\left(\frac{4\pi^2}{\beta},\frac{4\pi^2}{\bb}\right)\,.
\end{aligned}
\end{equation}
Then we define the crossing $\rho_*^{\text{Vir}}(h',\hb')=\rho_*^{\text{Vir}}(h')\rho_*^{\text{Vir}}(\hb')$ to reproduce the high temperature behavior of $Z_{\text{primary}}(\beta,\bb)$ i.e we have
\begin{equation}
\begin{aligned}
\int_{0}^{\infty} \text{d}h'\ e^{-\beta\left(h'-\frac{c-1}{24}\right)} \rho_*^{\text{Vir}}(h')&=\sqrt{\frac{2\pi}{\beta}}\left(\exp\left[\frac{\pi^2(c-1)}{6\beta}\right]-\exp\left[\frac{\pi^2(c-25)}{6\beta}\right]\right)\,,\\
\int_{0}^{\infty}\text{d}\hb'\ e^{-\bb\left(\hb'-\frac{c-1}{24}\right)} \rho_*^{\text{Vir}}(\hb')&=\sqrt{\frac{2\pi}{\bb}}\left(\exp\left[\frac{\pi^2(c-1)}{6\beta}\right]-\exp\left[\frac{\pi^2(c-25)}{6\bb}\right]\right)\,.\\
\end{aligned}
\end{equation}

Explicitly, $\rho_*^{\text{Vir}}$ would be given by the following function: 
\begin{equation}\label{defvir}
\rho_*^{\text{Vir}}(h')=\begin{cases}
0\ \text{if}\ h'<\frac{c-1}{24}\\
\frac{\sqrt{2}}{\sqrt{h-\frac{c-1}{24}}} \left[\cosh \left(4\pi\sqrt{\frac{(c-1)}{24} \left(h-\frac{c-1}{24}\right)}\right)-\cosh \left(4\pi\sqrt{\frac{(c-25)}{24} \left(h-\frac{c-1}{24}\right)}\right)\right]\,.
\end{cases}
\end{equation}
The analysis pertaining to the estimation of the heavy part presented before for the analysis of all the operators can be used as a stepping stone for a similar analysis for primaries for $c>1$ CFTs. We again use bandlimited functions and
%
%
we deduce that the support $\Lambda_{\pm}$ has to satisfy\footnote{Should we not assume twist gap, we would have $\Lambda_{\pm}< \frac{\sqrt{2}\pi}{\gamma}$, just like the analysis for all the operators without assuming twist gap.}:
\begin{align}\label{trickresultp}
\Lambda_{\pm}< \text{min}\left(\frac{\sqrt{2}\pi}{\zeta_p},\sqrt{2}\pi\right)\,,\quad \zeta_p^2\equiv \left(\frac{c-1}{12g}\right)\,.
\end{align}
The leading answer comes out to be 
\begin{equation}
\begin{aligned}
&\frac{1}{2}\frac{c_-}{\sqrt{h-\frac{c-1}{24}}\sqrt{\hb-\frac{c-1}{24}}}\exp\left[2\pi\left(\sqrt{\frac{(c-1)h}{6}}+\sqrt{\frac{(c-1)\hb}{6}}\right)\right]\\
&\leq\frac{1}{4\delta\db} \exp\left[S^{\text{Vir}}_{\delta,\db}\right] \leq \\
&\frac{1}{2}\frac{c_+}{\sqrt{h-\frac{c-1}{24}}\sqrt{\hb-\frac{c-1}{24}}}\exp\left[2\pi\left(\sqrt{\frac{(c-1)h}{6}}+\sqrt{\frac{(c-1)\hb}{6}}\right)\right]\,,
\end{aligned}
\end{equation}
where $c_\pm$ is defines as in the Eq.~\eqref{def:cpm}.

\paragraph{Asymptotic gap:} Now we come back to our discussion of asymptotic gap of primaries. We use the function given in Eq.~\eqref{lower} but now with constraint as given in Eq.~\eqref{trickresultp}. Thus the asymptotic binding square will have length $\frac{4\sqrt{3}\sigma}{\pi}$ and the binding circle would have radius $\frac{r\sigma}{\sqrt{2}}+\epsilon_g$ with $\epsilon_g>0$, and $\sigma$, $r$ are given by
\begin{align}
\sigma = \text{max}\left(1,\frac{c-1}{12g}\right)\,,\quad r=\frac{4\sqrt{3}}{\pi}\,.
\end{align}

If we consider tensoring the chiral Monster CFT with its antichiral avatar, we find $g=4$ and our result predicts that the asymptotic spectral gap involves a circle of radius $\frac{2\sqrt{6}}{\pi}$ irrespective of how infinity is approached. This is above the suspected optimal value $1$ (see fig.~\ref{fig:monster2}). In a unitary compact CFT without conserved currents, there is a bound on twist gap\cite{Collier:2016cls}:
\begin{align}
g\leq \frac{c-1}{12}\,.
\end{align}
In that scenario, we have 
\begin{align}
\Lambda_{\pm}< \text{min}\left(\frac{\sqrt{2}\pi}{\zeta_p},\sqrt{2}\pi\right)=\frac{\sqrt{2}\pi}{\zeta_p}\,,\quad \zeta_p^2\equiv \left(\frac{c-1}{12g}\right)\,.
\end{align}

As a result, we deduce the universal inequality satisfied by the ``areal" spectral gap $A$ and twist gap $g$:
\begin{align}
A g \leq \frac{\pi(c-1)r^2}{12} \,,
\end{align}
where we have shown $r=\frac{4\sqrt{3}}{\pi}\simeq 2.21>1$ and we suspect that it can be made to $1$.

}

\section{Lemma: density of states on strip of order one width}\label{lemma1}
In this section, we prove a lemma which is going to play a pivotal role in the next section, where we are going to prove an asymptotic result for the integrated density of states i.e number of states upto a large $(h,\hb)$ threshold. This also helps us to derive the asymptotic spectral gap via strip like regions as defined in \{\ref{stripping}\}. We start by defining the following functions
\begin{equation}
\begin{aligned}
Q(h,\bb)&\equiv \int_{h-\delta}^{h+\delta}\text{d}h^\prime\ \int_{0}^{\infty}\text{d}\hb'\ \rho(h',\hb') e^{-\bb\hb'}\,,\\
P(\hb,\beta)&\equiv \int_{\hb-\db}^{\hb+\db}\text{d}\hb^\prime\ \int_{0}^{\infty}\text{d}h'\ \rho(h',\hb') e^{-\beta h'} \,.
\end{aligned}
\end{equation}
The aim of this section is to prove the following lemma: 
\begin{align}
e^{\bb\hb}Q(h,\bb)\underset{\bb=\pi\sqrt{\frac{c}{6\hb}}}{=}&\ O\left(h^{-3/4}\exp\left[2\pi\left(\sqrt{\frac{ch}{6}}+\sqrt{\frac{c\hb}{6}}\right)\right]\right)\,,\\
e^{\beta h}P(\hb,\beta)\underset{\beta=\pi\sqrt{\frac{c}{6h}}}{=}&\ O\left(\hb^{-3/4}\exp\left[2\pi\left(\sqrt{\frac{ch}{6}}+\sqrt{\frac{c\hb}{6}}\right)\right]\right)\,.
\end{align}

Let us focus on the quantity $Q$, the argument for $P$ follows in a similar manner. In order to estimate $Q$, we write down the master inequality:
\begin{align}
\nonumber &e^{\beta(h-\delta)}\int_{0}^{\infty} dh^\prime\ \int_{0}^{\infty}\text{d}\hb'\ \rho(h',\hb') \phi_-(h')e^{-\beta h'-\bb\hb'} \\
\label{eq:goal1}& \leq Q(h,\bb)\\
\nonumber &\leq e^{\beta(h+\delta)}\int_{0}^{\infty} dh^\prime\ \int_{0}^{\infty}\text{d}\hb'\ \rho(h',\hb') \phi_+(h')e^{-\beta h'-\bb\hb'} \,,
\end{align}
where we have used 
\begin{align}
\phi_-(h') \leq \Theta\left(h'\in\left[h-\delta,h+\delta\right]\right) \leq \phi_+(h')\,.
\end{align}

Next we note that
\begin{equation}
\begin{aligned}
&\int_{0}^{\infty} \text{d}h^\prime\ \int_{0}^{\infty}\text{d}\hb'\ \rho(h',\hb') \phi_{\pm}(h')e^{-\beta h'-\bb\hb'} \\
&= e^{-\beta c/24-\bb c/24}\int_{-\infty}^{\infty} \text{d}t\ e^{-\imath tc/24}Z(\beta+\imath t,\bb) \hat{\phi}_{\pm}(t)\\
&=e^{-\beta c/24-\bb c/24}\int_{-\infty}^{\infty} \text{d}t\ e^{-\imath tc/24}Z\left(\frac{4\pi^2}{\beta+\imath t},\frac{4\pi^2}{\bb}\right) \hat{\phi}_{\pm}(t)\,.
\end{aligned}
\end{equation}

Now we separate the contribution to $Z\left(\frac{4\pi^2}{\beta+\imath t},\frac{4\pi^2}{\bb}\right)$ into two pieces $Z_L$ (contribution from the ``light" sector) and $Z_H$ (contribution from the ``heavy" sector). The inequality in \eqref{eq:goal1} can be written as
\begin{equation}
\begin{aligned}
&e^{\beta(h-c/24-\delta)-\bb c/24}\int_{-\infty}^{\infty} \text{d}t\ e^{-\imath tc/24}Z_L\left(\frac{4\pi^2}{\beta+\imath t},\frac{4\pi^2}{\bb}\right) \hat{\phi}_{-}(t)\\
&-e^{\beta(h-c/24-\delta)-\bb c/24}\bigg|\int_{-\infty}^{\infty} \text{d}t\ e^{-\imath tc/24}Z_H\left(\frac{4\pi^2}{\beta+\imath t},\frac{4\pi^2}{\bb}\right) \hat{\phi}_{-}(t)\bigg|\\
&\leq Q(h,\bb) \\
&\leq e^{\beta(h-c/24+\delta)-\bb c/24}\int_{-\infty}^{\infty} \text{d}t\ e^{-\imath tc/24}Z_L\left(\frac{4\pi^2}{\beta+\imath t},\frac{4\pi^2}{\bb}\right) \hat{\phi}_{+}(t)\\
&+e^{\beta (h-c/24+\delta)-\bb c/24}\bigg|\int_{-\infty}^{\infty} \text{d}t\ e^{-\imath tc/24}Z_H\left(\frac{4\pi^2}{\beta+\imath t},\frac{4\pi^2}{\bb}\right) \hat{\phi}_{+}(t)\bigg|\\
\end{aligned}
\end{equation}

For notational simplicity, let us name the terms
\begin{align}
I_{\pm}^{1}&=e^{\beta(h-c/24\pm\delta)-\bb c/24}\int_{-\infty}^{\infty} \text{d}t\ e^{-\imath tc/24}Z_L\left(\frac{4\pi^2}{\beta+\imath t},\frac{4\pi^2}{\bb}\right) \hat{\phi}_{\pm}(t)\,,\\
I_{\pm}^{2}&= e^{\beta (h-c/24\pm\delta)-\bb c/24}\bigg|\int_{-\infty}^{\infty} \text{d}t\ e^{-\imath tc/24}Z_H\left(\frac{4\pi^2}{\beta+\imath t},\frac{4\pi^2}{\bb}\right) \hat{\phi}_{\pm}(t)\bigg|\,.
\end{align}

{The idea is to show that $I^2_{\pm}$ is subleading with respect to $I^1_{\pm}$. The argument closely follows the argument presented in \S\ref{lemma0}. Let us concentrate on $I^2_{\pm}$ first. We have
\begin{equation}
I^2_{\pm} \leq e^{\beta (h-c/24\pm\delta)-\bb c/24}\int_{-\infty}^{\infty} \text{d}t\ Z_H\left(\frac{4\pi^2\beta}{\beta^2+t^2},\frac{4\pi^2}{\bb}\right) \bigg|\hat{\phi}_{\pm}(t)\bigg|\,.
\end{equation}

We notice that (for the heavy sector, $h'+\hb'>c/12$, thus one of them has to be greater than $c/24$)
\begin{equation}
\begin{aligned}
Z_H\left(\frac{4\pi^2\beta}{\beta^2+t^2},\frac{4\pi^2}{\bb}\right) &\ni e^{-\frac{4\pi^2\beta}{\beta^2+t^2}\left(h'-\frac{c}{24}\right)-\frac{4\pi^2}{\bb}\left(\hb'-\frac{c}{24}\right)}\\
&\leq \begin{cases}
e^{\frac{\pi^2c}{6\beta}}e^{-\frac{4\pi^2\beta}{\beta^2+\Lambda_{\pm}^2}h'-\frac{4\pi^2\bb}{\bb^2+\Lambda_{\pm}^2}\left(\hb'-\frac{c}{24}\right)}\,,&\hb'>c/24\\
e^{\frac{\pi^2c}{6\bb}}e^{-\frac{4\pi^2\beta}{\beta^2+\Lambda_{\pm}^2}\left(h'-\frac{c}{24}\right)-\frac{4\pi^2\bb}{\bb^2+\Lambda_{\pm}^2}\hb'}\,,&h'>c/24
\end{cases}\\
&\leq \begin{cases}
e^{\frac{\pi^2c}{6\beta}}e^{-\frac{4\pi^2\beta_*}{\beta_*^2+\Lambda_{\pm}^2}h'-\frac{4\pi^2\beta_*}{\beta_*^2+\Lambda_{\pm}^2}\left(\hb'-\frac{c}{24}\right)}\,,&\hb'>c/24\\
e^{\frac{\pi^2c}{6\bb}}e^{-\frac{4\pi^2\beta_*}{\beta_*^2+\Lambda_{\pm}^2}h'-\frac{4\pi^2\beta_*}{\beta_*^2+\Lambda_{\pm}^2}\left(\hb'-\frac{c}{24}\right)}\,,&h'>c/24
\end{cases}\\
&\leq \begin{cases}
e^{\frac{\pi^2c}{6\beta}}e^{\frac{-\pi^2c\beta_*}{6(\beta_*^2+\Lambda_{\pm}^2)}}e^{-\frac{4\pi^2\beta_*}{\beta_*^2+\Lambda_{\pm}^2}\left(h'+\hb'-\frac{c}{12}\right)}\,,&\hb'>c/24\\
e^{\frac{\pi^2c}{6\bb}}e^{-\frac{\pi^2c\beta_*}{6(\beta_*^2+\Lambda_{\pm}^2)}}e^{-\frac{4\pi^2\beta_*}{\beta_*^2+\Lambda_{\pm}^2}\left(h'+\hb'-\frac{c}{12}\right)}\,.&h'>c/24
\end{cases}
\end{aligned}
\end{equation}

Thus we have
\begin{equation}
\begin{aligned}
I^2_{\pm} &\leq e^{\beta (h+c/24\pm\delta)-\bb c/24}\int_{-\infty}^{\infty} \text{d}t\ Z_H\left(\frac{4\pi^2\beta}{\beta^2+t^2},\frac{4\pi^2}{\bb}\right) \bigg|\hat{\phi}_{\pm}(t)\bigg|\\
&\underset{\bb,\beta < 2\pi}{\leq} 
e^{\beta (h-c/24\pm\delta)-\bb c/24}\left(e^{\frac{\pi^2c}{6\bb}}+e^{\frac{\pi^2c}{6\beta}}\right)e^{-\frac{\pi^2\beta_* c}{6(\beta_*^2+\Lambda_{\pm}^2)}}Z_{H,\Delta}\left(\frac{4\pi^2\beta_*}{\beta_*^2+\Lambda_{\pm}^2}\right)\int_{-\infty}^{\infty} \text{d}t\  \bigg|\hat{\phi}_{\pm}(t)\bigg|\,.
\end{aligned}
\end{equation}
The above analysis is analogous to the one presented in \S\ref{lemma0}. Now we choose 
\begin{align}
\beta=\pi\sqrt{\frac{c}{6h}}\,,\bb=\pi\sqrt{\frac{c}{6\hb}}\,,
\end{align}
and use the HKS bound \cite{HKS} to estimate $Z_{H,\Delta}\left(\frac{4\pi^2\beta_*}{\beta_*^2+\Lambda_{\pm}^2}\right)$. This leads to the following inequality:
\begin{equation}
\begin{aligned}\label{i2}
I^2_{\pm}&\leq e^{\pi \sqrt{\frac{ch}{6}}} \left(e^{\pi \sqrt{\frac{c \hb}{6}}} +e^{\pi \sqrt{\frac{c h}{6}}}\right)\int_{-\infty}^{\infty} \text{d}t\  \bigg|\hat{\phi}_{\pm}(t)\bigg|
\begin{cases}
e^{2\pi \sqrt{\frac{c h}{6}}\frac{\Lambda_{\pm}^2}{4\pi^2}} &h\geq \hb\\
e^{2\pi \sqrt{\frac{c \hb}{6}}\frac{\Lambda_{\pm}^2}{4\pi^2}} &h<\hb
\end{cases}\,,
\end{aligned}
\end{equation}

To estimate $I^1_{\pm}$, we consider $\rho_*(h,\hb)$, the crossing kernel, defined as 
\begin{align}
\exp\left[\frac{\pi^2c}{6\beta}+\frac{\pi^2c}{6\bb}\right]=\int_{0}^\infty dh\ \int_{0}^{\infty}\ d\hb\ \rho_{*}(h,\hb) e^{-\beta (h-c/24)-\bb(\hb-c/24)}\,.
\end{align}
Here $\rho_*(h,\hb)=\rho_*(h)\rho_*(\hb)$, and $\rho_*(h),\rho_*(\hb)$ are given by 
\begin{equation}
\begin{aligned}
\rho_*(x)&=\pi\sqrt{\frac{c}{6}}\frac{I_{1}\left(2\pi\sqrt{\frac{c}{3}\left(x-\frac{c}{24}\right)}\right)}{\sqrt{x-\frac{c}{24}}}\theta\left(x-\frac{c}{24}\right)+\delta\left(x-\frac{c}{24}\right)\,,\\
&\underset{x\to\infty}{=} \left(\frac{c}{96x^3}\right)^{\frac{1}{4}}\exp\left[2\pi\sqrt{\frac{cx}{6}}\right]\,.
\end{aligned}
\end{equation}
Hence, we have
\begin{equation}
\begin{aligned}\label{i1}
I^{1}_{\pm}&=e^{\pi\sqrt{\frac{ch}{6}}}\int_{0}^\infty \text{d}h'\ \int_{0}^{\infty}\ \text{d}\hb'\ \rho_{*}(h',\hb') e^{-\sqrt{\frac{c}{6h}} h'-\sqrt{\frac{c}{6\hb}}\hb'}\phi_{\pm}(h')\\
&=2\delta c_{\pm} e^{\pi\sqrt{\frac{c\hb}{6}}}\rho_*(h)\,,
\end{aligned}
\end{equation}
where we have used separability of $\rho_*$ in the variable $h$ and $\hb$ and have defined
\begin{align}
c_{\pm}=\frac{1}{2}\int_{-\infty}^{\infty} dx\ \phi_{\pm}(h+\delta x)\,.
\end{align}

Comparing the inequalities \eqref{i1} and \eqref{i2}, we see that by choosing $\gamma\Lambda_{\pm}<\sqrt{2}\pi$, with $\text{max}(h,\hb)=\gamma^4\text{min}(h,\hb)$; one can make $I^{2}_{\pm}$ subleading,} consequently in the $h,\hb\to\infty$ limit we have
\begin{align}\label{lm1}
2\delta c_-\left(\frac{c}{96\hb^3}\right)^{-\frac{1}{4}} \rho_*(h,\hb)\leq e^{\pi\sqrt{\frac{c\hb}{6}}}Q(h) \leq 2\delta c_+\left(\frac{c}{96\hb^3}\right)^{-\frac{1}{4}} \rho_*(h,\hb)\,.
\end{align}

By symmetry we obtain
\begin{align}\label{lm2}
2\delta c'_-\left(\frac{c}{96h^3}\right)^{-\frac{1}{4}} \rho_*(h,\hb)\leq e^{\pi\sqrt{\frac{ch}{6}}}P(\hb) \leq 2\delta c'_+ \left(\frac{c}{96h^3}\right)^{-\frac{1}{4}} \rho_*(h,\hb)\,.
\end{align}

The best possible value of $c_{\pm}$ ($c'_{\pm}$) can be obtained from \cite{Ganguly:2019ksp}. For the verification purpose, here we choose the function given in \cite{Baur} for estimating $Q$
\begin{align}
\phi_{+}(h^\prime)&=\left(\frac{\sin\left(\frac{\Lambda_{+}\delta}{4}\right)}{\frac{\Lambda_{+}\delta}{4}}\right)^{-4} \left(\frac{\sin\left(\frac{\Lambda_{+}(h-h')}{4}\right)}{\frac{\Lambda_{+}(h-h')}{4}}\right)^4\,,\\
\phi_{-}(h^\prime)&=\left(1-\left(\frac{h-h'}{\delta}\right)^2\right)\left(\frac{\sin\left(\frac{\Lambda_{-}(h-h')}{4}\right)}{\frac{\Lambda_{-}(h-h')}{4}}\right)^4\,.
\end{align}
 The above function yields almost the same bound as found in \cite{Baur}, except for the fact that we have to take care of the constraint $\Lambda < \frac{\sqrt{2}\pi}{\gamma}$. In particular we find the following bounding function $s_\pm(\delta)=\log c_{\pm}$:
\begin{equation}
\begin{aligned}
c_+&=\begin{cases}
\frac{\pi}{3}\left(\frac{\pi\delta}{2\sqrt{2}\gamma}\right)^3\left(\sin\left(\frac{\pi\delta}{2\sqrt{2}\gamma}\right)\right)^{-4}\,&,\quad \delta<\frac{\gamma a_*}{\sqrt{2}\pi}\\
2.02\,&, \quad \delta\geq \frac{\gamma a_*}{\sqrt{2}\pi}
\end{cases}\\
c_-&=\begin{cases}
\frac{2\sqrt{2}\gamma}{3\pi\delta^3}\left(\delta^2-\frac{6\gamma^2}{\pi^2}\right)\,&,\quad \delta<\frac{6\gamma }{\sqrt{2}\pi}\\
0.46\,&, \quad \delta\geq \frac{6\gamma}{\sqrt{2}\pi}
\end{cases}
\end{aligned}
\end{equation}
where $a_*=3.38$ \cite{Baur}. We verify the eq.~\eqref{lm1} in the \S\ref{verify} using the above values of $c_\pm$ (in particular, see the fig.~\ref{fig:1}). Similar verification can be done for the eq.~\eqref{lm2} as well.\\

The analysis with the assumption of twist gap $g$ proceeds as in the end of \S\ref{lemma0}. We do not repeat the analysis here. We just state the result. In that scenario, one obtains
\begin{align}
\Lambda_{\pm}< \text{min}\left(\frac{\sqrt{2}\pi}{\zeta},\sqrt{2}\pi\right)\,,\quad \zeta^2\equiv \left(\frac{c}{12g}\right)\,.
\end{align}

If we make it specific for primaries, $c\mapsto c-1$ and we have
\begin{equation} 
\begin{aligned}
\Lambda_{\pm}&< \text{min}\left(\frac{\sqrt{2}\pi}{\zeta_p},\sqrt{2}\pi\right)\\
&=\frac{\sqrt{2}\pi}{\zeta_p}\,,\quad \zeta_p^2\equiv \left(\frac{c-1}{12g}\right)\geq 1\,.
\end{aligned}
\end{equation}
where the second equality follows only if there is no conserved currents because of the bound on twist gap.\\

We can use this lemma to prove the result about asymptotic spectral gap in terms of strips, as mentioned in \{\ref{stripping}\}. For this purpose, we use the magic function introduced in \cite{Ganguly:2019ksp}. We can let $\gamma=1$ in the above analysis. We have to keep in mind that now the support of the Fourier transform of $\phi_-$ satisfies $\Lambda < \frac{\sqrt{2}\pi}{\gamma}$, thus the minimal value of $\delta$ comes out to be $\frac{1}{\sqrt{2}}$ in stead of $1/2$ as in \cite{Ganguly:2019ksp}. 

\section{The integrated density of states}\label{mainproof}
\subsection{The main 2D Tauberian theorem}
We prove in this section
\begin{align}
\nonumber F(h,\hb)&\equiv \int_{0}^{h}\text{d}h' \int_{0}^{\hb}\text{d}\hb' \rho(h',\hb')\\
\nonumber&\underset{h,\hb\to\infty}{=}\frac{1}{4\pi^2}\left(\frac{36}{c^2h\hb}\right)^{1/4}\exp\left[2\pi\left(\sqrt{\frac{ch}{6}}+\sqrt{\frac{c\hb}{6}}\right)\right]\left[1+O\left(\tau^{\frac{\Upsilon}{4}-1/2}\right)\right]\,.
\end{align}
where $\tau$ is the twist of the state with $h,\hb$ and given by $\tau=2\text{min}\{h,\hb\}$ and $h=\hb^{\upsilon}$ with $1/2< \upsilon <2$ and $\Upsilon=\text{max}\left(\upsilon,1/\upsilon\right)$. When $\Upsilon=1$, this reduces to the eq~\eqref{masterequation}. In order to prove this we define
\begin{align}
\delta\rho(h,\hb)&=\rho(h,\hb)-\rho_*(h,\hb)\,,\\
\delta\mathcal{L}(\beta,\bb)&=\mathcal{L}_{\rho}(\beta,\bb)-\mathcal{L}_{\rho_*}(\beta,\bb)\,.
\end{align}
Since the leading term is already produced by $\rho_*(h,\hb)$, our job is to show that 
\begin{equation}\label{resulterror}
\begin{aligned}
&\int_{0}^{h}\text{d}h' \int_{0}^{\hb}\text{d}\hb' \delta\rho(h',\hb')\\
&=\frac{1}{4\pi^2}\left(\frac{36c^2}{h\hb}\right)^{1/4}\exp\left[2\pi\left(\sqrt{\frac{ch}{6}}+\sqrt{\frac{c\hb}{6}}\right)\right]O\left(\tau^{\Upsilon/4-1/2}\right)\,.
\end{aligned}
\end{equation}
In particular, we will be showing that
\begin{equation}
\begin{aligned}
&\int_{0}^{h}\text{d}h' \int_{0}^{\hb}\text{d}\hb' \delta\rho(h',\hb')\\
&=\frac{\sqrt{6c}}{4\pi^2}\exp\left[2\pi\left(\sqrt{\frac{ch}{6}}+\sqrt{\frac{c\hb}{6}}\right)\right]\left[O\left(h^{-3/4}\right)+O\left(\hb^{-3/4}\right)\right]\,.
\end{aligned}
\end{equation}
Thus if $\upsilon \in \left(1/2,2\right)$, the error term is suppressed by maximum of $\frac{h^{1/4}}{\sqrt{\hb}}$ and $\frac{\hb^{1/4}}{\sqrt{h}}$, arriving at \eqref{resulterror}.\\

In order to prove the above, we proceed as in \cite{Baur} and introduce the following kernel:
\begin{align}
G(\nu)&=\frac{1}{2\pi\imath}\int_{\beta-\imath\Lambda}^{\beta+\imath\Lambda}\frac{dz}{z}\frac{\Lambda^2+(z-\beta)^2}{\Lambda^2+\beta^2}e^{-\nu z}\,,\quad \nu=h'-h\,,\\
G(\bar\nu)&=\frac{1}{2\pi\imath}\int_{\bb-\imath\Lambda}^{\bb+\imath\Lambda}\frac{d\bar z}{\bar z}\frac{\Lambda^2+(\bar z-\bb)^2}{\Lambda^2+\bb^2}e^{-\bar\nu \bar z}\,,\quad \bar\nu=\hb'-\hb\,.
\end{align}
Here we have done slight abuse of notation. It is implicitly assumed that the function $G(\bar\nu)$ depends on $\bb$ instead of $\beta$. Now it can be shown that \cite{Baur}:
\begin{align}\label{kernel}
G(\nu)G(\bar\nu)=\left[\theta(-\nu)+G_+(\nu)\theta(\nu)+G_-(\nu)\theta(-\nu)\right]\left[\theta(-\bar\nu)+G_+(\bar\nu)\theta(\bar\nu)+G_-(\bar\nu)\theta(-\bar\nu)\right]\,,
\end{align}
where $G_{\pm}$ is defined exactly like in \cite{Baur} for both the variable $\nu$ and $\bar \nu$.
At this point, we use the kernel given in \eqref{kernel} and integrate it against $\delta\rho(\Delta)$. This yields us the following equation
\begin{align}
\nonumber &\int_{0}^{h}\text{d}h' \int_{0}^{\hb}\text{d}\hb' \delta\rho(h',\hb')=\int_{0}^{\infty}\text{d}h' \int_{0}^{\infty}\text{d}\hb' \delta\rho(h',\hb') G(\nu)G(\bar\nu)\\
&+\int_{0}^{\infty}\text{d}h' \int_{0}^{\infty}\text{d}\hb' \delta\rho(h',\hb') \left[-\theta(-\bar\nu)\theta(\nu)G_{+}(\nu)-\theta(-\bar\nu)\theta(-\nu)G_{-}(\nu)-(\nu\to\bar \nu)\right]\\
\label{line2}&+\int_{0}^{\infty}\text{d}h' \int_{0}^{\infty}\text{d}\hb' \delta\rho(h',\hb') \left[-\theta(-\bar\nu)G_-(\bar\nu)\theta(\nu)G_{+}(\nu)-\theta(-\bar\nu)G_-(\bar \nu)\theta(-\nu)G_{-}(\nu)\right]\\
\label{line3}&+\int_{0}^{\infty}\text{d}h' \int_{0}^{\infty}\text{d}\hb' \delta\rho(h',\hb') \left[-\theta(\bar\nu)G_+(\bar\nu)\theta(\nu)G_{+}(\nu)-\theta(\bar\nu)G_+(\bar \nu)\theta(-\nu)G_{-}(\nu)\right]\,.
\end{align}

Most of the terms can be estimated using techniques from \cite{Baur}. The new players in the game are the cross terms, for example the term: 
\begin{align}
Z=\int_{0}^{h}\text{d}h' \int_{0}^{\hb}\text{d}\hb' \delta\rho(h',\hb') \left[\theta(-\bar\nu)\theta(\nu)G_{+}(\nu)+\theta(-\bar\nu)\theta(-\nu)G_{-}(\nu)+(\nu\to\bar \nu)\right]\,.
\end{align}
Below we will illustrate how to handle these terms. We remark that this is what requires us to prove the lemma in the previous section. For concreteness, consider the following term $\theta(-\bar\nu)\theta(\nu)G_{+}(\nu)$ and analyze it carefully. The analysis for the other terms in $Z$ goes exactly in the same manner. In what follows, we will be using the inequalities for $\beta>0$:
\begin{align}
\theta(-\bar \nu)&\leq e^{-\bb\bar \nu}\,,\\
\label{b1}|G_{\pm}(\nu)|&\leq 2e^{-\beta\nu}\text{min}\left(1,(h-h')^{-2}\right)\,,\\
\label{b2}|G_{\pm}(\bar \nu)|&\leq 2e^{-\bb\bar \nu}\text{min}\left(1,(\hb-\hb')^{-2}\right)\,.
\end{align}
The inequalities \eqref{b1} and \eqref{b2} have been derived in the appendix of \cite{Baur}.

Consider the term 
\begin{align}
Z1=\int_{0}^{\infty}\text{d}h'\ \int_{0}^{\infty}\text{d}\hb'\ \theta(-\bar \nu)\Theta(\nu)G(\nu) \delta\rho(h',\hb')\,.
\end{align}
Clearly, we have
\begin{align}
|Z1| \leq 2e^{\beta h+\bb \hb}\int_{0}^{\infty}\text{d}h'\ \int_{0}^{\infty}\text{d}\hb'\ e^{-\beta h'-\bb \hb'} \left[\rho(h',\hb')+\rho_*(h',\hb')\right] \text{min}\left(1,(h-h')^{-2}\right)\,.
\end{align}
For the term with $\rho_*(h',\hb')$, the estimation procedure mimics the one presented in the section 5 of \cite{Baur}. In particular, we have
\begin{align}
\nonumber &e^{\beta h+\bb \hb}\int_{0}^{\infty}\text{d}h'\ \int_{0}^{\infty}\text{d}\hb'\ e^{-\beta h'-\bb \hb'}\rho_*(h',\hb')\text{min}\left(1,(h-h')^{-2}\right) \\
&= O\left(h^{-3/4}e^{2\pi\left(\sqrt{\frac{ch}{6}}+\sqrt{\frac{c\hb}{6}}\right)}\right)\,.
\end{align}
The $h'$ integral is done via saddle point method, note it is important to have the factor $\text{min}\left(1,(h-h')^{-2}\right)$ for the validity of saddle point approximation. This is why, the $\hb'$ integral can not be done using saddle, hence does not produce any polynomial suppression in $\hb$. Now consider the term 
\begin{align*}
2e^{\beta h+\bb \hb}\int_{0}^{\infty}\text{d}h'\ \int_{0}^{\infty}\text{d}\hb'\ e^{-\beta h'-\bb \hb'}\rho(h',\hb') \text{min}\left(1,(h-h')^2\right)\,,
\end{align*} 
and we divide it into three pieces
\begin{align}
a_1&=2e^{\beta h+\bb \hb}\int_{0}^{h-h^{3/8}}\text{d}h'\ \int_{0}^{\infty}\text{d}\hb'\ e^{-\beta h'-\bb \hb'}\rho(h',\hb') \text{min}\left(1,(h-h')^{-2}\right)\,,\\
a_2&=2e^{\beta h+\bb \hb}\int_{h-h^{3/8}}^{h+h^{3/8}}\text{d}h'\ \int_{0}^{\infty}\text{d}\hb'\ e^{-\beta h'-\bb \hb'}\rho(h',\hb') \text{min}\left(1,(h-h')^{-2}\right)\,,\\
a_3&=2e^{\beta h+\bb \hb}\int_{h+h^{3/8}}^{\infty}\text{d}h'\ \int_{0}^{\infty}\text{d}\hb'\ e^{-\beta h'-\bb \hb'}\rho(h',\hb') \text{min}\left(1,(h-h')^{-2}\right)\,.
\end{align}

The estimate for $a_1$ and $a_3$ again proceeds like in \cite{Baur} and we obtain
\begin{align}
a_1&=O\left(h^{-3/4}e^{2\pi\left(\sqrt{\frac{ch}{6}}+\sqrt{\frac{c\hb}{6}}\right)}\right)\,,\\
a_3&=O\left(h^{-3/4}e^{2\pi\left(\sqrt{\frac{ch}{6}}+\sqrt{\frac{c\hb}{6}}\right)}\right)\,.
\end{align}

The estimation of the term $a_2$ would require the lemma from the previous section \S\ref{lemma1}. We subdivide $a_2$ into three different parts:
\begin{align}
a_{21}&=2e^{\beta h+\bb \hb}\int_{h-h^{3/8}}^{h-1}\text{d}h'\ \int_{0}^{\infty}\text{d}\hb'\ e^{-\beta h'-\bb \hb'}\rho(h',\hb') \text{min}\left(1,(h-h')^{-2}\right)\,,\\
a_{22}&=2e^{\beta h+\bb \hb}\int_{h-1}^{h+1}\text{d}h'\ \int_{0}^{\infty}\text{d}\hb'\ e^{-\beta h'-\bb \hb'}\rho(h',\hb') \text{min}\left(1,(h-h')^{-2}\right)\,,\\
a_{23}&=2e^{\beta h+\bb \hb}\int_{h+1}^{h+h^{3/8}}\text{d}h'\ \int_{0}^{\infty}\text{d}\hb'\ e^{-\beta h'-\bb \hb'}\rho(h',\hb') \text{min}\left(1,(h-h')^{-2}\right)\,.
\end{align}

We have already estimated $a_{22}$ in the previous section  \S\ref{lemma1}. This is basically order one window. Since, the estimation of $a_{21}$ and $a_{23}$ proceeds in similar manner, we would demonstrate the estimation of the term $a_{21}$.
\begin{equation}
\begin{aligned}
a_{21}&=2e^{\beta h+\bb \hb}\int_{h-h^{3/8}}^{h-1}\text{d}h'\ \int_{0}^{\infty}\text{d}\hb'\ e^{-\beta h'-\bb \hb'}\rho(h',\hb') \text{min}\left(1,(h-h')^{-2}\right)\\
&=2e^{\bb \hb}\sum_{k=2}^{h^{3/8}}\int_{h-k}^{h-k+1}\text{d}h'\ \int_{0}^{\infty}\text{d}\hb'\ e^{\beta (h-h')-\bb \hb'}\rho(h',\hb') (h-h')^{-2}\\
&\leq 2e^{\bb \hb}\sum_{k=2}^{h^{3/8}} \frac{e^{\beta k}}{(k-1)^2}\int_{h-k}^{h-k+1}\text{d}h'\ \int_{0}^{\infty}\text{d}\hb'\ e^{-\bb \hb'}\rho(h',\hb')\\
&=O\left(h^{-3/4}e^{2\pi\left(\sqrt{\frac{ch}{6}}+\sqrt{\frac{c\hb}{6}}\right)}\sum_{k=2}^{h^{3/8}} \frac{e^{\beta k}}{(k-1)^2}\right)\\
&=O\left(h^{-3/4}e^{2\pi\left(\sqrt{\frac{ch}{6}}+\sqrt{\frac{c\hb}{6}}\right)}\right)\,,
\end{aligned}
\end{equation}
where going from the third line to the fourth line requires use of lemma proven in the previous section  \S\ref{lemma1}.

The estimation of the terms in \eqref{line2} and \eqref{line3} requires us to divide the $(h,\hb)$ plane into 9 regions (see figure~\ref{analysis}):
\begin{figure}[!ht]
\centering
\includegraphics[scale=0.25]{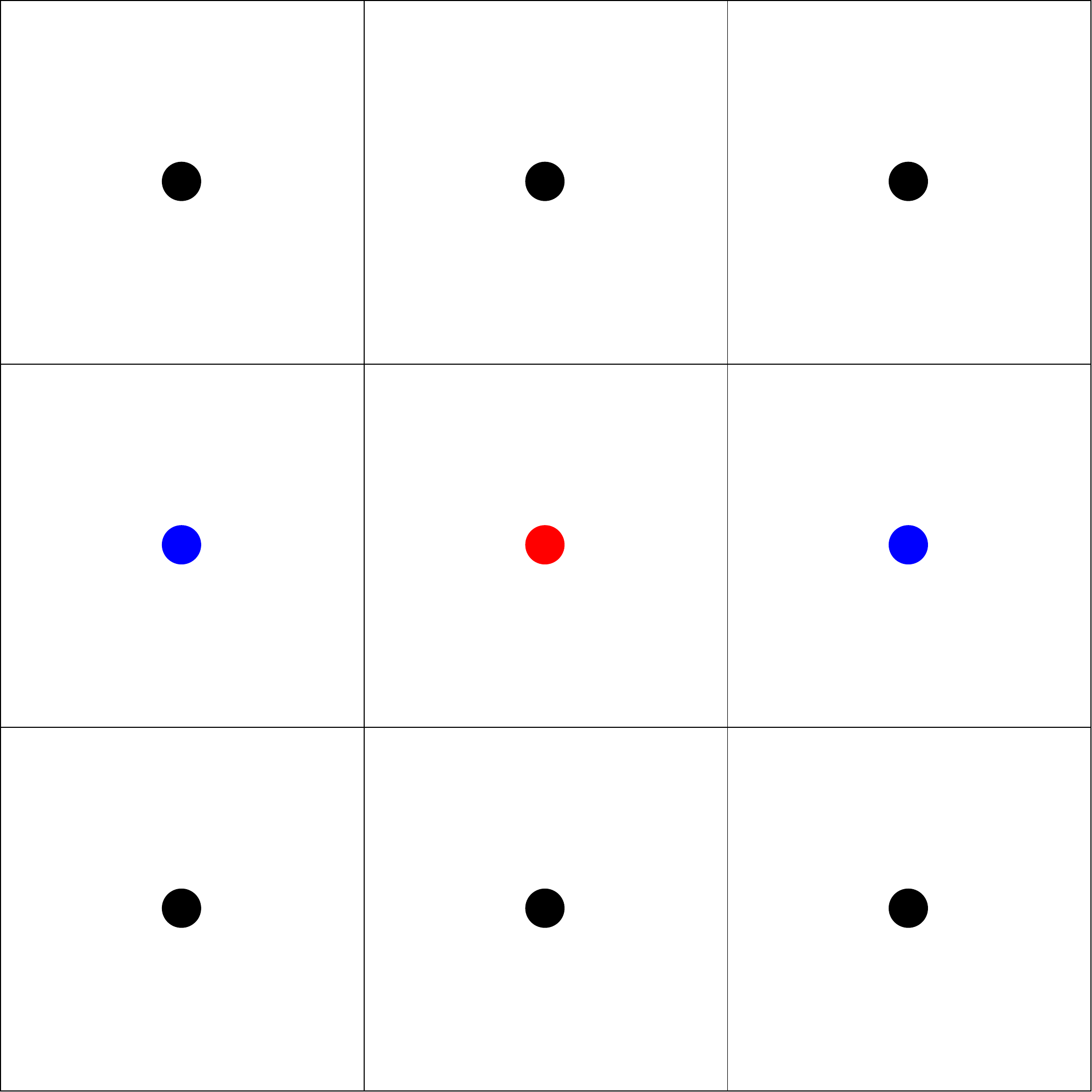}
\caption{The estimation of the terms in \eqref{line2} and \eqref{line3} requires us to divide the $(h,\hb)$ plane into 9 regions. In the figure, the horizontal lines are $h'=0,h-h^{3/8},h+h^{3/8},\infty$ while the vertical lines are $\hb'=0,\hb-\hb^{3/8},\hb+\hb^{3/8},\infty$. The different colors denote the different methods of treating them. As we will see, the region with red dot requires us to further subdivide it. We remark that the corner black ones can be colored blue as well, nonetheless in the main text, we have done the estimation in the ``blue" way.}
\label{analysis}
\end{figure}

\begin{equation}
\begin{aligned}
R_1&= \left\{(h',\hb'):\ h'\in [0,h-h^{3/8}], \hb'\in[0,\hb-\hb^{3/8}]\right\}\,,\\
R_2&=\left\{(h',\hb'):\ h'\in [0,h-h^{3/8}], \hb'\in[\hb-\hb^{3/8},\hb+\hb^{3/8}]\right\}\,,\\
R_3&=\left\{(h',\hb'):\ h'\in [0,h-h^{3/8}], \hb'\in[\hb+\hb^{3/8},\infty]\right\}\,,\\
R_4&=\left\{(h',\hb'):\ h'\in [h-h^{3/8},h+h^{3/8}], \hb'\in[0,\hb-\hb^{3/8}]\right\}\,,\\
R_5&=\left\{(h',\hb'):\ h'\in [h-h^{3/8},h+h^{3/8}], \hb'\in[\hb-\hb^{3/8},\hb+\hb^{3/8}]\right\}\,,\\
R_6&=\left\{(h',\hb'):\ h'\in [h-h^{3/8},h+h^{3/8}], \hb'\in[\hb+\hb^{3/8},\infty]\right\}\,,\\
R_7&=\left\{(h',\hb'):\ h'\in [h+h^{3/8},\infty], \hb'\in[0,\hb-\hb^{3/8}]\right\}\,,\\
R_8&=\left\{(h',\hb'):\ h'\in [h+h^{3/8},\infty], \hb'\in[\hb-\hb^{3/8},\hb+\hb^{3/8}]\right\}\,,\\
R_9&=\left\{(h',\hb'):\ h'\in [h+h^{3/8},\infty], \hb'\in[\hb+\hb^{3/8},\infty]\right\}\,.
\end{aligned}
\end{equation}
Basically, we have to estimate $\sum_i S_i$ where $S_i$ is given by
\begin{equation}
\begin{aligned}
&S_i=\int_{R_i}\text{d}h'\ \text{d}\hb' |\delta\rho(h',\hb')| \bigg[|G_{+}(\nu)G_{+}(\bar\nu)|+|G_-(\bar\nu)G_{+}(\nu)|+(+\leftrightarrow -)\bigg]\\
&\leq 4(S^{(1)}_{i}+S^{(2)}_{i})\,,
\end{aligned}
\end{equation}
where we have used $|\delta\rho|\leq \rho+\rho_*$, $|G_{\pm}(\nu)\leq 2e^{-\beta\nu}\text{min}\left(1,(h-h')^{-2}\right)$, and $|G_{\pm}(\bar\nu)\leq 2e^{-\bb\bar\nu}\text{min}\left(1,(\hb-\hb')^{-2}\right)$. The appearance of $4$ is due to the fact that there are $4$ terms in the integrand defining $S_i$ and each is suppressed in a same manner. Here $S^{(1)}_{i}$ and $S^{(2)}_{i}$ are defined as
\begin{equation}
\begin{aligned}
S^{(1)}_{i}=4\int_{R_i}\text{d}h'\ \text{d}\hb' \rho(h',\hb') e^{-\beta\nu-\bb\bar\nu}\text{min}\left(1,(h-h')^{-2}\right)\text{min}\left(1,(\hb-\hb')^{-2}\right)\,,\\
S^{(2)}_{i}=4\int_{R_i}\text{d}h'\ \text{d}\hb' \rho_*(h',\hb') e^{-\beta\nu-\bb\bar\nu}\text{min}\left(1,(h-h')^{-2}\right)\text{min}\left(1,(\hb-\hb')^{-2}\right)\,.
\end{aligned}
\end{equation}
 Now for $R_i$ with $i\neq 4,5,6$ we have
 \begin{equation}
\begin{aligned}
\frac{1}{4}S^{(1)}_{i\neq 4,5,6}&= e^{\beta h+\bb\hb}\int_{R_{i\neq 4,5,6}} \text{d}h' \ \text{d}\hb'\ \rho(h',\hb') e^{-\beta h'-\bb\hb'}\text{min}\left(1,(h-h')^{-2}\right)\text{min}\left(1,(\hb-\hb')^{-2}\right)\\
&\leq e^{\beta h+\bb\hb}\int_{R_{i\neq 4,5,6}} \text{d}h' \ \text{d}\hb'\ \rho(h',\hb') e^{-\beta h'-\bb\hb'}\text{min}\left(1,(\hb-\hb')^{-2}\right)\\
&=O\left(\hb^{-3/4}\exp\left[2\pi\left(\sqrt{\frac{ch}{6}}+\sqrt{\frac{c\hb}{6}}\right)\right] \right)\,.
\end{aligned}
\end{equation}

For $R_4$ and $R_6$ we observe the following
\begin{equation}
\begin{aligned}
\frac{1}{4}S^{(1)}_{i=4,6}=&e^{\beta h+\bb\hb}\int_{R_{i=4,6}} \text{d}h' \ \text{d}\hb'\ \rho(h',\hb') e^{-\beta h'-\bb\hb'}\text{min}\left(1,(h-h')^{-2}\right)\text{min}\left(1,(\hb-\hb')^{-2}\right)\\
&\leq e^{\beta h+\bb\hb}\int_{R_{i= 4,6}} \text{d}h' \ \text{d}\hb'\ \rho(h',\hb') e^{-\beta h'-\bb\hb'}\text{min}\left(1,(h-h')^{-2}\right)\\
&\leq e^{\beta h+\bb\hb}\hb^{-3/4}\int_{R_{i= 4,6}} \text{d}h' \ \text{d}\hb'\ \rho(h',\hb') e^{-\beta h'-\bb\hb'}\\
&=O\left(h^{-3/4}\exp\left[2\pi\left(\sqrt{\frac{ch}{6}}+\sqrt{\frac{c\hb}{6}}\right)\right] \right)\,.
\end{aligned}
\end{equation}

For analyzing the region $R_5$, we are required to subdivide it into $9$ regions again where each of the region is Cartesian product of order one interval in $h'$ and $\hb'$. Now one can use the lemma proven in the section \S\ref{lemma0} to show that
\begin{equation} 
\begin{aligned}
\frac{1}{4}S^{(1)}_{5}=&e^{\beta h+\bb\hb}\int_{R_{5}} \text{d}h' \ \text{d}\hb'\ \rho(h',\hb') e^{-\beta h'-\bb\hb'}\text{min}\left(1,(h-h')^{-2}\right)\text{min}\left(1,(\hb-\hb')^{-2}\right)\\
&=O\left(h^{-3/4}\hb^{-3/4}\exp\left[2\pi\left(\sqrt{\frac{ch}{6}}+\sqrt{\frac{c\hb}{6}}\right)\right] \right)\,.
\end{aligned}
\end{equation}
The estimation for $S^{(2)}=\sum_{i}S^{(2)}_{i}$ can be done by saddle point method: 
\begin{align}
\frac{1}{4}S^{(2)}&= e^{\beta h+\bb\hb}\int_{0}^{\infty}\int_{0}^{\infty} \text{d}h' \ \text{d}\hb'\ \rho_*(h',\hb') e^{-\beta h'-\bb\hb'}\text{min}\left(1,\frac{1}{(h-h')^{2}}\right)\text{min}\left(1,\frac{1}{(\hb-\hb')^{2}}\right)\\
&=O\left(h^{-3/4}\hb^{-3/4}\exp\left[2\pi\left(\sqrt{\frac{ch}{6}}+\sqrt{\frac{c\hb}{6}}\right)\right]\right)\,.
\end{align}

We are yet to estimate the following term 
\begin{align}
m\equiv \int_{0}^{\infty}\text{d}h' \int_{0}^{\infty}\text{d}\hb' \delta\rho(h',\hb') G(\nu)G(\bar\nu)\,.
\end{align}
which appears in the expression for $\int_{0}^{h}\text{d}h' \int_{0}^{\hb}\text{d}\hb' \delta\rho(h',\hb')$. Using the definition of $G(\nu)$ and $G(\bar\nu)$, we arrive at
\begin{align}
m=-\frac{1}{4\pi^2}\int_{\beta-\imath\Lambda}^{\beta+\imath\Lambda} \int_{\bb-\imath\Lambda}^{\bb+\imath\Lambda}\frac{dz d\bar z}{z \bar z}\frac{\Lambda^2+(z-\beta)^2}{\Lambda^2+\beta^2}\frac{\Lambda^2+(\bar z-\bb)^2}{\Lambda^2+\bb^2}e^{zh+\bar z\hb}\delta\mathcal{L}(z,\bar z)\,.
\end{align}
Thus we have
\begin{align}
|m|\leq \frac{1}{4\pi^2} \int_{-\Lambda}^{\Lambda}\int_{-\Lambda}^{\Lambda} \frac{\text{d}td\bar t}{|\beta+\imath t||\bb+\imath\bar t|} \left(\frac{\Lambda^2-t^2}{\Lambda^2+\beta^2}\right)\left(\frac{\Lambda^2-\bar t{}^2}{\Lambda^2+\bb^2}\right)e^{\beta h+\bb\hb} |\delta\mathcal{L}(z,\bar z)|\,.
\end{align}
 Now we use the inequality 
 \begin{align}
 |\delta\mathcal{L}(z,\bar z)|\leq e^{-(Re[z]+Re[\bar{z}])c/24}Z_{H}\left(\frac{4\pi^2Re[z]}{|z|^2},\frac{4\pi^2Re[\bar z]}{|\bar z|^2}\right)\,,
 \end{align}
and subsequently the method utilized in \S\ref{lemma0} to put a bound by $Z_{H,\Delta}$:
 \begin{align}
 |m|\leq \frac{4\Lambda^6}{9\beta\bb\pi^2}\frac{e^{\beta(h-c/24)}e^{\bb(\hb-c/24)}e^{\frac{\pi^2c}{6y}}}{\left(\Lambda^2+\beta^2\right)\left(\Lambda^2+\bb^2\right)}Z_{H,\Delta}\left(\frac{4\pi^2\beta_*}{\beta_*^2+\Lambda^2}\right)\,,
 \end{align}
 where $y=\beta\ \text{or}\ \bb$. Essentially this is exactly the same argument as in \S\ref{lemma0}. Then by choosing $\Lambda<\sqrt{2}\pi$ (when $h\simeq \hb$, otherwise we need to choose $\gamma\Lambda<\sqrt{2}\pi$, see the discussion in \S\ref{lemma0}) and using the HKS \cite{HKS} argument, one can show that the above term is exponentially suppressed compared to the leading answer coming from $\rho_*(h,\hb)$. When $h$ is not of the order of $\hb$, we need to assume existence of twist gap (as defined in \{\ref{def}\}) and proceed like we did in \S\ref{lemma0}. This concludes our analysis and hence the proof of the main theorem.
 
\subsection{Sensitivity of Asymptotics towards spin $J$} \label{spinsense}
The asymptotic formula given in eq.~\eqref{masterequation} and derived above can be rewritten in terms of dimension $\Delta=h+\hb$ and spin $J=|h-\hb|$: 
\begin{equation}\label{masterequationJ}
\begin{aligned}
&F(h\to\infty,\hb\to\infty)=\int_{0}^{h}\text{d}h' \int_{0}^{\hb}\text{d}\hb' \rho(h',\hb')\\
&=\frac{1}{4\pi^2}\left(\frac{36}{c^2\left(\Delta^2-J^2\right)}\right)^{1/4}\exp\left[2\pi\left(\sqrt{\frac{c(\Delta+J)}{12}}+\sqrt{\frac{c(\Delta-J)}{12}}\right)\right]\left[1+O\left(\Delta^{-1/4}\right)\right]\,,
\end{aligned}
\end{equation}
which is true when $1<\frac{\Delta}{J}=O(1)$. It turns out that even when $\Delta$ and $J$ is not of the same order, we can do order by order correction to this integrated density of states by spin. \\

First of all, when $J$ is of order one, we should just ignore $J$ dependence of the eq.~\eqref{masterequationJ}. In fact, $J=\Delta^{1/n}$ with $n>4/3$, one can ignore $J$ dependence. Thus in this regime, we have
\begin{equation}\label{masterequationJpp}
\begin{aligned}
&F(\Delta\to\infty,J=\Delta^{1/n}\to\infty)=\int_{0}^{h}\text{d}h' \int_{0}^{\hb}\text{d}\hb'\ \rho(h',\hb')\,,\quad 4/3<n \leq  \infty\\\
&=\frac{1}{4\pi^2}\left(\frac{6}{c\Delta}\right)^{1/2}e^{2\pi\sqrt{\frac{c\Delta}{3}}}\left[1+O\left(\Delta^{-1/4}\right)\right]\,.
\end{aligned}
\end{equation}

 When $8/7\leq n \leq 4/3$, the $J$ dependence is meaningful only within the exponential i.e. we have
\begin{equation}
\begin{aligned}
&F(\Delta\to\infty,J=\Delta^{1/n}\to\infty)=\int_{0}^{h}\text{d}h' \int_{0}^{\hb}\text{d}\hb'\ \rho(h',\hb')\,,\quad  8/7\leq n\leq 4/3\\
&=\frac{1}{4\pi^2}\left(\frac{6}{c\Delta}\right)^{1/2}e^{2\pi\sqrt{\frac{c\Delta}{3}}\left(1-\frac{J^2}{8\Delta^2}\right)}\left[1+O\left(\Delta^{-1/4}\right)\right]\,.
\end{aligned}
\end{equation}

 When $16/15<n<8/7$ ($n$ can not less than one because of unitarity bound), we have 
\begin{equation}\label{masterequationJp}
\begin{aligned}
&F(\Delta\to\infty,J=\Delta^{1/n}\to\infty)=\int_{0}^{h}\text{d}h' \int_{0}^{\hb}\text{d}\hb'\ \rho(h',\hb')\,,\quad  1<n<8/7\\
&=\frac{1}{4\pi^2}\left(\frac{6}{c\Delta}\right)^{1/2}\left(1+\frac{J^2}{4\Delta^2}\right)\exp\left[2\pi\left(\sqrt{\frac{c(\Delta+J)}{12}}+\sqrt{\frac{c(\Delta-J)}{12}}\right)\right]\left[1+O\left(\Delta^{-1/4}\right)\right]\,.
\end{aligned}
\end{equation}
We remark that not all the term in the exponential are meaningful, we have to do an $J/\Delta$ expansion and the only meaningful terms are of the form $\exp\left(\Delta^{-\ell}\right)$ with $\ell<1/4$ (since $\exp\left(\Delta^{-\ell}\right) \simeq 1+\Delta^{-\ell}$ becomes comparable to error term for $\ell<1/4$). For example, when $n=\frac{15}{14}$, it is meaningful to keep the following terms only:
\begin{equation*}
\begin{aligned}
&e^{2\pi\left(\sqrt{\frac{c(\Delta+J)}{12}}+\sqrt{\frac{c(\Delta-J)}{12}}\right)}=e^{2\pi\sqrt{\frac{c\Delta}{3}}\left(1-\frac{J^2}{8\Delta^2}-\frac{5 J^4}{128\Delta^4}-\frac{21 J^6}{1024\Delta^6}-\frac{429 J^8}{32768\Delta^8}-\frac{2431 J^{10}}{262144\Delta^{10}}\right)}\left[1+O\left(\Delta^{-3/10}\right)\right]\,.
\end{aligned}
\end{equation*}

One can generalize the equation \eqref{masterequationJp} for $J=\Delta^{1/n}$ with $\frac{2^{m+3}}{2^{m+3}-1}<n<\frac{2^{m+2}}{2^{m+2}-1}\leq 8/7$ where $m\geq 1$ is a fixed integer.  We define $f(m)=\frac{2^{m}}{2^{m}-1}$ and we have :
\begin{equation}\label{masterequationJpg}
\begin{aligned}
&F(\Delta\to\infty,J=\Delta^{1/n}\to\infty)=\int_{0}^{h}\text{d}h' \int_{0}^{\hb}\text{d}\hb'\ \rho(h',\hb')\,,\quad  f(m+3)<n<f(m+2)\\
&=\frac{1}{4\pi^2}\left(\frac{6}{c\Delta}\right)^{1/2}\left(\sum_{k=0}^{m} a_{k} \left(\frac{J^2}{\Delta^2}\right)^{k} \right)\exp\left[2\pi\left(\sqrt{\frac{c(\Delta+J)}{12}}+\sqrt{\frac{c(\Delta-J)}{12}}\right)\right]\left[1+O\left(\Delta^{-1/4}\right)\right]\,.
\end{aligned}
\end{equation}
where $a_k$'s are defined as
\begin{align}
\left(1-\frac{J^2}{\Delta^2}\right)^{1/4}= \sum_{k=0}^{\infty} a_{k} \left(\frac{J^2}{\Delta^2}\right)^{k} \,,\ \Delta>J\,.
\end{align}

\section{Verification: $2$D Ising model}\label{verify}
\begin{figure}[ht!]
\centering
\includegraphics[scale=0.6]{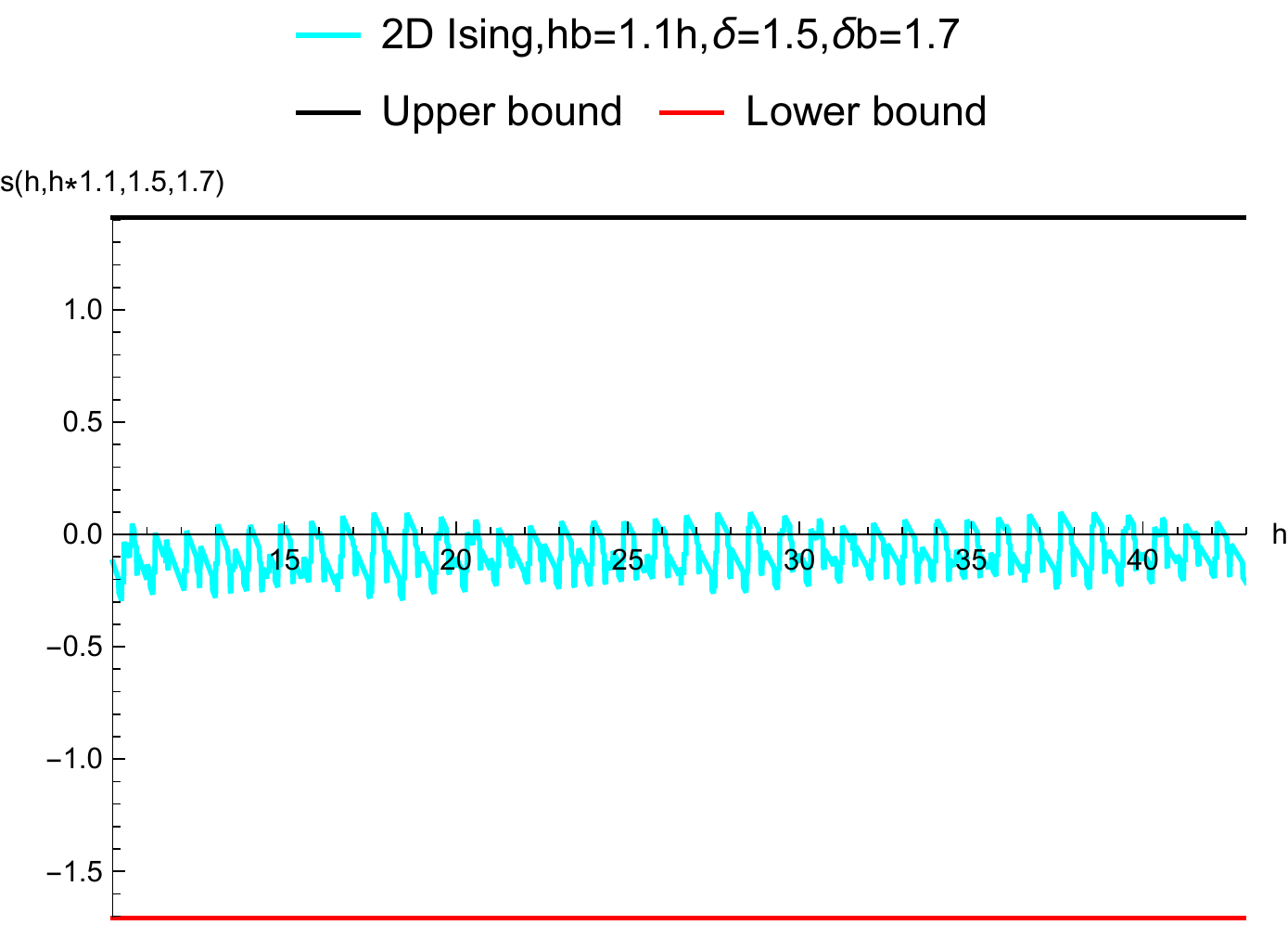}
\caption{Verifying the bounds on $O(1)$ correction to entropy ($s(\delta,\db,h,\hb)$) associated with order one window centered at some large $h$ and $\hb$, the black line is the upper bound while the red line below is the lower bound. The curvy cyan line is the difference between the actual number of states lying within the window and leading answer coming from the Cardy formula. Here $h$ and $\hb$ are of the same order.}
\label{fig:01}
\end{figure}

\begin{figure}[ht!]
\centering
\includegraphics[scale=0.6]{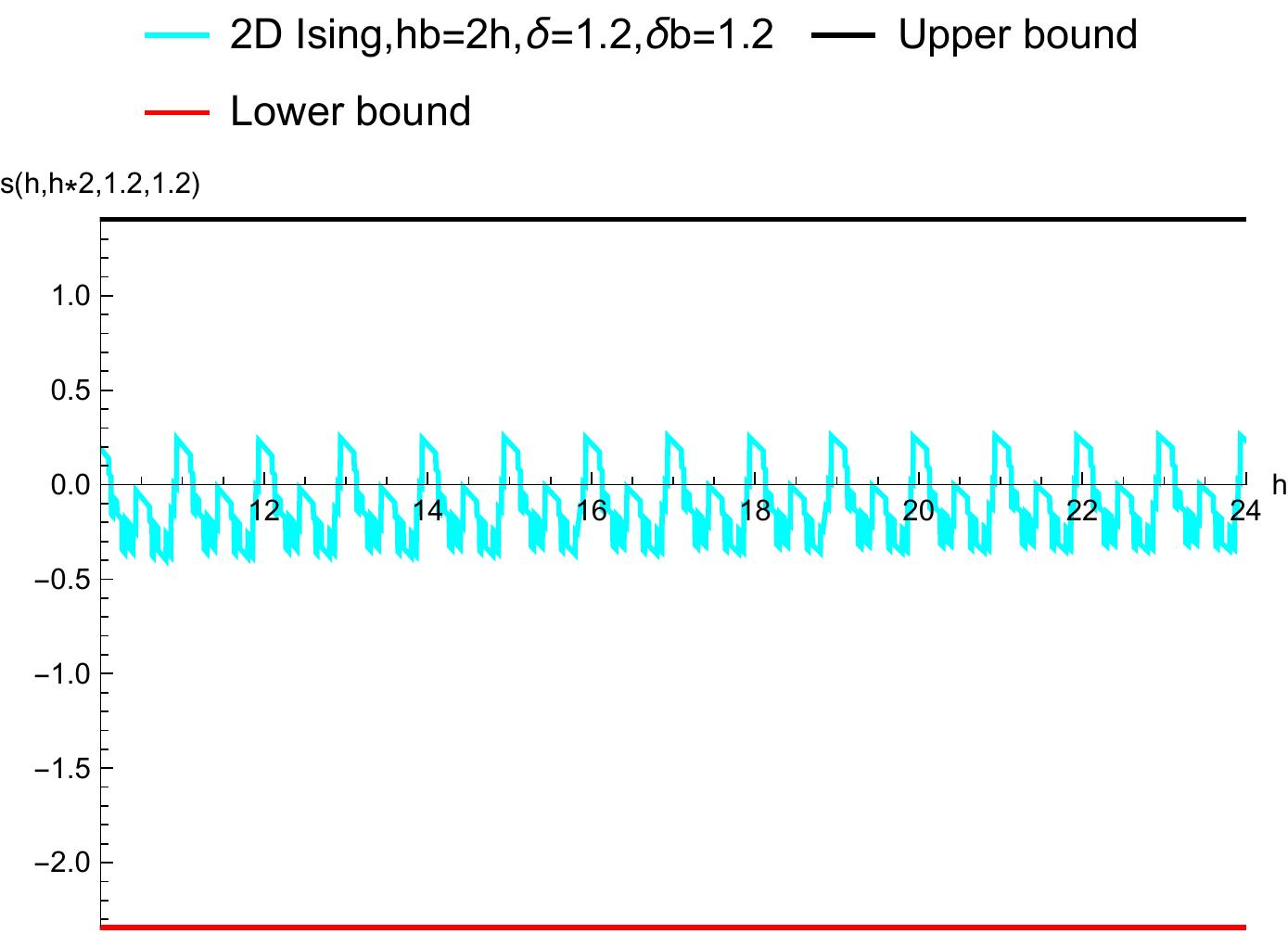}
\caption{Verifying the bounds on $O(1)$ correction to entropy ($s(\delta,\db,h,\hb)$) associated with order one window centered at some large $h$ and $\hb$, the black line is the upper bound while the red line below is the lower bound. The curvy cyan line is the difference between the actual number of states lying within the window and leading answer coming from the Cardy formula. Here $h$ and $\hb$ are of the same order.}
\label{fig:02}
\end{figure}

\begin{figure}[ht!]
\centering
\includegraphics[scale=0.6]{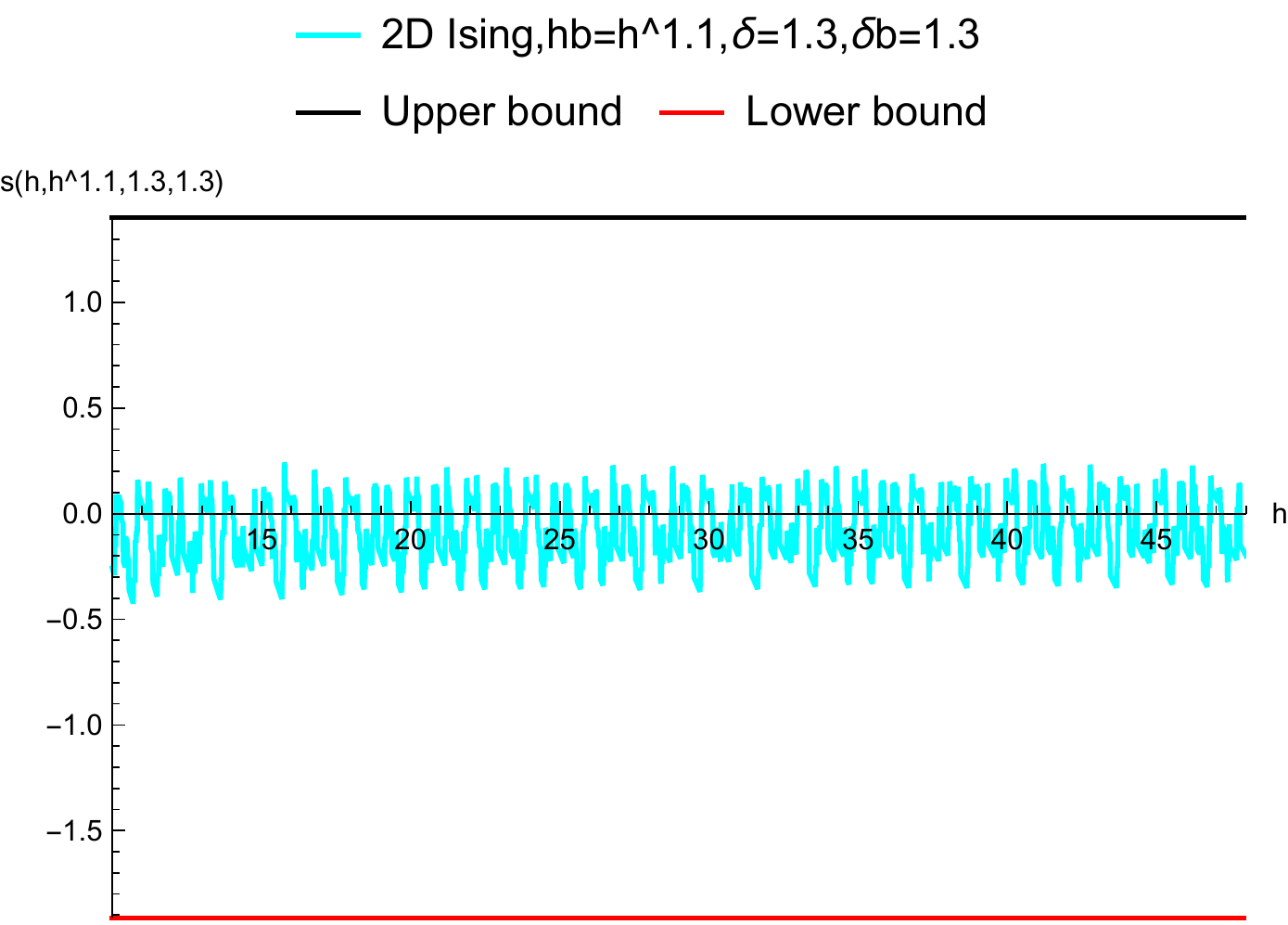}
\caption{Verifying the bounds on $O(1)$ correction to entropy ($s(\delta,\db,h,\hb)$) associated with order one window centered at some large $h$ and $\hb$, the black line is the upper bound while the red line below is the lower bound. The curvy cyan line is the difference between the actual number of states lying within the window and leading answer coming from the Cardy formula. Here $h$ and $\hb$ are not of the same order.}
\label{fig:03}
\end{figure}

We verify the bounds on $O(1)$ correction to entropy associated with order one window centered at some large $h$ and $\hb$, proven in \S\ref{lemma0} as shown in the figure~\ref{fig:01}, \ref{fig:02} and \ref{fig:03}. In \S\ref{lemma0}, we have obtained two kinds of bounds, one without assuming any twist gap while the other one assumes existence of a twist gap. For $2$D Ising model, it turns out that the bound coming from assuming a twist gap (indeed $2$D Ising model has a twist gap) is stronger than the one without using the information about twist gap. Thus in the figures below, we verify the bounds that uses the information about twist gap. In fact, as we have mentioned in \S\ref{lemma0}, the use of twist gap actually enables us to probe regions where $h$ and $\hb$ are not of the same order. The figure~\ref{fig:03} elucidates such a scenario. 

\begin{figure}[ht!]
\centering
\includegraphics[scale=0.75]{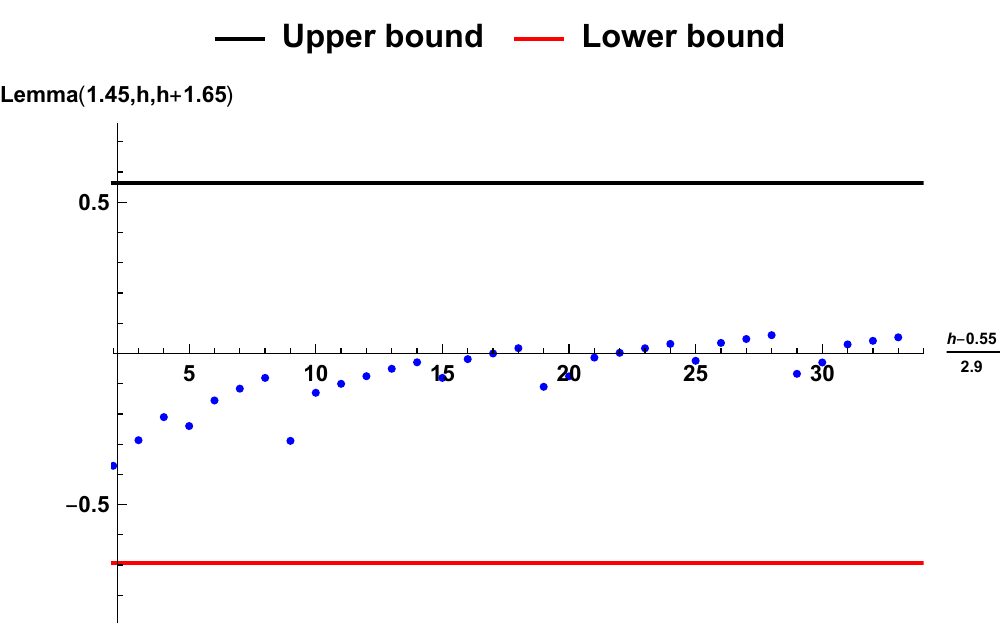}
\caption{The blue dots denotes the values of $\text{Lemma}(\delta,h,\hb)\equiv \log\left[\frac{1}{2\delta}Q\left(h,\bb=\pi\sqrt{\frac{c\hb}{6}}\right)\right]-2\pi\sqrt{\frac{ch}{6}}-\pi\sqrt{\frac{c\hb}{6}}$, here we have $\delta=1.45$ and $\hb=h+1.65$, for different values of $h$. The blue dots are bounded by an order one i.e. $h,\hb$ independent number, denoted by the red and the black curve. The values of $c_{\pm}$ found in \S\ref{lemma1} are used as the bounds. }
\label{fig:1}
\end{figure}

\begin{figure}[ht!]
\centering
\includegraphics[scale=0.75]{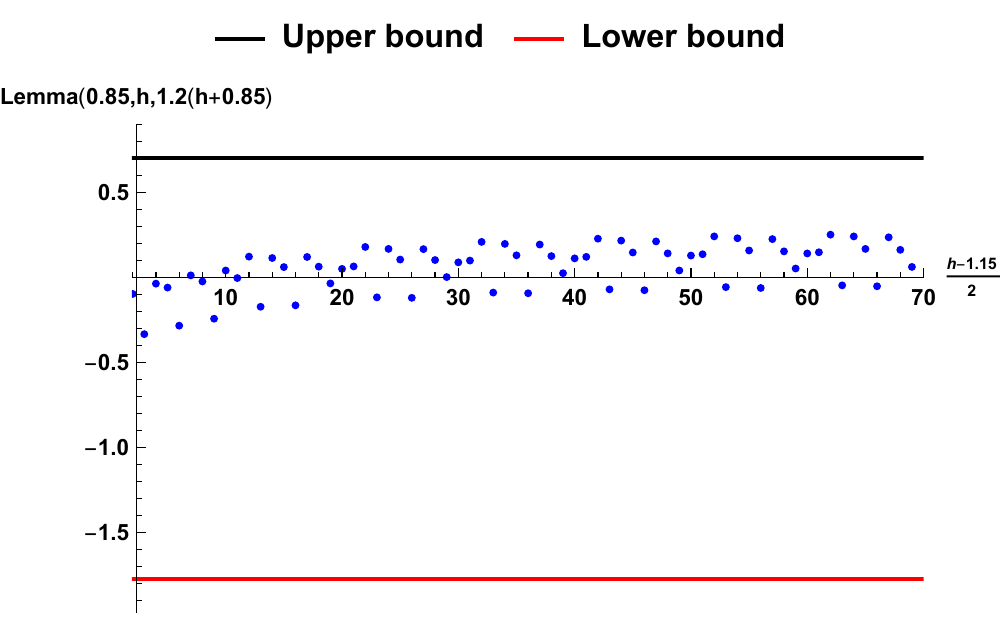}
\caption{The blue dots denotes the values of $\text{Lemma}(\delta,h,\hb)\equiv\log\left[\frac{1}{2\delta}Q\left(h,\bb=\pi\sqrt{\frac{c\hb}{6}}\right)\right]-2\pi\sqrt{\frac{ch}{6}}-\pi\sqrt{\frac{c\hb}{6}}$, where $\delta=1.7$ and $\hb=1.2(h+0.85)$, for different values of $h$. The blue dots are bounded by an order one i.e. $h,\hb$ independent number, denoted by the red and the black curve. The $c_{\pm}$ found in the \S\ref{lemma1} are used as the bounds.}
\label{fig:2}
\end{figure}

We verify the lemma proven in \S\ref{lemma1} as shown in the figure~\ref{fig:1} and figure~\ref{fig:2} using $2$ dimensional Ising CFT. The partition function for $2$ dimensional Ising CFT is given by 
\begin{align}
Z_{\text{Ising}}(\beta,\bb)=\frac{1}{2}\left(\sqrt{\frac{\theta_2(\beta)}{\eta(\beta)}}\sqrt{\frac{\theta_2(\bb)}{\eta(\bb)}}+\sqrt{\frac{\theta_3(\beta)}{\eta(\beta)}}\sqrt{\frac{\theta_3(\bb)}{\eta(\bb)}}+\sqrt{\frac{\theta_4(\beta)}{\eta(\beta)}}\sqrt{\frac{\theta_4(\bb)}{\eta(\bb)}}\right)\,.
\end{align}

\begin{figure}[!ht]
\centering
\includegraphics[scale=0.5]{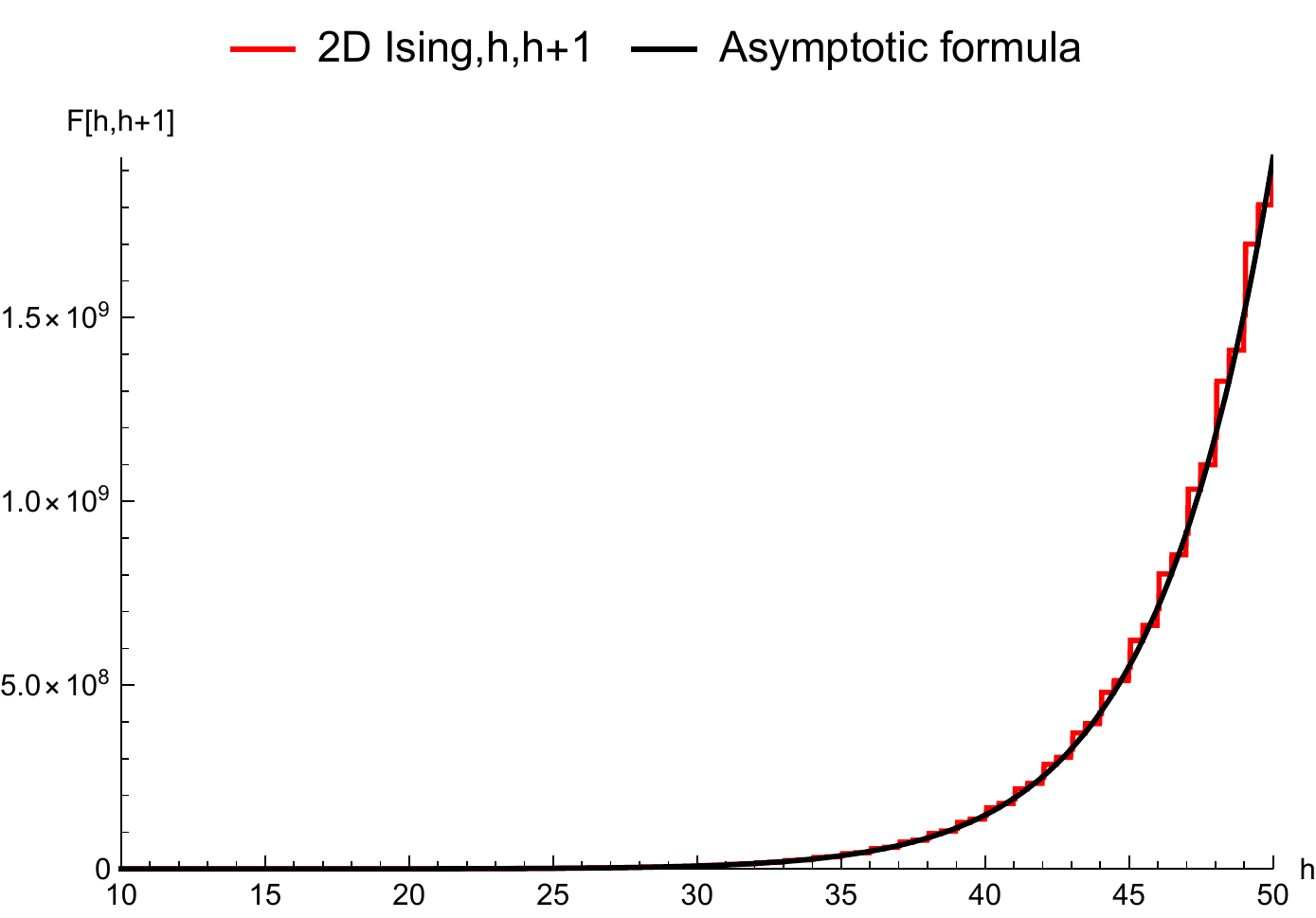}
\includegraphics[scale=0.5]{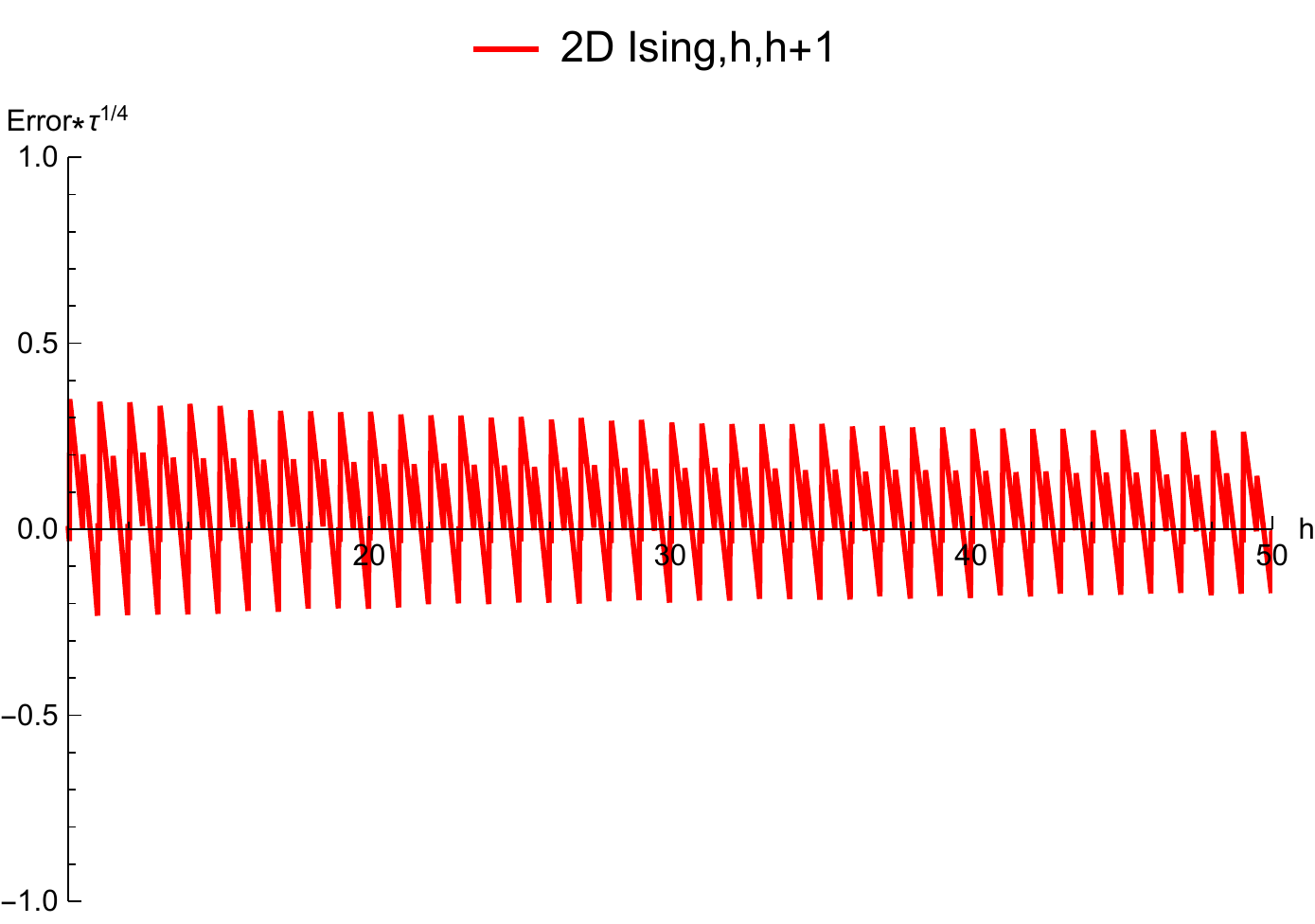}
\caption{In the left figure, the black curve is the asymptotic Cardy formula while the red curve is the actual number of operators till $(h,\hb)$ present in the Ising model. Here we have taken $h=\hb-1$. The picture on the right hand side plot the relative error times $h^{1/4}$.The error times $h^{1/4}$ is bounded above and below by an order one number.}
\label{fig:3}
\end{figure}

\begin{figure}[ht!]
\centering
\includegraphics[scale=0.5]{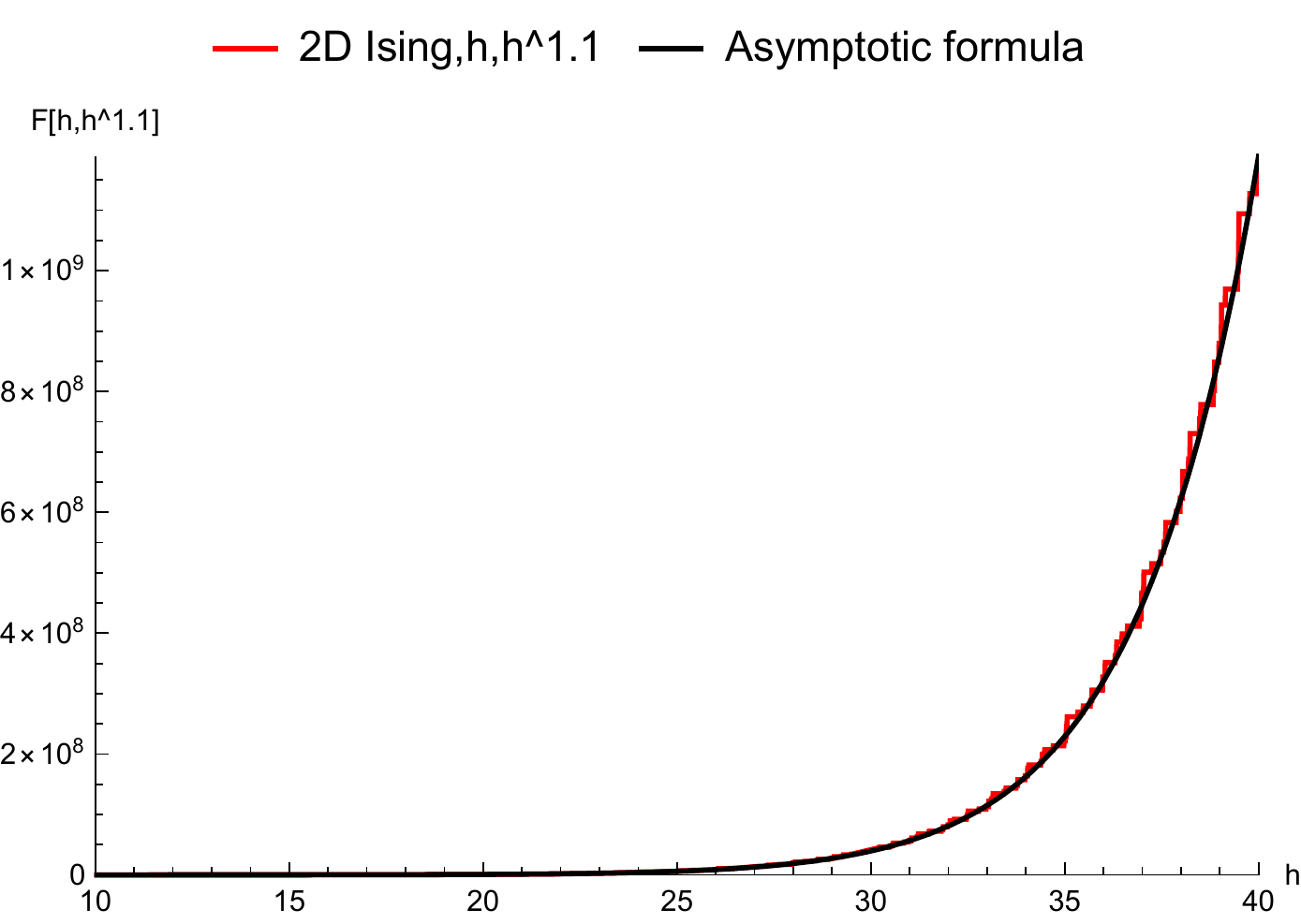}
\includegraphics[scale=0.5]{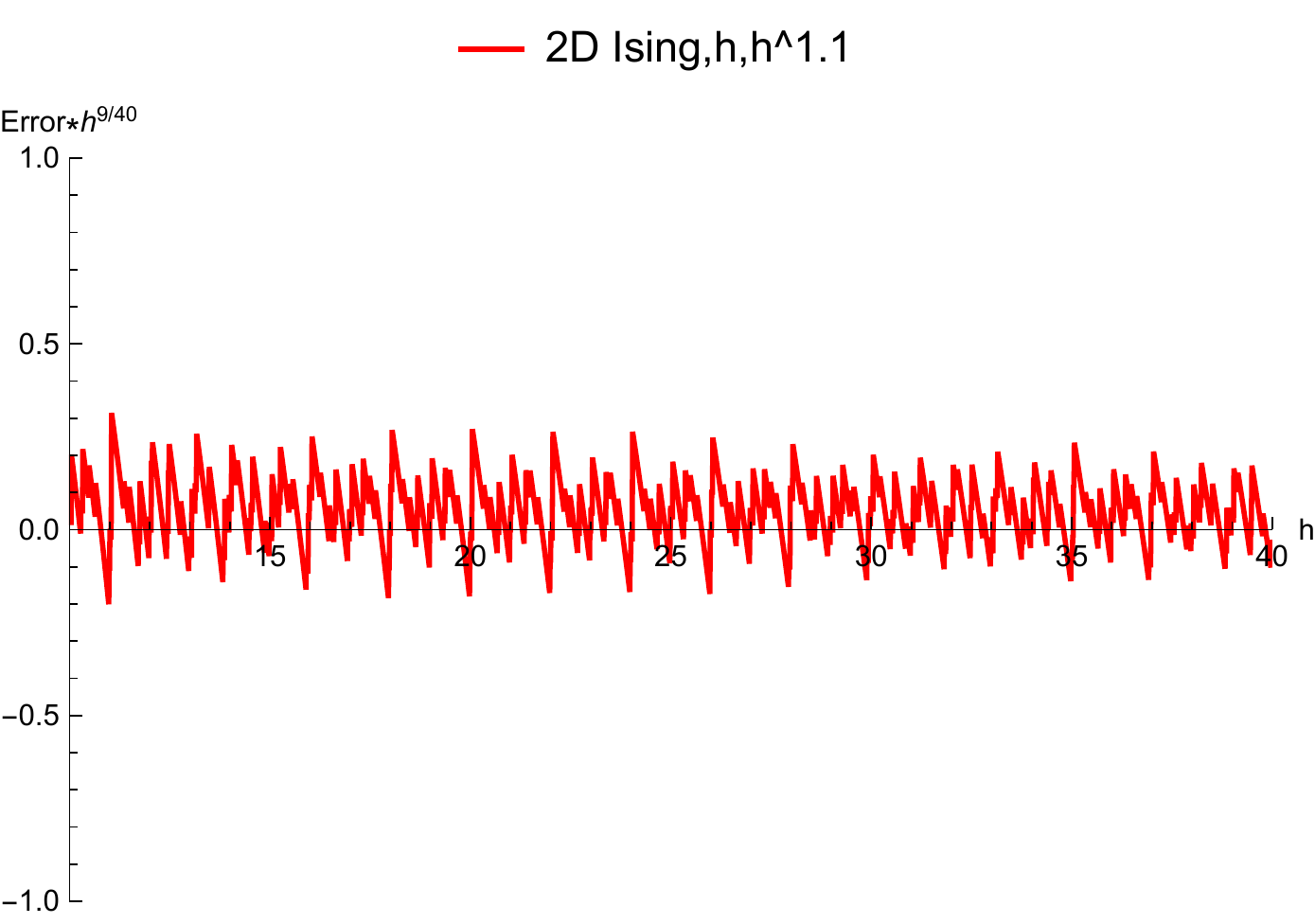}
\caption{In the left figure, the black curve is the asymptotic Cardy formula while the red curve is the actual number of operators till $(h,\hb)$ present in the Ising model. Here we have taken $h^{1.1}=\hb$. The picture on the right hand side plot the relative error times $h^{9/40}$. The error times $h^{9/40}$ is bounded above and below by an order one number.}
\label{fig:4}
\end{figure}

We verify the main theorem proven in \S\ref{mainproof} in fig.~\ref{fig:3} and \ref{fig:4}. We plot the total number of states upto some $(h,\hb)$ as a function of $h$. We have considered two different cases where the asymptote is approached along different curves. We compare it against the asymptotic formula we derive in \S\ref{mainproof}. 

{
\section{Finite twist-Large spin}\label{lsft}
In this section, our aim is to estimate the finite twist and large spin sector of the density of states. Without loss of generality, we will assume $h$ is finite and $\hb\to\infty$. We will probe the following quantity $U_h(\hb)$ in the limit $\hb\to\infty$:
\begin{align}
U_h(\hb)=\int_{0}^{h}\ \text{d}h'\int_{\hb-\db}^{\hb+\db}\ \text{d}\hb'\ \rho(h',\hb') \,.
\end{align}
This is because we have
\begin{align}
\int_{h-\delta}^{h+\delta}\ \text{d}h'\int_{\hb-\db}^{\hb+\db}\ \text{d}\hb'\ \rho(h',\hb') \leq U_{h+\delta}(\hb)\,.
\end{align}
Intuitively, it is clear that one can not make both $\beta$ and $\bb$ to approach zero while looking at the partition function, as this would probe the regime $h,\hb\to\infty$. This suggests that we should let $\bb \to 0$ and keep $\beta$ fixed.

We can prove an upper bound on density of states in this sector. Let us write down the following inequality:
\begin{equation}
\begin{aligned}
U_h(\hb) \leq e^{\beta (h-c/24)} \int_{0}^{\infty}\ \text{d}h'\int_{\hb-\db}^{\hb+\db}\ \text{d}\hb'\ \rho(h',\hb') e^{-\beta (h'-c/24)}\,,
\end{aligned}
\end{equation}
from which we can write 

\begin{equation}
\begin{aligned}
 U_h(\hb)\leq e^{\beta \left(h-\frac{c}{24}\right)+\bb\hb+\bb\db} \int_{0}^{\infty}\ \text{d}h'\int_{\hb-\db}^{\hb+\db}\ \text{d}\hb'\ \rho(h',\hb') \phi_+(\hb') e^{-\beta \left(h'-\frac{c}{24}\right)-\bb\hb'}\,.
\end{aligned}
\end{equation}

At this point we choose
\begin{align}
\beta =2\pi \,, \quad \bb=\pi\sqrt{\frac{c}{6\hb}}\,,\ \hb\to\infty\,.
\end{align}
Again we separate the partition function into two pieces; the light sector, where the contribution comes from two kinds of states: a) all the states with conformal weight $(h',0)$ with $h'\geq 0$ and b) the states such that $h'+\hb'<\frac{c}{12}$, and the heavy sector, which is defined to be the complement of this. We remark the heavy sector does not contain any operator with conformal weight $(h',0)$. In the usual asymptotic analysis as done previously, the operators with $(h',0)$ such that $h'>\frac{c}{12}$ is put into the heavy sector, but here we can not do that since $\beta$ is finite, there is no separation between light or heavy in the sense of HKS \cite{HKS} i.e. the operators with $(h',0)$ contribute on equal footing as the operator $(0,0)$. The upshot of the above discussion is that one can define a kernel $\rho^{\text{ft}}_*$ (``ft" stands for finite twist) such that 
\begin{align}
\int_0^{\infty}\text{d}h'\ \int_0^{\infty}\text{d}\hb'\ \rho^{\text{ft}}_*(h',\hb') e^{-\beta \left(h'-\frac{c}{24}\right)-\bb\hb'}=e^{\frac{\pi^2c}{6\bb}}\sum_{h_i}e^{-\beta\left(h_i-\frac{c}{24}\right)}\,,
\end{align}
where the sum on the right hand side is over all the states with weight $(h_i,0)$ and we have used $\beta=\frac{4\pi^2}{\beta}=2\pi$. As a result $\rho^{\text{ft}}_*$ is given by
\begin{align}
\rho^{\text{ft}}_*(h',\hb')=\rho_*(\hb')\sum_{h_i}\delta\left(h'-h_i\right)\,.
\end{align}

Once we have defined $\rho_*$, we follow the usual method and our next aim is to show that the heavy part has a subleading contribution. The estimation of the heavy part is done depending on whether both $h'$ and $\hb'$ is greater than $c/24$ or one of them is less than $c/24$. In the later case, we have to assume existence of twist gap. Last but not the least, we have to estimate the contribution from the states with $(0,\hb')$ with $\hb'>c/12$. Following the methods in \S\ref{lemma0}, we can show that they are indeed subleading. The leading answer is then given by
\begin{equation}
\begin{aligned}
U_h(\hb) \leq e^{2\pi \left(h-\frac{c}{24}\right)+\pi\sqrt{\frac{c\hb}{6}}}\int\ \text{d}h'\int \text{d}\hb' \rho^{\text{ft}}_*(h',\hb') e^{-2\pi  \left(h'-\frac{c}{24}\right)-\pi\sqrt{\frac{c}{6\hb}}\hb'}\Phi_{+}(\hb')\,.
\end{aligned}
\end{equation}
The $\hb'$ integral can be done using saddle point approximation and we obtain
\begin{equation}
\begin{aligned}
\frac{1}{2\db}U_h(\hb) &\leq e^{-\frac{\pi c}{12}}e^{2\pi h}c_+ \rho_*(\hb') \sum_{h_i} e^{-2\pi  \left(h_i-\frac{c}{24}\right)}\,, \\
\frac{1}{2\db}U_h(\hb) &\leq e^{-\frac{\pi c}{12}}e^{2\pi h}c_+ \rho_*(\hb') \chi_{0}(e^{-2\pi})\,.
\end{aligned}
\end{equation}
where $\chi_0(q)$ is the vacuum character and $q=e^{-\beta}$ where we have assumed that there is no nontrivial conserved current. Here $c_+$ is defined as
\begin{align}
c_+=\int_{-\infty}^{\infty}dx\ \Phi_{+}(\hb'+\db x)\,.
\end{align} One can extend this argument to a scenario where we have nontrivial conserved currents, then we would have a sum over characters for all the conserved currents 
\begin{equation}
\begin{aligned}
\frac{1}{2\db}U_h(\hb) \leq c_+ \rho_*(\hb') e^{2\pi \left(h-\frac{c}{24}\right)}\sum_{\tilde{h}}\chi_{\tilde{h}}(e^{-2\pi})\,.
\end{aligned}
\end{equation}
This sum over $\tilde{h}$ is convergent as the absolute value of the sum is bounded above by partition function evaluated at $\beta=\bb=2\pi$. Similar result applies to $h,\hb$ getting swapped. Thus for finite $h$ and $\hb\to\infty$ limit we have 
\begin{align}
S_{\delta,\db}\leq \mathbb{S}_{h,\delta,\db} \leq 2\pi\sqrt{\frac{c\hb}{6}}-\frac{3}{4}\log\left(\hb\right)+ 2\pi \left(h+\delta-\frac{c}{24}\right)+ \log\left[2c_+\db\sum_{\tilde{h}}\chi_{\tilde{h}}(e^{-2\pi})\right]\,,
\end{align}
where $\chi_{\tilde{h}}$ is the character for the conserved current with weight $(\tilde{h},0)$. 


The similar result specific for the primaries with finite $h$ and $\hb\to\infty$ limit, would read: 
\begin{align}
S^{\text{Vir}}_{\delta,\db}\leq \mathbb{S}^{\text{Vir}}_{h,\delta,\db} \leq 2\pi\sqrt{\frac{(c-1)\hb}{6}}+2\pi \left(h+\delta-\frac{c-1}{24}\right)+ \log\left[c_+\frac{2\db}{\sqrt{\hb}}\sum_{\tilde{h}}e^{-2\pi\left(\tilde{h}-\frac{c-1}{24}\right)}\right]\,,
\end{align}
where the zero twist primaries have weight $(\tilde{h},0)$. Here again, the sum over $\tilde{h}$ is convergent as the absolute value of the sum is bounded above by partition function evaluated at $\beta=\bb=2\pi$.\\
 
\paragraph{CFT without nontrivial zero twist primary:}In case the CFT does not have any conserved current, we do not need to worry about $(h',0)$ operators anymore and we can do much better as it is possible to choose $\beta\neq 2\pi$ and still define $\rho^{\text{ft-Vir}}_*(h',\hb')=\rho^{\text{ft-Vir}}_*(h')\rho^{\text{ft-Vir}}_*(\hb')$ as a solution to the following equality:
\begin{align}
\int_{0}^{\infty}\text{d}\hb'\ \rho^{\text{ft-Vir}}_*(\hb') e^{-\bb\left(\hb'-\frac{c-1}{24}\right)}=  \sqrt{\frac{2\pi}{\bb}} \left(1-e^{-\frac{4\pi^2}{\bb}}\right)\,.
\end{align}
This parallels the analysis for Virasoro primary in \S\ref{lemma0} (see \eqref{defvir}).\\

In the present case, using the kernel $\rho^{\text{ft-Vir}}_*(h',\hb')$ the leading answer turns out to be
\begin{equation}
\begin{aligned}
\frac{1}{2\db}U^{\text{Vir}}_{h}(\hb) \leq c_+\rho_*(\hb) e^{\beta \left(h-\frac{c-1}{24}\right)}e^{\frac{\pi^2(c-1)}{6\beta}}\sqrt{\frac{2\pi}{\beta}}\left(1-e^{-\frac{4\pi^2}{\beta}}\right)\,.
\end{aligned}
\end{equation}

Now by appropriately choosing $\beta$ we can recover the ``square-root" edge present in the analysis in \cite{Maxfield:2019hdt}. The square root edge in $h$ dependence of the density of states should produce a factor of $(h-(c-1)/24)^{3/2}$. In particular,  we choose $\beta=\frac{1}{(h-\frac{c-1}{24})}$ and let $h-\frac{c}{24}$ to be very small\footnote{This is analogous to the condition written down in \cite{Maxfield:2019hdt} as $0<\hb-\frac{c-1}{24}<<1/c$, there $\hb$ is finite and $h$ is let to infinity.} 
\begin{equation}\label{edgeresult}
\begin{aligned}
\frac{1}{2\db}U^{\text{Vir}}_{h}(\hb) &\leq c_+ e \rho_*(\hb) e^{\frac{\pi^2(c-1)\left(h-\frac{c-1}{24}\right)}{6}}\sqrt{2\pi\left(h-\frac{c-1}{24}\right)}\left(1-e^{-4 \pi^2\left(h-\frac{c-1}{24}\right)}\right)\\
& \underset{h-c/24 << \frac{1}{c-1}}{\simeq} c_+ \left(4\sqrt{2}\pi^{5/2}e\right) \rho_*(\hb) \left(h-\frac{c-1}{24}\right)^{3/2}\,,
\end{aligned}
\end{equation}
The above result is consistent with the leading result as reported in \cite{Kusuki:2018wpa,Kusuki:2019gjs,Maxfield:2019hdt,Benjamin:2019stq}. 

}

\section{Holographic CFTs}\label{largec}
Holographic CFTs are the ones characterized by a sparse low lying spectrum and large central charge. The sparseness condition is first derived in \cite{HKS}, then rederived in \cite{Baur}, where it emerges naturally out of the Tauberian formalism. In the context of asymptotic behavior of OPE coefficients in large $c$ CFTs, a stronger sparseness condition appears as elucidated in \cite{Pal:2019yhz, Michel:2019vnk}.  In this section we will be exploring such CFTs with large central charge and a low lying sparse spectra. In particular, we derive an expression for the density of states in the limit $h,\hb \sim c \rightarrow \infty$. Following \cite{Baur}, we parameterize $h,\hb$ as 
\begin{equation}
	h = c\left(\epsilon + \frac{1}{24}\right), \,\,\, \hb = c\left(\eb + \frac{1}{24}\right), \,\,\, c \rightarrow \infty, \epsilon-\text{fixed}, \eb-\text{fixed}.
\end{equation}

In this limit the asymptotic of the vacuum crossing kernel is given by
\begin{equation}
	\rho_*(h,\hb) = \frac{\sqrt{6}}{24} c^{-1} \left(\frac{1}{\epsilon\eb}\right)^{\frac{3}{4}} e^{2\pi c \big(\sqrt{\frac{\epsilon}{6}}+ \sqrt{\frac{\eb}{6}}\big)}  \theta(\epsilon)\theta(\eb) + \cdots\,.
\end{equation}
As done in \S\ref{lemma0} we separate out the contribution to the partition function into two pieces: the light $Z_L$, and the heavy $Z_H$.Here we choose
\begin{equation}
	\beta = \frac{\pi}{\sqrt{6\epsilon}}\,, \,\,\, \bb = \frac{\pi}{\sqrt{6\eb}}\,.
\end{equation}
We are required to show that  $Z_H$ term is sub-leading. $Z_{H}$ term gets contribution whenever $h'$ or $\hb'$ is greater than $c/24$ and $\Delta'>c/12$. 
It can be shown that $Z_H$ contribution is sub-leading (the method is exactly similar as in \S\ref{lemma0}, one has to be careful about $e^{-\beta c/24-\bb c/24}$ factor, since $c$ is not finite in the analysis.). The result of the analysis is summarized below: 

\begin{equation}
	\Lambda^2_\pm < \bigg(\frac{\sqrt{2}\pi}{\gamma}\bigg)^2	\bigg(1-\frac{\gamma^2}{12\epsilon^*}\bigg)\,,
\end{equation}
where $\gamma^4 = \frac{\epsilon^*}{\epsilon_*} \geq 1, \epsilon^* = \max{(\epsilon,\eb)}, \epsilon_* = \min{(\epsilon,\eb)}$.
The above requires that $1-\frac{\gamma^2}{12\epsilon^*} > 0$, that is, $\epsilon^* \epsilon_* = \epsilon \eb > \frac{1}{12^2}$.\\

In fact, it turns out that we will be requiring much more stronger condition on $\epsilon,\bar\epsilon$:
\begin{align}\label{epsilonvalidity}
\epsilon_*>\frac{1}{6}\,,&\ \epsilon^*>\text{max}\left(\frac{\gamma^4}{6},\frac{\gamma^2}{12}\right)=\frac{\gamma^4}{6}>\frac{1}{6}\,,\\
\tau &> \frac{5c}{12}\,.
\end{align}
This condition justifies the assumption that the first term is dominated by the vacuum.  We compare this with the result in \cite{HKS}, where the Cardy formula is reported to be applicable for $\epsilon_*\epsilon^*=\epsilon \eb > \frac{1}{24^2}$. This implies that there is further scope to improve our result and reach the HKS threshold rigorously. If we assume existence of a twist gap $g$, we can push the regime of validity to following:
\begin{equation}\label{modifiedvalidity}
\begin{aligned}
\epsilon_*>\frac{1}{6}\left[\text{max}\left\{\frac{1}{2}\,,\left(1-\frac{6g}{c}\right)^2\right\}\right]\,. 
\end{aligned}
\end{equation}
We will come back to the derivation of the bound on $\epsilon,\bar\epsilon$ at the end of this section. For now, with this assumption of vacuum dominance, we find :

\begin{equation}
	e^{-2\pi c \big(\sqrt{\frac{\epsilon}{6}}\delta + \sqrt{\frac{\eb}{6}} \db\big)}\rho_*(\epsilon,\eb) c_- \leq \frac{1}{4\delta\db} \int_{h-\delta}^{h+\delta}\int_{\hb-\db}^{\hb+\db} \text{d}F(h,\hb') \leq e^{2\pi c \big(\sqrt{\frac{\epsilon}{6}}\delta + \sqrt{\frac{\eb}{6}} \db\big)}\rho_*(\epsilon,\eb) \tilde{c}_+.	
\end{equation}

As a consequence,  we find that for fixed $\delta,\db > \delta_{gap}$,

\begin{equation}
	 	S_{h,\hb}(\delta,\db)
	 	 = 2\pi \sqrt{\frac{c}{6}\bigg(h-\frac{c}{24}\bigg)} + 2\pi \sqrt{\frac{c}{6}\bigg(\hb-\frac{c}{24}\bigg)} - \log c+ O(1),  \,\, c\rightarrow \infty\,.
\end{equation}
We can extend the above result to the case where $\delta,\db \sim c^\alpha$ where $0 < \alpha < 1$ by splitting the integral domain into squares of unit area:
\begin{equation}
	\int_{h-\delta}^{h+\delta} \int_{\hb-\db}^{\hb+\db} \text{d}h' \ \text{d}\hb' \rho(h',\hb') = \sum_{m=1}^{2\delta}\sum_{n=1}^{2\db} \int_{h-\delta + m - 1}^{h-\delta + m} \int_{\hb - \db + n - 1}^{\hb - \db + n} \text{d}h' \ \text{d}\hb' \rho(h',\hb')\,,
\end{equation}

and then, using the previous bound, we find

\begin{equation}\label{resultlargec}
	S_{h,\hb}(\delta,\db) = 2\pi \sqrt{\frac{c}{6}\bigg(h+\delta- \frac{c}{24}\bigg)} + 2\pi \sqrt{\frac{c}{6}\bigg(\hb+\db-\frac{c}{24}\bigg)} - \log c + O(1), \,\,\, c \rightarrow \infty.	
\end{equation}

Now we show that the vacuum contribution dominants the contribution from the light sector. This is straightforward for the sector where $h_{L}<c/24$ and $\hb_{L}<c/24$. For this region, we consider the sum

\begin{equation}
	A = e^{\beta h + \bb \hb} \sum_{h_L,\hb_L \leq c/24} \int \text{d}h' \ \text{d}\hb' \rho_{h_L,\hb_L}(h',\hb') e^{-\beta h' -\bb \hb'}\Phi_+(h',\hb')\,,	
\end{equation}
where the crossing kernel of the operator with conformal dimension $(h_L,\hb_L)$ is given by
\begin{equation}
	\begin{aligned}
	&\rho_{h_L}(h) = 2\pi \sqrt{\frac{\frac{c}{24}-h_L}{h-\frac{c}{24}}} I_1\bigg(4\pi \sqrt{\big(\frac{c}{24}-h_L\big)\big(h-\frac{c}{24}\big)}\bigg)\theta\left(h-\frac{c}{24}\right) + \delta\left(h-\frac{c}{24}\right), \\
		& \rho_{h_L,\hb_L}(h,\hb) = \rho_{h_L}(h)\rho_{\hb_L}(\hb),
	\end{aligned}
\end{equation}
and the above reproduces the contribution of this operator in the dual channel i.e. at high temperature:
\begin{equation}
	\int dh d\hb\ \rho_{h_L,\hb_L}(h,\hb)\ e^{-\beta h - \bb \hb} = e^{-\frac{4\pi^2}{\beta}(h_L - c/24) - \frac{4\pi^2}{\bb}(\hb_L - c/24)}.
\end{equation}
The asymptotic of the function $\rho_{h_L}(h)$ is given by
\begin{equation}
	\rho_{h_L}(\epsilon)\sim \frac{1}{2}\frac{1}{6^{1/4}}c^{-1/2}\epsilon^{-3/4}\bigg(1-\frac{24h_L}{c}\bigg)^{1/4} e^{2\pi c\sqrt{\frac{\epsilon}{6}\big(1-\frac{24h_L}{c}\big)}}.
\end{equation}
Evaluating the each integral by saddle point approximation, we find 
\begin{equation}
	A = O\bigg(c^{-1}e^{2\pi c\sqrt{\frac{\epsilon}{6}} + 2\pi c\sqrt{\frac{\eb}{6}}}\sum_{h_L+\hb_L\leq \Delta_H}e^{-4\pi\sqrt{6\epsilon}h_L - 4\pi\sqrt{6\eb}\hb_L}\bigg)\,.	
\end{equation}
Subsequently, using the following sparseness condition
\begin{equation}
	\sum_{h_L + \hb_L \leq \Delta_H} e^{-\beta h_L - \bb \hb_L} = O(1), \,\,\, \beta,\bb > 2\pi, \,\,\, c \rightarrow \infty,	
\end{equation}
we find that
\begin{equation}
	A \sim O(\rho_*(h,\hb)).	
\end{equation}
where the condition $\beta,\bb > 2\pi$ translates to $\epsilon,\bar{\epsilon}>1/24$. Below we will see that we need to have a more stronger condition i.e $\epsilon,\bar{\epsilon}>1/6$.

{
We are left to investigate the region where either of the $h_L$ or $\hb_{L}$ is greater than $c/24$. This is basically given by the brown region in fig.~\ref{brown}. 
\begin{figure}[!ht]
\centering
\includegraphics[scale=0.65]{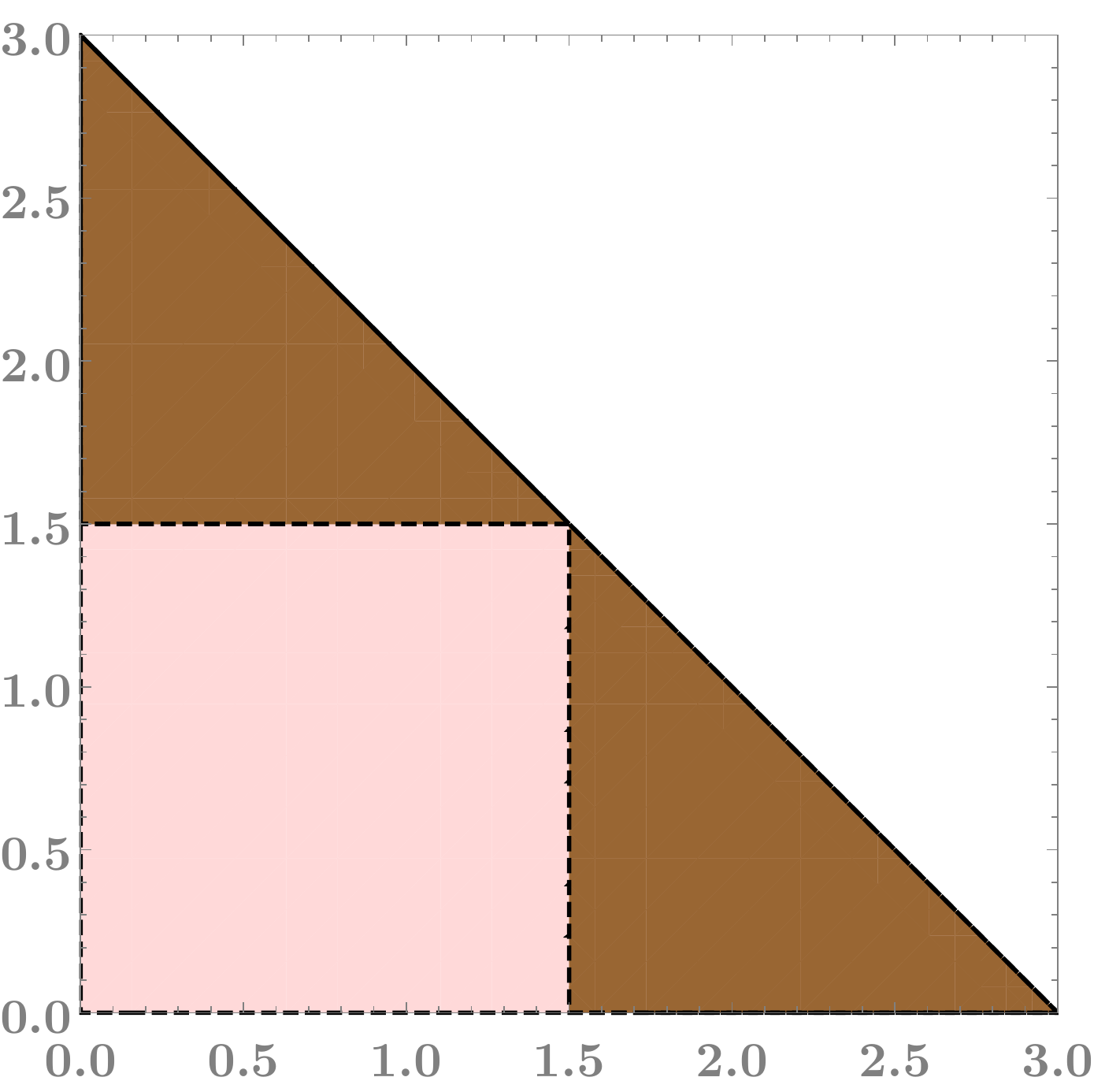}
\caption{The light region is the union of the region colored with pink and the region colored with brown. The brown regions requires a more careful treatment.}
\label{brown}
\end{figure}
Without loss of generality, let us look at the region where $h_L>c/24$. Now the crossing would be given by 
\begin{equation}
	\begin{aligned}
	&\rho_{h_L}(h) = 2\pi \sqrt{-\left(\frac{h_L-\frac{c}{24}}{h-\frac{c}{24}}\right)} I_1\bigg(4\pi\sqrt{-\left(h_L-\frac{c}{24}\right)\left(h-\frac{c}{24}\right)}\bigg)\theta\left(h-\frac{c}{24}\right) + \delta\left(h-\frac{c}{24}\right), \\
		& \rho_{h_L,\hb_L}(h,\hb) = \rho_{h_L}(h)\rho_{\hb_L}(\hb),
	\end{aligned}
\end{equation}
We are going to estimate the following:
\begin{equation}
\begin{aligned}
B&=e^{\beta h+\bb\hb}\int \text{d}h' \ \text{d}\hb'\ \rho_{h_L,\hb_{L}} e^{-\beta h'-\bb\hb'} \Phi_{+}(h',\hb')\\
&=e^{\beta (h-c/24)+\bb(\hb-c/24)}\int \text{d}h' \ \text{d}\hb'\ \rho_{h_L,\hb_{L}} e^{-\beta (h'-c/24)-\bb(\hb'-c/24)} \Phi_{+}(h',\hb')\,.
\end{aligned}
\end{equation}

The integral over $\hb'$ proceeds in usual manner since $\hb_{L}<c/24$. The integral over $\hb'$ proceeds in usual manner since $\hb_{L}<c/24$. The integral over $h'$ requires bit of care. 
Let us focus on 
\begin{equation}\label{analysishl}
\begin{aligned}
&\bigg|e^{\beta(h-c/24)} \int\ \text{d}h' \rho_{h_L}(h') e^{-\beta (h'-c/24)}\Phi_{+}(h',\bar h')\bigg| \\
&=\bigg|e^{\beta(h-c/24)} \int_{c/24}^{\infty}\ \text{d}h' \rho_{h_L}(h') e^{-\beta (h'-c/24)}\Phi_{+}(h',\bar h')\bigg| \\
&=e^{\beta(h-c/24)} \int_{0}^{\infty}\ dE' \bigg|\rho_{h_L}(h')e^{-\beta E'} \Phi_{+}(h',\bar h')\bigg| \\
& \leq  Me^{\pi c\sqrt{\frac{\epsilon}{6}}} \int\ dE' \frac{1}{\sqrt{E'}}e^{-\beta E'}\\
& \leq M\sqrt{\left(h_L-\frac{c}{24}\right)}  \exp\left[\pi c\sqrt{\frac{\epsilon}{6}}\right]\\
&\underset{c\to\infty}{\leq} M\sqrt{\frac{1}{c}} \exp\left[2\pi c\sqrt{\frac{\epsilon}{6}}\right]\exp\left[-2\pi\sqrt{6\epsilon}h_L\right]\,,
\end{aligned}
\end{equation}
where in the second line we have used the $\Theta$ function present in the expression for $\rho_{h_L}$. In the penultimate line, we have used $h_L<c/12$. The strict inequality is important to make sure replacing $\sqrt{\left(h_L-\frac{c}{24}\right)}$ with $\frac{1}{\sqrt{c}}$ does not spoil the inequality because of the presence of exponential term. Now for the rest of the argument, we will require that 
\begin{equation}
\begin{aligned}
&\text{min}\left(4\epsilon,\bar{\epsilon}\right)>\frac{1}{6}\ \,\&\,\ \text{min}\left(\epsilon,4\bar{\epsilon}\right)>\frac{1}{6}\, \Rightarrow \epsilon,\bar{\epsilon} > \frac{1}{6}\,.
\end{aligned}
\end{equation}
The above leads to \eqref{epsilonvalidity}. Now we sum over all the light states in the brown region to obtain
\begin{equation}
\begin{aligned}
B&=O\left(\frac{e^{2\pi c\sqrt{\frac{\epsilon}{6}}+2\pi c\sqrt{\frac{\bar \epsilon}{6}}}}{c}\left[ \sum_{h_L>\hb_L} e^{-2\pi\sqrt{6\epsilon}h_L-4\pi\sqrt{6\bar\epsilon}\hb_L}+\sum_{h_L\leq \hb_L} e^{-4\pi\sqrt{6\epsilon}h_L-2\pi\sqrt{6\bar\epsilon}\hb_L} \right]\right)\\
&=O\left(\rho_*(h,\hb)\right)\,.
\end{aligned}
\end{equation}

\paragraph{Twistgap:} If we assume a finite twist gap $g$, the \eqref{analysishl} can be revisited in the light of the twist gap:
\begin{equation}
\begin{aligned}
&\bigg|e^{\beta(h-c/24)} \int\ \text{d}h' \rho_{h_L}(h') e^{-\beta (h'-c/24)}\Phi_{+}(h',\bar h')\bigg| \\
&=\bigg|e^{\beta(h-c/24)} \int_{c/24}^{\infty}\ \text{d}h' \rho_{h_L}(h') e^{-\beta (h'-c/24)}\Phi_{+}(h',\bar h')\bigg| \\
&=e^{\beta(h-c/24)} \int_{0}^{\infty}\ dE' \bigg|\rho_{h_L}(h')e^{-\beta E'} \Phi_{+}(h',\bar h')\bigg| \\
& \leq  Me^{\pi c\sqrt{\frac{\epsilon}{6}}} \int\ dE' \frac{1}{\sqrt{E'}}e^{-\beta E'}\\
& \leq M\sqrt{\left(h_L-\frac{c}{24}\right)}  \exp\left[\pi c\sqrt{\frac{\epsilon}{6}}\right]\\
&\underset{c\to\infty}{\leq} M\sqrt{\frac{1}{c}} \exp\left[2\pi c\sqrt{\frac{\epsilon}{6}}\right]\exp\left[-2\left(1-\frac{6g}{c}\right)^{-1}\pi\sqrt{6\epsilon}h_L\right]\,,
\end{aligned}
\end{equation}
where now in the penultimate step we use $h_{L}\leq c/12-g/2$. And this leads to the modified validity regime as given in \eqref{modifiedvalidity}.
}

\section{Open problems}
We end with a list of open problems which would be nice to figure out:
\begin{enumerate}
\item One can hope to use these techniques to investigate the asymptotic OPE coefficients \cite{KM,dattadaspal,dyer,Collier:2018exn,Maloney,Cardy:2017qhl,Das:2017cnv,Brehm:2018ipf,Hikida:2018khg,Romero-Bermudez:2018dim,Michel:2019vnk} and make it spin sensitive. A richer structure in such scenarios is expected as well. 
 
\item For the large spin, finite twist, we have derived an upper bound on the windowed entropy. It would be nice to put a lower bound as well and match up to the results of \cite{Kusuki:2018wpa,Kusuki:2019gjs,Collier:2018exn,Maxfield:2019hdt,Benjamin:2019stq}. On a conservative note, we remark that even if one can prove a lower bound, it seems hard to distinguish the order one error coming from considering a bin of width $2\delta$ in spin and the dependence of the extended Cardy formula on finite twist. At present, all our attempts to prove a lower bound provided us with a Cardy like growth multiplied by a negative number, which is trivially true, hence we omitted the details of the trivial lower bound in the text. Furthermore, it would be nice to find out the asymptotic formula for the integrated version (integrated from spin $0$ upto some large spin) of the density of states.
\item It would be nice to have an expression for integrated density of states upto some particular $\Delta$ within a specified range of spin. To be specific, we want to have an estimate of the following quantity:
\begin{equation}
\begin{aligned}
\int_{0}^{\Delta} \text{d}\Delta^\prime\ \int_{J_1}^{J_2} \text{d}J^\prime\ \rho(\Delta^\prime,J^\prime) \,,\quad \text{where}\ J_1, J_2\in \left[-\Delta^\prime,\Delta^\prime\right]\,.
\end{aligned}
\end{equation}
\item It would be nice to improve on the value of $r$ and possibly prove that $r=1$ (the parameter appearing in the asymptotic ``areal" spectral gap) either by some suitable choice of magic functions or by better estimate of the heavy sector of the partition function. The naive generalization from  \cite{Ganguly:2019ksp} would not suffice. So one needs to be more creative. And this might shed light on the twist gap and provide a way to expound on the proposed gap in \cite{Benjamin:2019stq}.
\end{enumerate}

Our work should be thought of a part of modular bootstrap program\cite{Hellerman:2009bu,Friedan:2013cba,Collier:2016cls,Cho:2017fzo,Anous:2018hjh,Afkhami-Jeddi:2019zci,Afkhami-Jeddi:2017idc,Cardy:2017qhl,KM,dattadaspal,Das:2017cnv,Brehm:2018ipf,Kraus:2018pax,Ashrafi:2019ebi,Brehm:2019pcx,Alday:2019vdr}. On a more general ground, it would be interesting to see whether Tauberian theorems and/or Modular bootstrap program can say anything about the chaotic, irrational CFTs. An approach borrowing ideas from Tauberian techniques and that of extremal functionals appearing in \cite{Mazac:2016qev,Mazac:2018ycv,Mazac:2018mdx,Hartman:2019pcd,Mazac:2019shk,Paulos:2019gtx,Carmi:2019cub} might be useful in this regard. Furthermore, for holographic CFTs, we can only achieve a reduced regime of validity of Cardy formula compared to what is reported in \cite{HKS}. It might be possible to improve our result. Albeit, we remark that if the twist gap is greater than $c/12$, it is possible to achieve  the regime of validity of Cardy formula as predicted in \cite{HKS}. We hope to come back to these problems in future.

\section*{Acknowledgements}
The authors acknowledge Ken Intriligator and John McGreevy for fruitful discussions and encouragement. SP thanks Nathan Benjamin, Diptarka Das, Baur Mukhametzhanov, Shu-Heng Shao for several illuminating discussions. SP thanks Raghu Mahajan, Ben Michael, Baur Mukhametzhanov for several suggestions regarding improving readability of the draft. SP thanks Simons Center for Geometry and Physics, where a part of this work was presented, for hospitality, and Simeon Hellerman, Zohar Komargodski, Dalimil Mazac for discussions that followed the talk. This work was in part supported by the US Department of Energy (DOE) under cooperative research agreement DE-SC0009919 and Simons Foundation award  \#568420. SP also acknowledges the support from Inamori Fellowship, Ambrose Monell Foundation and DOE grant DE-SC0009988.

{\bibliographystyle{bibstyle2017}
\bibliography{refs}
\hypersetup{urlcolor=RoyalBlue!60!black}
}

\end{document}